\documentclass[twocolumn, superscriptaddress, preprintnumbers,amsmath,amssymb,aps]{revtex4}

\allowdisplaybreaks
\allowdisplaybreaks[4]
\usepackage{CJK}
\usepackage{graphicx}
\usepackage{dcolumn}
\usepackage{bm}
\usepackage{mathptmx}
\usepackage[
            pdfstartview=FitH,
            CJKbookmarks=true,
            bookmarksnumbered=true,
            bookmarksopen=true,
            colorlinks,
            linkcolor=blue,
            anchorcolor=blue,
            citecolor=blue
            ]{hyperref}

\begin{document}

\begin{CJK}{GBK}{song}

\title{Entanglement and coherence of the wobbling mode}

\author{Q. B. Chen}
\email[Corresponding author:~]{qbchen@phy.ecnu.edu.cn}
\affiliation{Department of Physics, East China Normal University,
Shanghai 200241, China}

\author{S. Frauendorf}
\email[Corresponding author:~]{sfrauend@nd.edu}
\affiliation{Physics Department, University of Notre Dame, Notre
Dame, IN 46556, USA}

\date{\today}

\begin{abstract}

The entanglement and coherence of the wobbling mode are studied in the framework of
the particle plus triaxial rotor model for the one-quasiparticle nucleus $^{135}$Pr
and the two-quasiparticles nucleus $^{130}$Ba. The focus lies on the coupling between
the total and the particle angular momenta. Using the Schmidt decomposing, it is
quantified in terms of the von Neumann entropy of the respective sub-systems, which
measures their mutual entanglement. The entropy and the entanglement increase with
spin $I$ and number of wobbling quanta $n$. The coherence of the wobbling mode is
studied by means of the eigenstate decomposition of its reduced density matrix. To
a good approximation, the probability distributions of the total angular momentum
can be interpreted as the incoherent combination of the coherent contributions from
the first two pairs of eigenvectors with the largest weight of the reduced density
matrix. Decoherence measures are defined, which, in accordance, scatter between
0.1 to 0.2 at low spin and between 0.1 and 0.3 at high spin. Entanglement in the
framework of the adiabatic approximation is further analyzed. In general, the coherent
eigenstates of the effective collective Hamiltonian approximate the reduced density
matrix with the limited accuracy of its pair of eigenstates with the largest weight.
As the adiabatic approximation becomes more accurate with decreasing
excitation energy, the probability distribution of the angle of the total
angular momentum around a principal axis approaches the one of the
full reduced density matrix. The  $E2$ transition probabilities and
spectroscopic quadrupole moments reflect this trend.

\end{abstract}

\maketitle


\section{Introduction}
\label{intro}

The wobbling motion is a distinctive rotational mode indicating triaxiality
within the nucleus~\cite{Bohr1975}. A triaxial rotor prefers to rotate
about the axis with the largest moment of inertia. When slightly excited,
the rotational axis deviates from this principal axis, causing precession
oscillations around the fixed angular momentum vector and
resulting in the observed wobbling motion. This wobbling
energy, characterized by the wobbling frequency, increases
with the nuclear spin~\cite{Bohr1975}.

In the study of triaxial rotors coupled with high-$j$ quasiparticles,
Frauendorf and D\"{o}nau~\cite{Frauendorf2014PRC} categorized wobbling
motions into two types: longitudinal wobbling (LW) and transverse
wobbling (TW). LW occurs when the quasiparticle is mid-shell occupied,
with its angular momentum $\bm{j}$ parallel to the principal axis of largest
moment of inertia. TW, on the other hand, happens when the quasiparticle
is bottom or top occupied, with its angular momentum $\bm{j}$ perpendicular
to this axis. Both types show enhanced $I \to I-1$ electric
quadrupole ($E2$) transitions between wobbling bands, characterized
by rotational $E2$ bands corresponding to different oscillation
quanta ($n$). LW's wobbling frequency increases with spin,
resembling the originally predicted wobbler behavior, while
TW's wobbling frequency decreases with spin.
Later on, Chen and Frauendorf~\cite{Q.B.Chen2022EPJA} proposed a
comprehensive classification of wobbling motion using spin coherent
state (SCS) maps, which show the probability distribution for the
orientation of angular momentum on the unit sphere projected onto
the polar angle ($\theta$) and azimuthal angle ($\phi$) plane.
In this scheme, LW involves the total angular momentum $\bm{J}$
revolving around the axis with the largest moment of inertia,
while TW involves $\bm{J}$ revolving around an axis perpendicular
to it. This classification is further validated by spin squeezed
state (SSS) plots, linking discrete $\bm{J}$-space representation
with the continuous coordinate $\phi$'s wave
function~\cite{Q.B.Chen2024PRC_v1}.

Experimental evidence of wobbling motion was first reported in
triaxial strongly deformed nucleus $^{163}\textrm{Lu}$ in
2001~\cite{Odegaard2001PRL}, later interpreted as TW~\cite{Frauendorf2014PRC}.
The TW was also identified in the normally deformed nucleus
$^{135}\textrm{Pr}$~\cite{Matta2015PRL}. The LW was observed
in $^{187}\textrm{Au}$~\cite{Sensharma2020PRL}. As of now,
wobbling candidate bands have been detected in over 15 nuclei
across various mass regions~\cite{R.J.Guo2024PRL, Timar2019PRL,
Matta2015PRL, Sensharma2019PLB, Biswas2019EPJA, Petrache2019PLB,
Q.B.Chen2019PRC_v1, Chakraborty2020PLB, Devi2021PLB,
B.F.Lv2022PRC, Prajapati2024PRC, Odegaard2001PRL, Jensen2002PRL,
Bringel2005EPJA, Schonwasser2003PLB, Amro2003PLB, Hartley2009PRC,
Mukherjee2023PRC, Sensharma2020PRL, Nandi2020PRL}. A recent review is
available in Refs.~\cite{Frauendorf2024arXiv, B.W.Sun2024NPR}.

Wobbling motion was initially predicted for even-even nuclei using the
triaxial rotor model (TRM)~\cite{Bohr1975}. After experimental confirmation
in odd-$A$ nuclei~\cite{Odegaard2001PRL}, the particle triaxial rotor
(PTR) model became the primary framework for describing wobbling motion,
including energy spectra and electromagnetic transitions~\cite{Hamamoto2002PRC,
Hamamoto2003PRC, Frauendorf2014PRC, Streck2018PRC, Q.B.Chen2019PRC_v1,
Q.B.Chen2020PLB_v1, Broocks2021EPJA, L.Hu2021PRC, Q.B.Chen2022EPJA,
W.C.Li2022EPJA, H.M.Dai2023PRC, S.H.Li2024CPC, Q.B.Chen2024PRC_v1,
H.M.Dai2024PRC}. Additionally, several approximate methods
have also been employed to study wobbling motion~\cite{Tanabe2017PRC,
Raduta2017PRC, Budaca2018PRC, Lawrie2020PRC, Budaca2021PRC, Raduta2021JPG,
Raduta2022JPG, Budaca2022PRC_v1}. Different explanations for transverse
wobbling motion have been proposed within these approaches~\cite{Tanabe2017PRC,
Lawrie2020PRC}. Moreover, the cranking model plus random phase
approximation (RPA)~\cite{Marshalek1979NPA, Shimizu1995NPA, Matsuzaki2002PRC,
Matsuzaki2004PRC, Matsuzaki2004PRCa, Matsuzaki2004EPJA, Shimizu2004arXiv,
Shimizu2005PRC, Almehed2006Phys.Scr., Shimizu2008PRC,
Shoji2009PTP, Frauendorf2015PRC} or collective
Hamiltonian~\cite{Q.B.Chen2014PRC, Q.B.Chen2016PRC_v1}, as
well as the triaxial projected shell model~\cite{Shimada2018PRC,
Y.K.Wang2020PLB, F.Q.Chen2021PRC} have been utilized to
investigate and explain wobbling motion.

Within the framework of the PTR model, the phenomena of TW
and LW have been interpreted as the interplay of one (or more)
high-$j$ quasiparticle, which act as gyroscopes, and the remaining
nucleons, which are modeled as a triaxial rotor. The PTR model describes
this composite system using bases with good particle and total angular
momentum and taking full account of the coupling between particle and rotor
subsystems. In this work we will study the topological classification scheme
suggested in Refs.~\cite{Frauendorf2014PRC, Q.B.Chen2022EPJA, Q.B.Chen2024PRC_v1}
from the fundamental perspective of composite quantum systems. We will address
the entanglement of the particle-rotor subsystems and the resulting limits
of the interpretations based on the individual subsystems. Compared to
other approaches, PTR is best suited for such a study because
it is a simple two-component system. Previous studies have shown
that the TW becomes unstable with increasing angular momentum~\cite{Frauendorf2014PRC,
Matta2015PRL, Streck2018PRC, Q.B.Chen2022EPJA, S.H.Li2024CPC, Q.B.Chen2024PRC_v1,
Frauendorf2024arXiv, H.M.Dai2024PRC}. It changes into LW via a transition
region. The variation is caused by the strong coupling between the angular
momenta of the particles and the rotor. These are examples of
strong entanglement between the particle and core angular momenta,
well suited  to quantify the entanglement and the resulting information
loss in the case of wobbling motion.

Entanglement is a fundamental concept in quantum mechanics, which
characterizes the correlations between particles or partitions
within a composite system that cannot be described in terms of
independent subsystems. Quantum many-body systems show specific
signatures of entanglement, which are of great interest for condensed
matter physics and quantum field theory~\cite{Calabrese2004JSM,
Amico2008RMP, Peschel2009JPA, Horodecki2009RMP, Nishioka2009JPA,
Eisert2010RMP, Lin2020NPB}. Recent advances in quantum
information science and quantum computing have renewed interest
in exploring entanglement in nuclear systems~\cite{Enyo2015PRC, Legeza2015PRC,
Enyo2015PTEP, Enyo2015PTEP_v1, Beane2019PRL, Robin2021PRC, Faba2021PRA, Kruppa2022PRC,
Pazy2023PRC, D.Bai2022PRC, Lacroix2022PRD, Tichai2023PLB, Bulgac2023PRC,
Bulgac2023PRC_v1, Johnson2023JPG, C.Y.Gu2023PRC}.

The reduced density matrices for the total and particle angular
momenta are the centerpieces of the tools introduced in
Refs.~\cite{Frauendorf2014PRC, Q.B.Chen2022EPJA, Q.B.Chen2024PRC_v1}
in order to elucidate the physics of the PTR system.
The concept of the density matrix was introduced by
John von Neumann~\cite{Neumann1927GN} in order to establish a sound
basis for the measuring process in quantum mechanics and by Lev Landau
to extend the concept of Gibbs entropy from classical
to quantum statistical mechanics~\cite{Landau1927ZP}.
There are several entanglement measures
based on the density matrix, which are commonly
employed to quantify many-body correlations in quantum many-body
systems. One such measure is the entropy of entanglement, also
von Neumann  (vN) entropy. This measure quantifies the
extent of quantum entanglement between two subsystems within
a composite quantum system. The vN entropy has been extensively employed
in studies related to entanglement in the
fields of condensed matter physics and quantum field
theory~\cite{Calabrese2004JSM, Amico2008RMP, Peschel2009JPA,
Horodecki2009RMP, Nishioka2009JPA,  Eisert2010RMP, Lin2020NPB} and
the atomic nuclei~\cite{Enyo2015PRC, Legeza2015PRC, Enyo2015PTEP,
Enyo2015PTEP_v1, Beane2019PRL, Robin2021PRC, Faba2021PRA, Kruppa2022PRC,
Pazy2023PRC, D.Bai2022PRC, Lacroix2022PRD, Tichai2023PLB, Bulgac2023PRC,
Bulgac2023PRC_v1, Johnson2023JPG, C.Y.Gu2023PRC}.
In our study of entanglement in the wobbling motion, we will
utilize the vN entropy to gain insights into the entanglement
properties of the wobbling motion.

Spatial coherence is another important characteristics of entangled quantum
systems~\cite{Born1999book}. It measures the capability of waves propagating
in a medium to generate interference pattern, which appear as a consequence
of a definite relation between of the phase of the wave
and the location in space~\cite{Hecht1998book}. A familiar example is
the two-slit experiment monochromatic light. The two-dimensional waves
behind the slits generate the interference pattern. For infinite narrow slits
the interference fringes extend to infinity. In real systems they
become weaker with the distance from the slits and eventually disappear
over a distance called the coherence length. The reason for the damping is
dephasing, which is the gradual loss of the definite relation
between location and phase of the wave. There are various reasons for dephasing.
The finite width of the slits is one of them. An incoherent mixture of
different wave lengths is another. In analogy, the quantum wave function defines
a unique phase difference between two points in space. A quantum system
is called coherent when it can be described by wave function. Its
``pure" density matrix is the product of the wave function with its complex conjugate.
The general case are the ``mixed" systems which are ensembles of such pure systems
with different wave lengths such that the coherence
is partially lost. We will address the loss of coherence in
the particle-rotor system.

It has to be stressed that, although the entanglement causes the loss of coherence,
the degree of it depends on the specific system. For example,
the coupling of the electromagnetic field of the light
 wave with the degrees of freedom of the medium causes dephasing,
the degree of which depends not only on the strength of the entanglement but
also on the time scales of the involved degrees of freedom. Optical interference
in glass with a high refraction index is an example for good coherence and substantial
entanglement of the electric field with the atomic degrees of freedom of the glass.
The phenomena can be very well described by the Maxwell equations in matter,
which is an effective theory where the entanglement appears only in form of the
permittivity and permeability of the material. In analogy, we address the question
to what extend can the particle-rotor system understood in terms of an effective
Hamiltonian in the orientation degrees of freedom of the total angular momentum only,
so to speak a quasi-rotor. Such an approach is used in molecular physics to account
for the coupling between the rotational and vibrational degrees of freedom, where
effective rotor Hamiltonians with higher than two powers of the total angular
momentum are introduced.

In this paper, we will investigate the entanglement entropy and coherence
properties of the PTR model, using the nuclei $^{135}$Pr studied
in Refs.~\cite{Frauendorf2014PRC, Matta2015PRL, Streck2018PRC,
Sensharma2019PLB, Q.B.Chen2022EPJA, Q.B.Chen2024PRC_v1} and
$^{130}$Ba studied in Refs.~\cite{Petrache2019PLB, Q.B.Chen2019PRC_v1,
Q.B.Chen2024PRC_v1} as examples for one- and two-quasiparticles
coupled to a triaxial rotor.


\section{Theoretical framework}

\subsection{Particle triaxial rotor model}\label{sec:plots}

The PTR model is utilized to describe the coupling of a high-$j$
particle to a triaxial rotor core. The corresponding Hamiltonian
can be expressed as~\cite{Bohr1975}
\begin{align}\label{eq:HPTR}
 H_{\textrm{PTR}}
  &=\hat{h}_{p}(\gamma)
  +\sum\limits_{i=1,2,3}\frac{(\hat J_i-\hat j_i)^2}{2{\cal J}_i},\\
 \label{eq:hproton}
 \hat{h}_p(\gamma)&=\kappa\left[\left(3\hat{j}_3^2-\bm{j}^2\right)\cos\gamma
     +\sqrt3\left(\hat{j}_1^2-\hat{j}_2^2\right)\sin\gamma\right],
\end{align}
Here, $\hat{J}_i=\hat{R}_i+\hat{j}_i$ represents the total angular
momentum, where $\hat{j}_i$ corresponds to the angular momentum
of the particle and $\hat{R}_i$ represents the angular momentum
of the triaxial rotor. The ${\cal J}_i$ denotes the
moment of inertia along the $i$ axis, which is dependent on
the deformation parameters $\beta$ and $\gamma$. The single-particle
Hamiltonian $h_p(\gamma)$ takes into account the coupling strength to the
deformed potential, and is expressed as a function of $\gamma$,
with $\kappa$ representing the coupling strength constant.

The PTR Hamiltonian (\ref{eq:HPTR}) can be decomposed into four parts
\begin{align}\label{eq:HPTR1}
 \hat{H}_{\textrm{PTR}}=\hat{h}_{p}(\gamma)
  +\hat{H}_{\textrm{rot}}+\hat{H}_{\textrm{rec}}+\hat{H}_{\textrm{cor}},
\end{align}
with the pure rotational operator of the rotor
\begin{align}
 \hat{H}_{\textrm{rot}}=\sum_{i=1,2,3}\frac{\hat J_i^2}{2{\cal J}_i},
\end{align}
the recoil term
\begin{align}
 \hat{H}_{\textrm{rec}}=\sum_{i=1,2,3}\frac{\hat{j}_i^2}{2{\cal J}_i},
\end{align}
and the Coriolis interaction term
\begin{align}\label{eq:Cor}
 \hat{H}_{\textrm{cor}}=-\sum_{i=1,2,3}\frac{\hat{J}_i \hat{j}_i}{{\cal J}_i},
\end{align}
of which $H_{\textrm{rot}}$ acts only on the degrees of freedom of the rotor and
$H_{\textrm{rec}}$ acts in the coordinates of the valence particle only, whereas
$H_{\textrm{cor}}$ couples the degrees of freedom of the rotor to the degrees of
freedom of the valence particle. Therefore, the entanglement between the rotor
and the valence particle is generated only by the $H_{\textrm{cor}}$.

The PTR Hamiltonian is diagonalized in the product basis
$\vert IIK\rangle\vert jk\rangle$, where $\vert IIK\rangle$
represents rotor states with half-integer $I$ and $\vert jk\rangle$
the high-$j$ particle states in good spin $j$ approximation.
The eigenstates of the PTR Hamiltonian are expressed as
\begin{equation}
 |II, \nu\rangle=\sum_{K,k} C_{IKk}^{(I\nu)} \vert IIK\rangle\otimes\vert jk\rangle,
\end{equation}
in terms of the coefficients $C_{IKk}^{(I\nu)}$. Here, $K$ and
$k$ run respectively from $-I$ to $I$ and from $-j$ to $j$.
The coefficients are  restricted by the requirement that collective
rotor states must be symmetric representations of the D$_2$ point group.
This implies that the difference $K-k$ must be even and one-half of all coefficients
is fixed by the symmetric relation
\begin{align}\label{eq:D2symmetry}
C_{I-K-k}^{(I\nu)}=(-1)^{I-j}C_{IKk}^{(I\nu)}.
\end{align}

The generalization to the two-quasiparticle triaxial rotor model is
straightforward,
\begin{equation}
 |II, \nu\rangle=\sum_{K,k_1,k_2} C_{IKk_1k_2}^{(I\nu)}
 \vert IIK\rangle\otimes\vert jk_1\rangle
 \otimes\vert jk_2\rangle,
\end{equation}
where $\vert IIK\rangle$ represents rotor states with integer $I$.
Here, both $k_1$ and $k_2$ run from $-j$ to $j$. The D$_2$ symmetry
implies that the difference $K-k_1-k_2$ must be even and one-half of all
coefficients is fixed by the symmetric relation
\begin{align}\label{eq:D2symmetry2}
C_{I-K-k_1-k_2}^{(I\nu)}=(-1)^{I-2j}C_{IKk_1k_2}^{(I\nu)},
\end{align}
and the Pauli exclusion principle
\begin{align}
 C_{IKk_1k_2}^{(I\nu)}=-C_{IKk_2k_1}^{(I\nu)}.
\end{align}

From the amplitudes of the eigenstates $C_{IKk}^{(I\nu)}$, we can
calculate the reduced density matrices, as described in
Refs.~\cite{Q.B.Chen2022EPJA, Q.B.Chen2024PRC_v1}.
The reduced density matrices provide valuable information about
the particle angular momentum $\bm{j}$ and the total angular
momentum $\bm{J}$. Specifically, we have the reduced density
matrices for the particle angular momentum states $|j\rangle$
\begin{align}\label{eq:TWrhoj}
 \rho_{kk'}^{(I\nu)}=\sum_K C_{IKk}^{(I\nu)}C_{IKk'}^{(I\nu)*},
\end{align}
and for the total angular momentum states $|II\rangle$
\begin{align}\label{eq:TWrhoJ}
 \rho_{KK'}^{(I\nu)}=\sum_k C_{IKk}^{(I\nu)}C_{IK'k}^{(I\nu)*}.
\end{align}

Similarly, for the two-quasiparticle triaxial rotor model,
we can also calculate the reduced density matrix from the
eigenstates $C_{IKk_1k_2}^{(I\nu)}$ for the total particle
angular momentum states $|j_{12}\rangle =|j\rangle \otimes |j\rangle$
\begin{align}\label{eq:RDM2qp_proton}
 \rho_{k_{12}, k_{12}^\prime}^{(I\nu)}
  &=\sum_{j_{12}, K} C_{j_{12}, k_{12}, K}^{(I\nu)}
  C_{j_{12}, k_{12}^\prime, K}^{(I\nu)*},\\
  C_{j_{12}, k_{12}, K}^{(I\nu)}
  &=\sum_{k_1, k_2} \langle j k_1 j k_2|j_{12}k_{12}\rangle
  C_{IKk_1k_2}^{(I\nu)},
\end{align}
and for the total angular momentum states
$|II\rangle$
\begin{align}\label{eq:RDM2qp_total}
 \rho_{KK'}^{(I\nu)}=\sum_{k_1k_2}
  C_{IKk_1k_2}^{(I\nu)}C_{IK'k_1k_2}^{(I\nu)*}.
\end{align}

The reduced density matrices contain the information
about the distribution and correlations of angular momenta
within the system, which are the basis for interpreting the properties
and behavior of triaxial nuclei by means of the PTR
model~\cite{Q.B.Chen2022EPJA, Q.B.Chen2024PRC_v1}. They
describe ``what the particle and the total angular momentum
vectors are doing". The reduced density matrices represent
mixed states. In the following we analyse the mutual entanglement
of the particle and rotor subsystems, which  results in a partial loss
coherence in each partition.

\subsection{Schmidt decomposition}

The PTR model introduces a bipartition of the of the system into the two
subsystems of the orientation of the total angular momentum $\vert IIK\rangle$
and of the orientation of the particle $\vert jk\rangle$. The Hilbert
space of the PTR model is the direct product of the Hilbert spaces of
the two subsystems, $\mathcal{H}_{Kk}=\mathcal{H}_K\otimes\mathcal{H}_k$.
The Schmidt decomposition (equivalent to the singular value
decomposition of a matrix)~\cite{Nielsen2010book} diagonalizes
the two reduced density matrices in their respective subspaces,
\begin{align}\label{eq:SD-J}
    \rho_{KK'}=\sum_m p_m C^{(m)}_{IK}C^{(m)}_{IK'},
\end{align}
where $p_m$ are the eigenvalues and $C^{(m)}_{IK}$
are the normalized eigenvectors;
\begin{align}\label{eq:SD-j}
    \rho_{kk'}=\sum_m p_m C^{(m)}_{k}C^{(m)}_{k'},
\end{align}
where $p_m$ are the eigenvalues and $C^{(m)}_{k}$ are the normalized
eigenvectors. For the case that the dimension $d_j$ of the $\bm{j}$-space
is smaller than dimension $d_J$ of the $\bm{J}$-space, the
first $d_j$ eigenvalues  $p_m$ of $\rho_{KK'}$ are the same as the $p_m$
for $\rho_{kk'}$, and the remaining $p_m=0$. For $d_J<d_j$ it
is the other way around. The $p_m$ satisfy the normalization condition
\begin{align}
 \sum_m p_m=1.
\end{align}

If $p_1=1$ and $p_{m>1}=0$ the system is in a pure state. The
particle and the total angular momentum are described by
the eigenstates
\begin{align}
|II, 1\rangle=\sum_{K} C_{IK}^{(1)} \vert IIK\rangle,\quad
|j, 1\rangle=\sum_{k} C_{k}^{(1)} \vert jk\rangle.
\end{align}
The two subsystems are separate, the combined state is the product of
the substates. This corresponds to the frozen alignment (FA)
approximation introduced in Ref.~\cite{Frauendorf2014PRC}.

If some  $p_m\ne 1$ the two subsystems are no longer
independent. They are said to be entangled, where $p_m$ represents
the probability for the subsystems to be in one of the pure states
$|II, m\rangle$ and $|j, m\rangle$ and the total
system being in the product state of both. In the following we
discuss various measures that quantify the degree and
characteristics of the entanglement.

\subsection{Entropy}

The vN entropy \cite{Neumann1927GN} proved to be an appropriate
entanglement measure in quantum many-body systems (see,
for instance, Refs.~\cite{Amico2008RMP, Tichy2011JPB} and
references therein). It characterizes the lack of complete
information about a subsystem when the total system is a
known pure state. In our case the bipartition concerns
the total ($\bm{J}$) and proton ($\bm{j}$) angular momenta.

The vN entropy, also  called  ``entanglement entropy", is defined
by either of the reduced density matrices
\begin{align}\label{eq:eq1}
 S_{j(J)}=-\textrm{Tr}~\rho_{j(J)} \ln \rho_{j(J)}=-\sum_m p_m \ln p_m.
\end{align}
It quantifies the ignorance due to the correlation between
subsystems $\bm{j}$ and $\bm{J}$~\cite{Vedral1997PRL}. As it characterizes
the relation between the subsystems, it is the same for either
of them. In information theory the last expression in (\ref{eq:eq1})
is also called the Shannon entropy. They use the $\log$ in base 2, which
is the natural unit for the binary digit.

The entanglement entropy measures how correlated (``entangled") the
two sectors are. If there is no entanglement between the two subsystems,
we find $S=0$; and $S > 0$ if there is entanglement. Any subsystem eigenstate
which has $p_m = 0$ or $p_m = 1$ will not contribute to the entropy.

Without the Coriolis interaction, $\mathcal{H}_K$ and $\mathcal{H}_k$ are
decoupled. The entropy takes its minimum
\begin{align}
S^{(\textrm{min})}=
\left\{\begin{array}{ll}
  0, & \textrm{for even-even}, \\
   \ln{2}, &\textrm{for odd-}A,\\
 \ln{4}, &\textrm{for odd-odd}.
\end{array}
\right.
\end{align}
The reason for the difference is Kramer's degeneracy.
The single-particle Hamiltonian $\hat{h}_{p}(\gamma)$ couples only
states  that differ by $\Delta k=2$. For the even-even system
the eigenstates have either even-$k$ or odd-$k$. The energies of
the two sets are different, though not very. That is, the eigenstates
are pure and their entropy is $S^{(\textrm{min})}=0$.
In the case of odd-$A$, the eigenstates of $\hat{h}_{p}(\gamma)$
have $k=j+\textrm{even}$ (labeled as $|\varphi\rangle$) or
$k=-j+\textrm{even}$ (labeled as $|\bar{\varphi}\rangle$),
which have the same energy (Kramer's degeneracy).
Coupling matrix elements between the two classes will generate
superpositions $(|\varphi\rangle\pm|\bar{\varphi}\rangle)/\sqrt{2}$.
This has the consequence that the eigenvalues of the $\bm{j}$-density
matrix come in doublets. Approaching the limit of zero Coriolis
interaction, there are two eigenvalues of 1/2 and the rest is zero,
because any whatsoever small coupling will generate the superposition.
Thus we consider $S^{(\textrm{min})}=\ln 2$ as the minimal entropy,
which is also the result of the numerical calculations.

If the reduced $\bm{j}$-density matrix is given by the normalized
identity matrix $\rho_{kk'}=\delta_{kk'}/d_j$,
then the vN entropy reaches the maximal value
\begin{align}
 S_j^{(\textrm{max})}=\ln~d_j.
\end{align}
In this limit the subsystems $\bm{j}$ and $\bm{J}$ are maximally
entangled. Each state $\vert jk\rangle$ is realized with the same
probability $1/d_j$, that is, there is maximal randomness. The
SSS plot~\cite{Q.B.Chen2024PRC_v1} is the straight line $P(\phi)=1/2\pi$.

The entropy of the $\bm{J}$-subsystem is $\ln~d_j$ as well, because
the $p_m$ are the same or 0. However, in units of the maximal possible
$\bm{J}$-entropy (i.e., complete randomness), $S_J^{(\textrm{max})}=\ln~d_J$,
it is smaller than one ($\ln~d_j/\ln~d_J<1$). Accordingly,
the $\bm{J}$-density matrix (\ref{eq:SD-J}) with $p_m=1/d_j$ for $ m\leq d_j$
and 0 $ m> d_j$ is not diagonal in $K,~K'$. The pertaining SSS
plot~\cite{Q.B.Chen2024PRC_v1} shows quantum fluctuations.
The random motion of the particle cannot completely
randomize the total angular momentum orientation with respect to
the principal axes.

The limit of complete randomness of a small subsystem is  approached
by strongly coupling it to a large subsystem, which acts as a randomizing
heat bath. This scenario differs from the $\bm{j}$-$\bm{J}$ subsystems
of the PTR. As discussed below, $S$ stays well below its maximum
for all states.

\subsection{Purity}

Furthermore, we calculate the purity, which is a measure
on quantum states on how much a state is mixed.
It is defined as
\begin{align}\label{eq:eq2}
 \mathcal{P}=\textrm{Tr}~\rho^2=\sum_m p_m^2,
\end{align}
with $p_m$ being the eigenvalues of $\rho$. When it is 1,
the state is pure. Otherwise, it will be smaller than 1,
with the minimum $1/d_j$ for the particle or total angular momenta.
Because of Kramer's degeneracy, the maximal purity is 1/2 in
the case of odd-$A$.

\subsection{Coherence}

``Coherence" specifies to which extend the system can be described
by wave function with the characteristic interference phenomena.
If the system is pure, i.e., it is represented by only one wave function
in $K$-space, then one has
\begin{align}\label{eq:TWrhoJK}
 \left[\rho^{(I\nu)}\right]^2_{KK'}
 &=\sum\limits_{K''=-I}^I
 \rho^{(I\nu)}_{KK''}\rho^{(I\nu)}_{K''K'}\notag\\
 &=\sum\limits_{K''=-I}^I
 C_{IK}^{(I\nu)}C_{IK''}^{(I\nu)*} C_{IK''}^{(I\nu)} C_{IK'}^{(I\nu)*}\notag\\
 &=C_{IK}^{(I\nu)} C_{IK'}^{(I\nu)*}\notag\\
 &=\rho^{(I\nu)}_{KK'},
\end{align}
which indicates complete coherence. For mixed states, this relation no longer
holds. In order to measure the loss of coherence, it is natural to ask how the
square of the density matrix deviates from the matrix. One thus considers
the matrix square as a modified density matrix and asks how it deviates
from the original one. However, the trace of the squared matrix is smaller
than 1, where the reduction measures the purity. For a proper comparison,
the squared matrix must be normalized by dividing it by its trace.
We compare the $K$-probability distributions and use the absolute
deviation $\Delta_K$ as a measure for decoherence.  It turns out that
the deviations depend to some extend on the representation of
the density matrix. The corresponding formula are given as
\begin{align}\label{DK}
 P(K) &=\rho^{(I\nu)}_{KK}, \\
 P_2(K) &=\left[\rho^{(I\nu)}\right]^2_{KK}
 /\mathrm{Tr}\left\{\left[\rho^{(I\nu)}\right]^2\right\},\\
\label{eq:delta_K}
 \Delta_K &=\sum_{K=-I}^I \Big|P_2(K)-P(K)\Big|.
\end{align}
Here, we use the absolute value instead of the root-mean-squared of
the difference between the $P(K)$ and $P_2(K)$ in calculating
$\Delta_K$, because the latter does not fulfill the ``continuity
criterion" for a coherence measure~\cite{Baumgratz2014PRL}.

The analogue comparison can be based on the spin squeezed
state (SSS) plots~\cite{Q.B.Chen2024PRC_v1}. They represent the
$K$-space in terms of a continuous coordinate $\phi$, which
represents the angle between the projection of $\bm{J}$ onto
the short-medium ($sm$) plane with the short ($s$) axis of the triaxial
nuclear shape. Mathematically, one changes to the Fourier
representation of the density matrix
\begin{align}\label{eq:rhophiphi}
    \rho_{\phi, \phi'}^{(I\nu)} &= \frac{1}{2\pi}\sum\limits_{K, K'=-I}^I
 \rho^{(I\nu)}_{KK'}e^{-i(K \phi -K' \phi')}, \\
  \left[\rho^{(I\nu)}\right]^2_{\phi, \phi'} &= \frac{1}{2\pi}\int_{-\pi}^{\pi}d\phi''
 \rho_{\phi, \phi''}^{(I\nu)}\rho_{\phi'', \phi'}^{(I\nu)} \notag \\
 &= \frac{1}{2\pi}\sum_{K,K', K''=-I}^I
 \rho^{(I\nu)}_{KK''}\rho^{(I\nu)}_{K''K'}e^{-i(K \phi -K' \phi')}.
\end{align}
As a result, the decoherence is quantified by
\begin{align}\label{eq:DSSS}
    P(\phi)&=\rho_{\phi, \phi}^{(I\nu)}, \\
    P_2(\phi)&=\left[\rho^{(I\nu)}\right]^2_{\phi, \phi}
    /\mathrm{Tr}\left\{\left[\rho^{(I\nu)}\right]^2\right\},\\
\label{eq:delta_SSS}
    \Delta_{\textrm{SSS}}&=\int_{-\pi}^{\pi}d\phi~\Big|P_2(\phi)-P(\phi)\Big|.
\end{align}

The normalized $P_2$ distributions have the eigenvalues $p_m^2/\sum_{m'} p_{m'}^2$
$(m \geq 1)$. The ratio $p_m/p_1$ determines the probabilities of the incoherent
terms for the distribution $P$. The ratio $(p_m/p_1)^2$ determines
the probability of the incoherent terms for the distribution $P_2$.
As $(p_m/p_1)^2<p_m/p_1$, the distribution $P_2$ is purer than $P$.
The expressions (\ref{eq:delta_K}) and (\ref{eq:delta_SSS})
measure how the admixture of the $m>1$ modify the functions $P(K)$
and $P(\phi_J)$. It depends on the relative
phases of the eigenstates $C_{IK}^{(m)}$ of the reduced density
matrix to which degree their combination suppresses the quantal
oscillations, that is causes decoherence.

Furthermore, the Ref.~\cite{Baumgratz2014PRL} introduced the $l_1$ norm
\begin{align}\label{CI1}
C_{l_1}=\sum_{K\neq K'}\vert\rho^{(I\nu)}_{KK'}\vert
\end{align}
as a measure of coherence, which quantifies how sparse the density
matrix is. This quantity has a simple intuitive form which is directly
connected to the off-diagonal elements of $\rho$ in the basis $|K\rangle$
of interest. The rational behind it is that adding the contributions from
eigenstates $|m\rangle$ with very different phases will reduce the absolute
values the non-diagonal matrix elements. According to this measure,
a diagonal density matrix is completely incoherent.
The maximum of the $C_{l_1}^{(\textrm{max})}
=d-1=2I$~\cite{Baumgratz2014PRL}, which corresponds
to the absolute value of all matrix elements being equal to $1/d$.
The density matrix of the pure state with the amplitudes $1/\sqrt{d}$
is an example. As for the above discussed
expressions, the value of the $C_{l_1}$ norm depends on the chosen basis.

Since the coherence measures depend on the representation,
we also calculate the $C_{l_1}$ norm for the reduced
density matrix in the discrete SSS representation
\begin{align}\label{eq:Cl1_SSS}
C_{l_1}&=\sum_{m\neq n}\vert\rho^{(I\nu)}_{\phi_m,\phi_n}\vert, \\
\rho^{(I\nu)}_{\phi_m,\phi_n}&=\sum_l p_l C_{Im}^{(l)*}C_{In}^{(l)},
\end{align}
in which $C_{Im}^{(l)}$ denotes the normalized eigenvectors
of the reduced density matrix in the discrete SSS representation.
It is related to $C_{IK}^{(l)}$ by the transformation
\begin{equation}\label{eq:K-SSS}
C_{Im}^{(l)}=\frac{1}{\sqrt{2I+1}}
 \sum\limits_{K=-I}^I e^{i K \phi_m}C_{IK}^{(l)},
\end{equation}
with $\phi_m=2\pi m/(2I+1)$, where $m=-I$, $-I+1$, ..., $I$.

In this work, we will compare the qualities of coherence
obtained from the $K$-plots and SSS-plots.

\section{Numerical details}

First, we will discuss the entanglement entropy and coherence of the PTR
model using $^{135}$Pr studied in Refs.~\cite{Frauendorf2014PRC,
Matta2015PRL, Streck2018PRC, Sensharma2019PLB, Q.B.Chen2022EPJA,
Q.B.Chen2024PRC_v1} as the first example for TW of triaxial
nuclei with normal deformation. The parameters of the PTR are
$\beta=0.17$ (corresponds to $\kappa=0.038$), $\gamma=-26^\circ$,
and $\mathcal{J}_{m,s,l}=21$, 13, $4~\hbar^2/\textrm{MeV}$ for
medium ($m$), short ($s$), long ($l$) axes, respectively.
A comparison of the PTR results with the experimental energies
and transition probabilities can be found in Refs.~\cite{Frauendorf2014PRC,
Matta2015PRL, Streck2018PRC, Sensharma2019PLB, Q.B.Chen2022EPJA,
Q.B.Chen2024PRC_v1}. Then, we will discuss the entanglement
entropy and coherence for a two-quasiparticle triaxial rotor
system using $^{130}$Ba studied in Ref.~\cite{Q.B.Chen2019PRC_v1}
as example. The deformation parameters are $\beta=0.24$ (corresponds
to $\kappa=0.055$), $\gamma=21.5^\circ$ and the three spin-dependent
moments of inertia $\mathcal{J}_i=\Theta_i(1+cI)$ are determined by
the parameters $\Theta_{s,m,l}=1.09$, 1.50, and 0.65~$\hbar^2/\textrm{MeV}$
and $c=0.59$. A comparison of the PTR results with the experimental energies
and transition probabilities can be found in Ref.~\cite{Q.B.Chen2019PRC_v1}.


\section{Wobbling modes}

\subsection{Excitation energy}

\begin{figure}[!th]
  \begin{center}
    \includegraphics[width=0.80\linewidth]{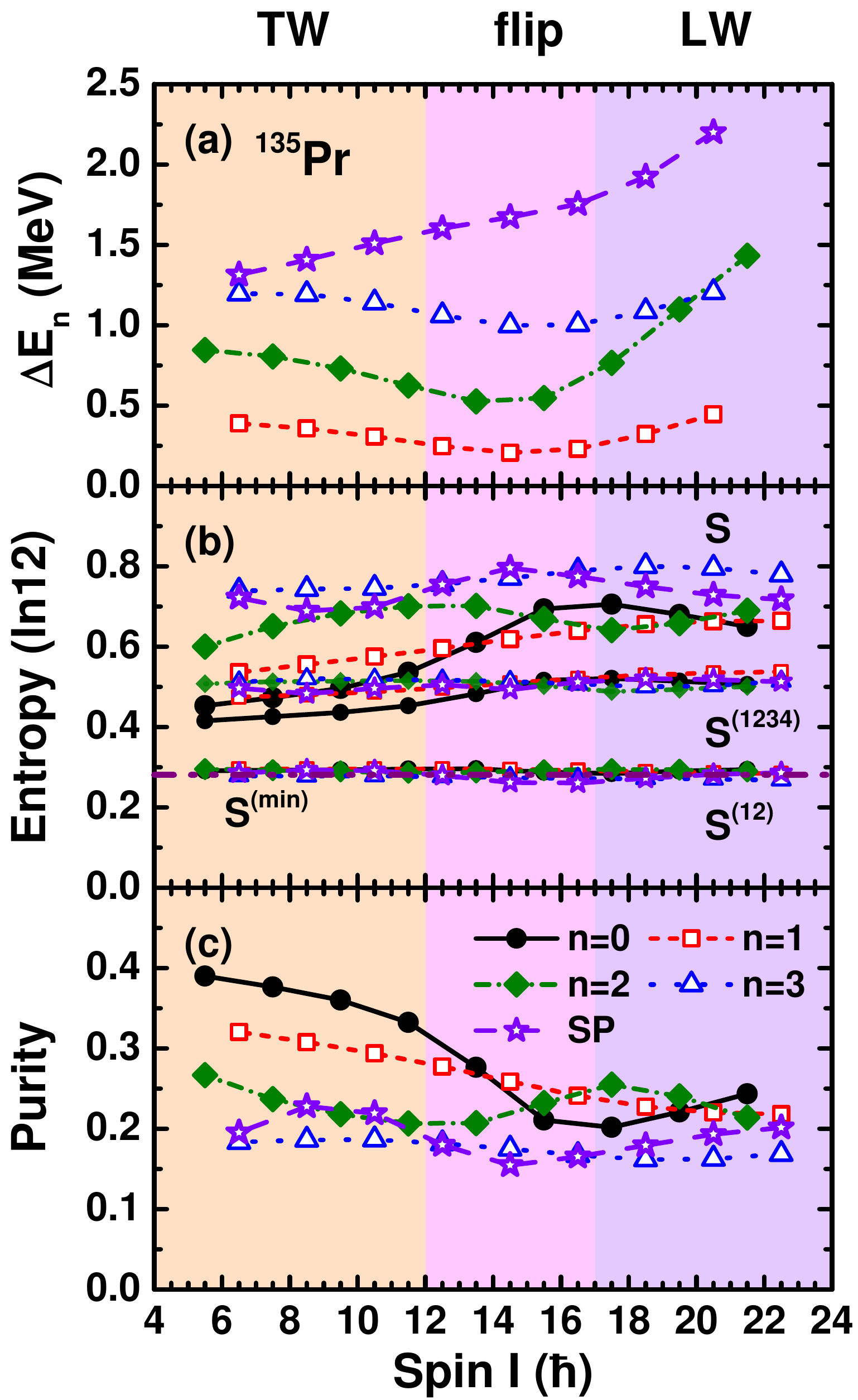}
    \caption{(Color online) (a) Excitation energies of the wobbling bands with
    wobbling numbers $n=1$, 2, and 3 and signature band with respect to the
    zero wobbling energy $\bar{E}_{\textrm{yrast}}(I)$ calculated by the PTR
    for $^{135}$Pr. Note that the states in the wobbling bands are labelled by
    the wobbling quantum number $n$. (b) The entropy of the states $S$, $S^{(12)}$,
    and $S^{(1234)}$ in units of the maximal entropy $S_j^{(\textrm{max})}=\ln 12$.
    The dash-dot-dot line gives minimal entropy
    $S^{\textrm{(min)}}=\ln 2/\ln12 \approx 0.28$. (c) The purity of
    the states. The different backgrounds delineate the
    regions of the TW, flip, and LW modes. }\label{f:Ew_S_Pu_135Pr}
  \end{center}
\end{figure}

Figure~\ref{f:Ew_S_Pu_135Pr}(a) compares the excitation energies $\Delta
E_n=E_n-\bar{E}_{n=0}$ of the wobbling bands with different wobbling numbers
$n=1$, 2, and 3 in $^{135}$Pr. The $n=0$ zero-wobbling energies $\bar{E}_{n=0}$
are equal to $E_{\textrm{yrast}}(I)$ for the signature $I=11/2+\textrm{even}$
and $[E_{\textrm{yrast}}(I-1)+E_{\textrm{yrast}}(I+1)]/2$ for the
signature $I=13/2+\textrm{even}$. The states are labelled by
the wobbling number $n$. However, this labelling does not imply that the
wobbling motion is harmonic. In our previous studies~\cite{Q.B.Chen2022EPJA,
Q.B.Chen2024PRC_v1}, we have classified the structure of the states as follows
\begin{itemize}
    \item $I\leq 23/2$  is the transverse wobbling (TW) region.
    The total angular momentum of the nucleus precesses  around the $s$ axis.
    \item $I=27/2$-$35/2$ is the flip region.  The
    total angular momentum is tilted into the $sm$ plane about halfway
    between the two axes and jumps
    between the orientations of $s$ and $m$ axes.
    \item $I\geq 37/2$ is the longitudinal wobbling (LW) region. The total
    angular momentum of the nucleus precesses around the medium axis.
\end{itemize}
The three rotational modes are delineated by different
background colors in Fig.~\ref{f:Ew_S_Pu_135Pr}.

\begin{figure*}[!th]
  \begin{center}
    \includegraphics[width=0.92\linewidth]{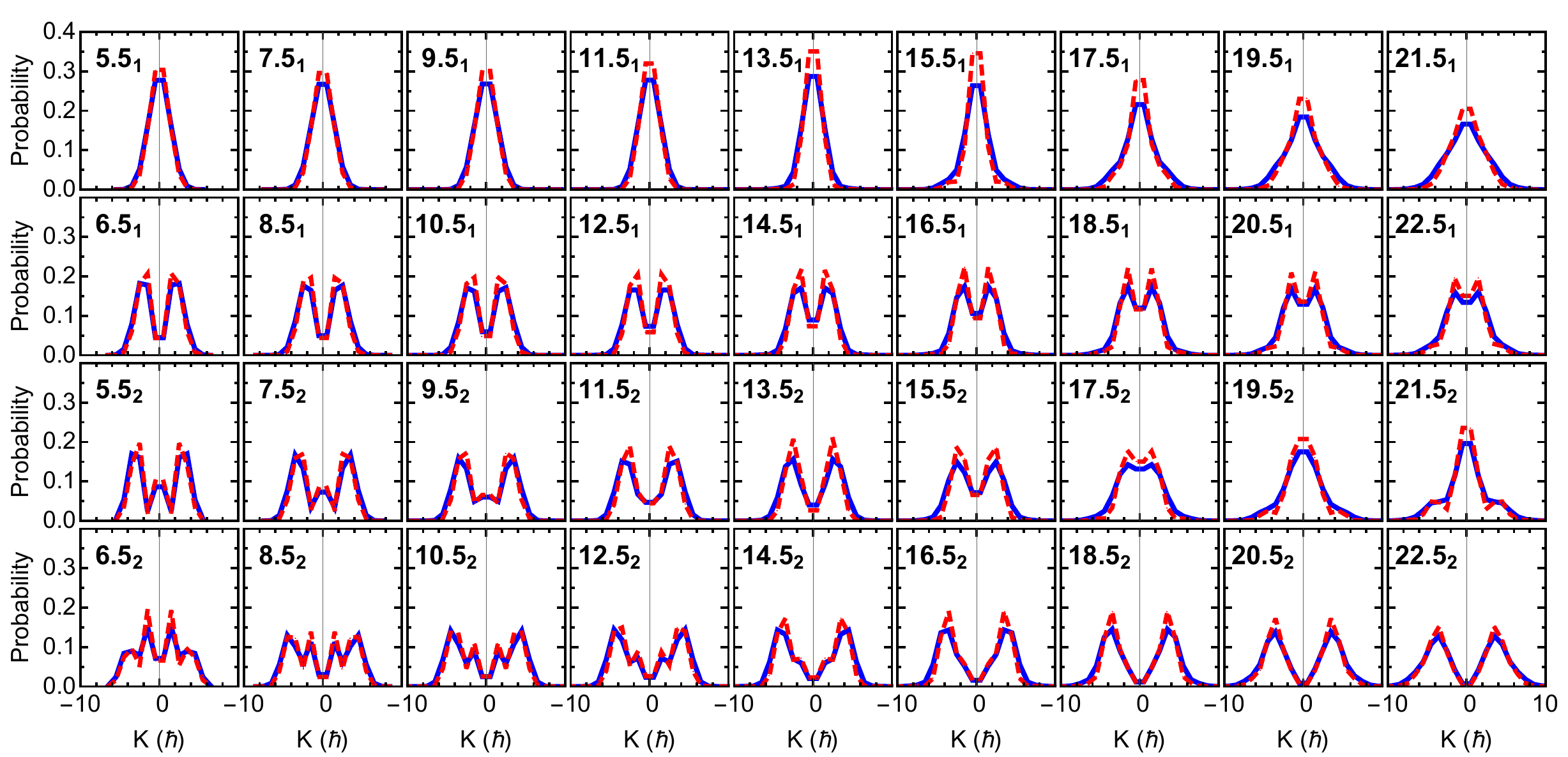}
    \caption{(Color online) Probability distribution of the
    total angular momentum projection on the $l$ axis for the $n=0$-$3$
    states in $^{135}$Pr calculated by PTR. Full curves show the one calculated by $P(K)$,
    while the dashed curves display the one by $P_2(K)$. Only the region
    $-9.5\leq K\leq 9.5$ is shown.}\label{f:P_P2_135Pr}
  \end{center}
\end{figure*}

\begin{figure*}[!th]
  \begin{center}
    \includegraphics[width=0.92\linewidth]{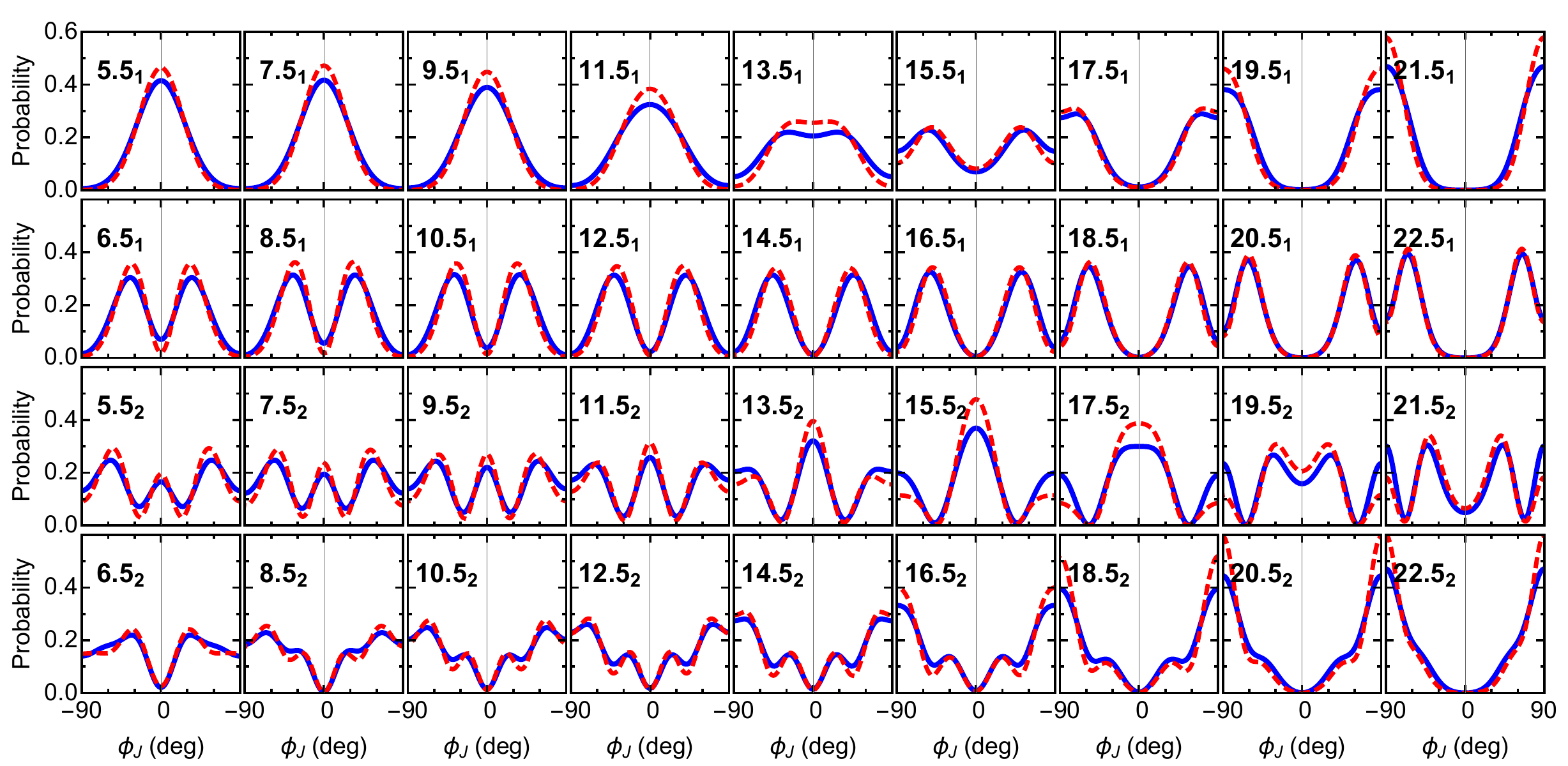}
    \caption{(Color online) Full curves show the probability density
    of the SSS $P(\phi_J)$ for the $n=0$-$3$ states in $^{135}$Pr calculated by
    PTR model, while the dashed curves display the $P_2(\phi_J)$.}\label{f:SSS_135Pr}
  \end{center}
\end{figure*}

Figures~\ref{f:P_P2_135Pr} and \ref{f:SSS_135Pr} illustrate the
structure of the PTR states in $^{135}$Pr by by showing, respectively,
the probability distributions $P(K)$ of the projection of the total
angular momentum onto the $l$ axis and $P(\phi_J)$ of the angle of
its projection onto the $sm$ plane with the $s$ axis. Both visualizations
of the structure were introduced in our previous
work~\cite{Q.B.Chen2022EPJA, Q.B.Chen2024PRC_v1}, where
the details were given.

As discussed in detail there, the $n = 0$ and 1 bands display a well defined
wobbling structure. In the TW regime, its excitation energy $\Delta E_{n=1}$
decreases as the angular momentum $I$ increases up to $I=29/2$, which indicates
that the collective wobbling motion becomes less and less stable. Subsequently,
the TW regime transitions into a flip mode  and then into a LW regime.
This evolution is accompanied by a change to an increase of the excitation
energy with increasing spin, which indicates that the collective motion
becomes more stable again, resulting from a reduction in
the amplitude of the wobbling motion.

The $n=2$ states exhibit significant deviations from the expected
distributions of the harmonic limit, but still show the three peaks,
which  qualify them as distorted
$n = 2$ TW states~\cite{Q.B.Chen2022EPJA, Q.B.Chen2024PRC_v1}.
As seen in Fig.~\ref{f:Ew_S_Pu_135Pr}(a), their excitation energies
deviate from being twice the ones of the $n=1$ ones at high spin.
As the angular momentum $I$ increases, the $n = 2$ states undergo
a transition into unharmonic two-wobbling LW structures.

The structure of the  $n=3$ states structure differs qualitatively from
the pattern expected for the unharmonic $n = 3$ wobbling
motion~\cite{Q.B.Chen2022EPJA, Q.B.Chen2024PRC_v1}. It has
a flip structure already at the beginning of the band.
The total angular momentum in the $n=3$ band mainly aligns along
the $m$ axes. In addition, in the low spin region, it has
admixture with the signature partner (SP) band. In the high
spin region, it crosses with the $n=2$ band.

In addition, the excitation energy of SP band is
also presented in Fig.~\ref{f:Ew_S_Pu_135Pr}(a). It increases
with spin and is larger than the wobbling energies.

\subsection{Angular momentum}

\begin{figure}[!th]
  \begin{center}
    \includegraphics[width=0.80\linewidth]{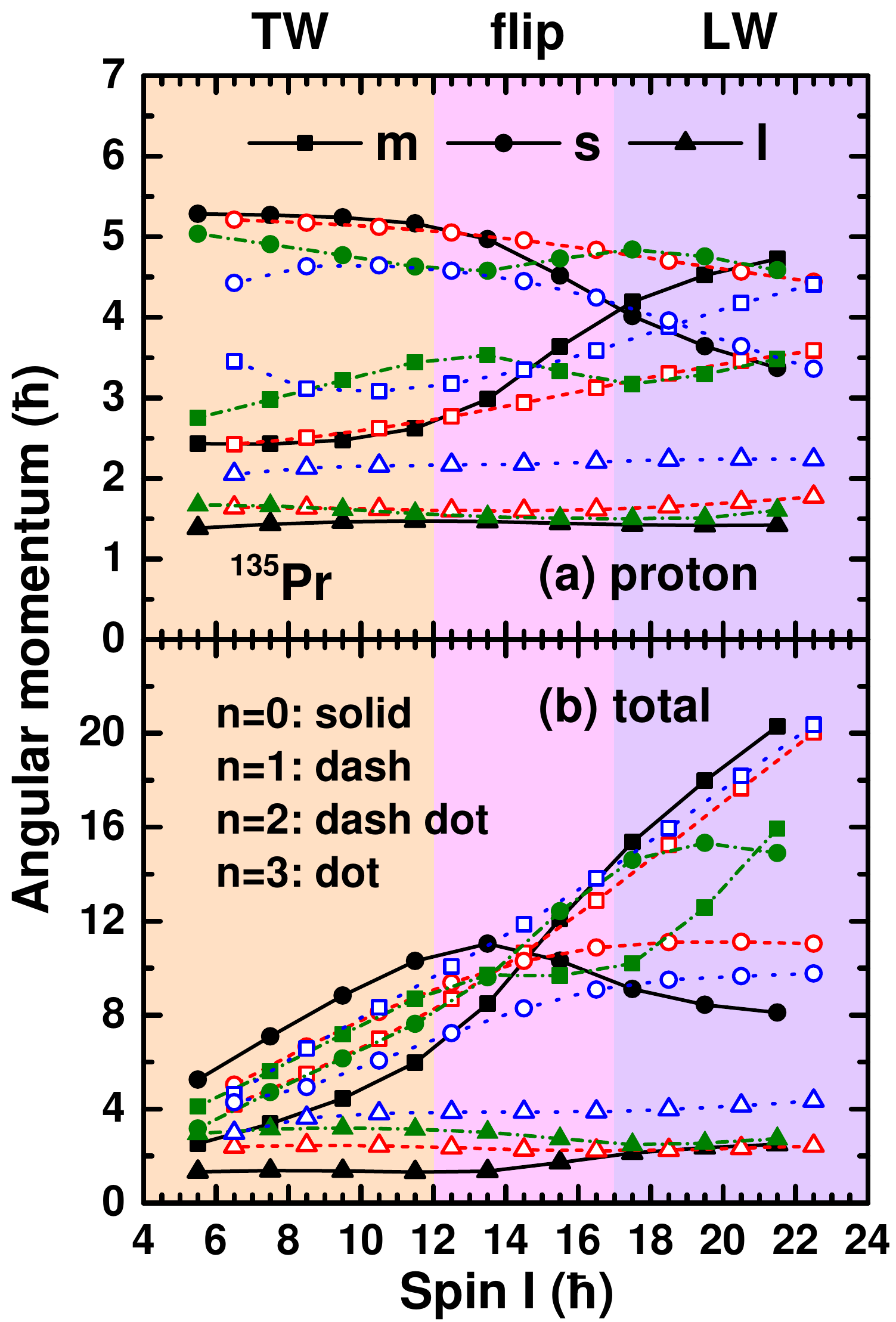}
    \caption{(Color online) Root mean square expectation values of the angular
    momentum components of the (a) proton and (b) total nucleus
    for the $n=0$, $1$, $2$, and $3$ bands in $^{135}$Pr calculated by
    the PTR model.}\label{f:ANG_135Pr}
  \end{center}
\end{figure}

From the $k$-distribution and $K$-distribution plots, one can
calculate the square roots of expectation values of the squares
of angular momentum components for the particle and the total nucleus,
respectively~\cite{Q.B.Chen2022EPJA}. Figure~\ref{f:ANG_135Pr} shows
corresponding results for the $n=0$, 1, 2, and 3 wobbling bands in
$^{135}$Pr. The triaxial rotor is coupled with an $h_{11/2}$ proton.
The $s$ axis is the preferred orientation, because it maximizes the overlap
of the particle orbit with the triaxial core~\cite{Frauendorf1996ZPA}.
The rotation energy of the rotor core prefers the medium axis with the
largest moment of inertia. At low spin $I$, the torque of the quasiparticle
wins. The orientation of $\bm{J}$ and $\bm{j}$ along the $s$ axis represents
the stable configuration. The growth of total angular momentum
$\bm{J}$ is generated by an increase of the rotor angular momentum
along the $s$ axis. Above the critical angular momentum
the torque of the rotor core takes over. The total angular momentum
moves into the $sm$ plane. The growth of total angular momentum
$\bm{J}$ is essentially generated by an increase the rotor angular momentum
along the medium axis. The particle angular momentum  $\bm{j}$ is pulled
toward to mdium axis because the Coriolis force tries to minimize
the angle between $\bm{j}$ and $\bm{I}$. However, it does not align with the
medium axis for the considered values of $I$~\cite{Q.B.Chen2022EPJA}.

\begin{figure}[!th]
  \begin{center}
    \includegraphics[width=0.80\linewidth]{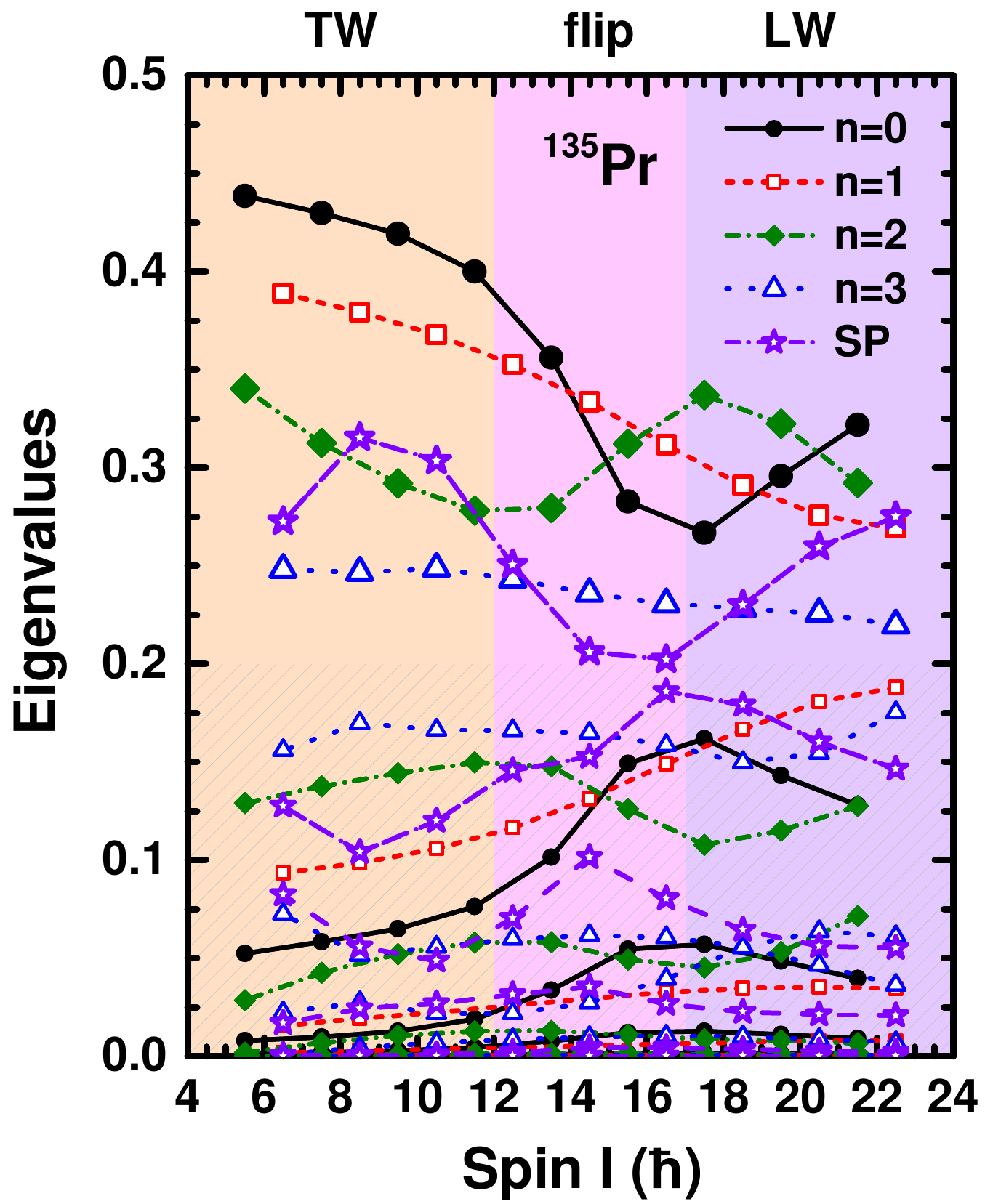}
    \caption{(Color online) Eigenvalues of the reduced density
    matrices of $^{135}$Pr as functions of spin for the
    $n=0$, $1$, $2$, and $3$ and SP states. The slant shadow denotes
    the eigenvalues below 0.2.}\label{f:Eigenvalue_135Pr}
  \end{center}
\end{figure}

\begin{figure}[!th]
  \begin{center}
    \includegraphics[width=0.90 \linewidth]{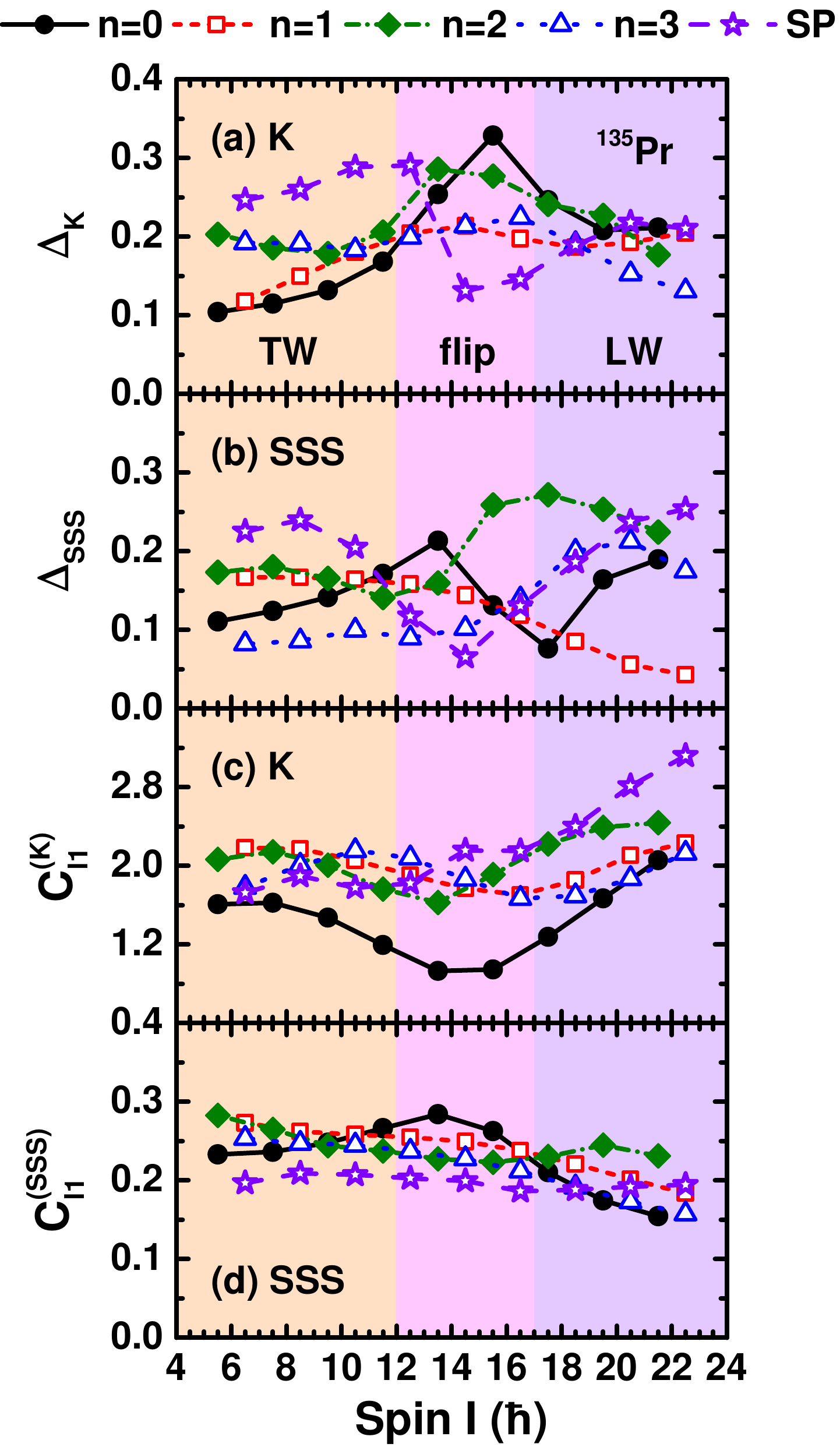}
    \caption{(Color online) The absolute deviations $\Delta_K$ (a) and
    $\Delta_{\textrm{SSS}}$ (b) as well as the $l_1$ norm
    $C_{l_1}^{(K)}$ (c) and $C_{l_1}^{(\textrm{SSS})}$ (d)
    of the $K$ distributions $P(K)$ and SSS plots $P(\phi_J)$ as
    measures of decoherence for the $n=0$-$3$ states and signature
    partner band in $^{135}$Pr calculated by PTR model.}\label{f:DeltaC_135Pr}
  \end{center}
\end{figure}

\begin{figure*}[!th]
  \begin{center}
    \includegraphics[width=0.92\linewidth]{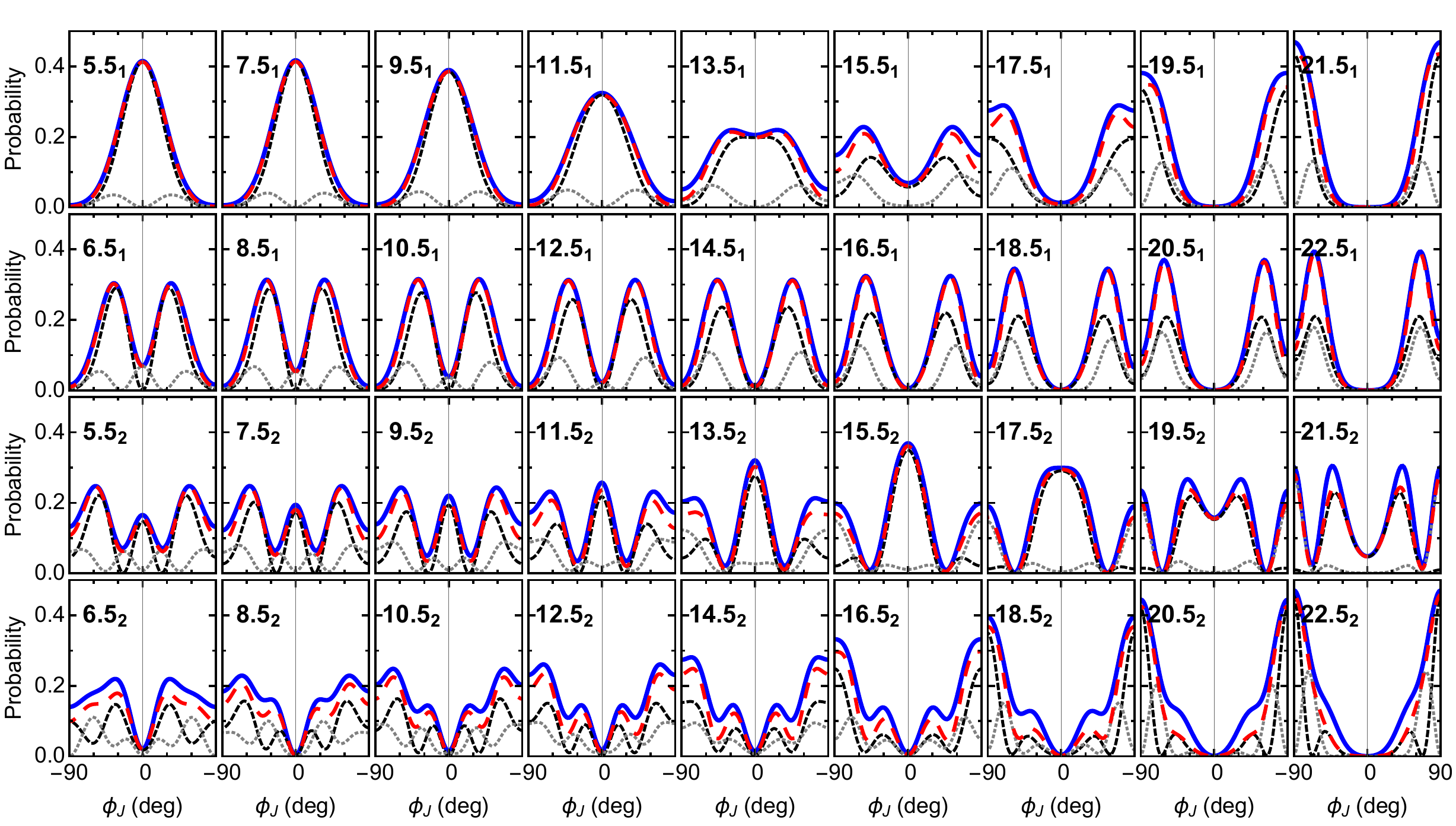}
    \caption{(Color online) Eigenstate decomposition of the SSS probability density
     $P(\phi_J)$ (thick blue full curves) of the total angular momentum
    for the PTR $n=0$-$3$ states in $^{135}$Pr.
    The individual terms are color coded as $P_{12}(\phi_J)$ (black short dash),
    $P_{34}(\phi_J)$ (gray dotted)
    and $P_{1234}(\phi_J)$ (thick red long dash).}\label{f:SSS_DM_total_135Pr}
  \end{center}
\end{figure*}

\begin{figure*}[!th]
  \begin{center}
    \includegraphics[width=0.92\linewidth]{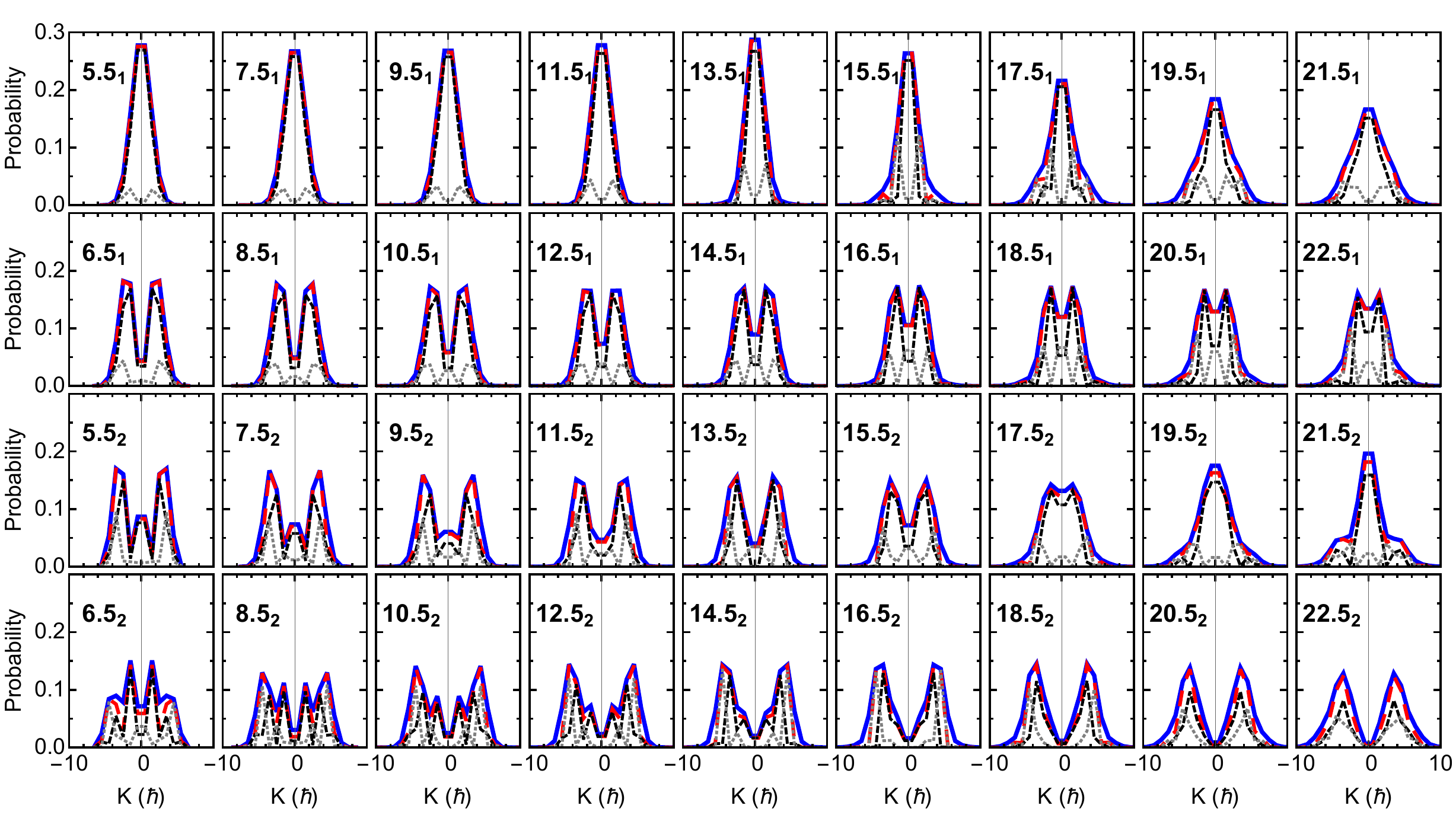}
    \caption{(Color online) Eigenstate decomposition of the
    probability distributions of the total angular momentum
    projection on the $l$ axis $P(K)$, $P_{12}(K)$, $P_{34}(K)$,
    and $P_{1234}(K)$ for the $n=0$-$3$ states in $^{135}$Pr calculated by PTR.
    Only the region $-9.5\leq K\leq 9.5$ is shown. Same color coding
    as in Fig.~\ref{f:SSS_DM_total_135Pr}.}\label{f:PK_DM_total_135Pr}
  \end{center}
\end{figure*}

\section{Entropy and coherence}

\subsection{Eigenvalue spectra of the reduced density matrix}\label{dm}

As discussed above one can quantify the degree of entanglement
and coherence by diagonalizing the reduced density matrix.
Figure~\ref{f:Eigenvalue_135Pr} displays the eigenvalues
of the reduced density matrices for the PTR states of
$^{135}$Pr as functions of spin for the $n=0$, $1$, $2$,
and $3$ states. For a pure, completely coherent state one eigenvalue
is 1 and all other are zero. The pure triaxial rotor states are examples,
which appear as the limiting case of zero Coriolis coupling in even-even
nuclei. For partial coherence one has one large eigenvalue, the eigenvector
of which represents the coherent wave function and the eigenvalue its
probability. The remaining eigenfunctions, which appear with the probability
of their small eigenvalues, distort the coherence.

As discussed above for the odd-$A$, the  eigenvalues of the $\bm{j}$-density
matrix are two-fold degeneracy due to Kramer's degeneracy. The same
holds for the non-zero eigenvalues of the $\bm{J}$-density matrix,
which are identical. That is, the limit of complete coherence corresponds
to two eigenvalues of 1/2. For partial coherence one has one large pair
of eigenvalues, the eigenvector of which represents the coherent
wave function and the eigenvalues their probabilities. The remaining
eigenfunctions, which appear with the probabilities of their
small pairs of eigenvalues, distort the coherence.

From Fig.~\ref{f:Eigenvalue_135Pr}, one finds that the $I=11/2,~n=0$ state
is dominated with the probability of 0.44 by one pair of eigenvectors.
For the $I=13/2,~n=1$ state the largest probability is 0.38,
for the $I=11/2,~n=2$ state it is 0.34, and for the $I=13/2,~n=3$
state it has fallen to  0.25. The dominance of one pair
decreases with the excitation energy. For the $n=1,~3$ sequences,
the probability for the strongest eigenstate  decreases smoothly with $I$.
For the $n=2,~4$ sequences the $I$ dependence is more complex.
The eigenvalues of the most likely and second likely pair of eigenvectors
approach each other in the TW region, stay close in  the flip region,
and depart from each other in the LW region. The two pairs seem
to exchange their character with increasing $I$. The eigenvalues of
SP band shows similar trend as those of $n=0$ band. In Sec.~\ref{deco}
we will discuss how the $I$ dependence of the eigenvectors of
the $\bm{j}$- and $\bm{J}$-reduced density matrices is reflected
by the structure of the PTR states.

\subsection{Entropy and purity}\label{enpu}

Using the eigenvalues of the reduced density matrix, we can easily
calculate the entropy $S$ and purity $\mathcal{P}$ using Eqs.~(\ref{eq:eq1})
and (\ref{eq:eq2}), respectively. In Figs.~\ref{f:Ew_S_Pu_135Pr}(b) and (c),
we present $S$ and $\mathcal{P}$ as functions of spin for
the $n=0$, 1, 2, and $3$ states in $^{135}$Pr calculated by the PTR.
Naturally, the purity exhibits an opposite behavior compared to the
entropy. A large purity corresponds to a small entropy, while a small
purity corresponds to a large entropy. Therefore, in our discussions,
we will focus on the behavior of the vN entropy.

For fundamental reasons, one expects that the vN entropy should
increase with excitation energy, that is, the particle and total angular
momenta should become more entangled with the wobbling number $n$.
One could also expect that $S$ growths with the value
of $I$, because the Coriolis interaction becomes stronger.
Figure~\ref{f:Ew_S_Pu_135Pr}(b) demonstrates the expected
behaviour is only seen in the TW region. Above, $S$ does
not change much with $I$, and the states group around
0.7 with no specific order. We attribute this to the smallness
of the system.

In accordance with the entropy, the purity decreases in the TW region.
Above, the states  group around 0.2, which is more than twice the minimal
value of $1/d_j=1/(2j+1)=0.083$.

The entropy and purity reflect the $I$ dependence of the eigenstates
of the density matrices, which were discussed in Sec.~\ref{dm}.
In particular the avoided crossing of two largest
eigenvalues of the $n=0$ and 2 bands is seen
as the exchange of the order.

Figure~\ref{f:Ew_S_Pu_135Pr}(b) includes the minimum of
the entropy $S^{\textrm{(min)}}=\ln 2/\ln 12 \approx 0.28$,
which corresponds to the strong-coupling limit of the PTR
model~\cite{Ring1980book}, where the Coriolis Hamiltonian
$\hat{H}_{\textrm{cor}}$ (\ref{eq:Cor}) is set to zero.
The realistic entropy values of the $n=0$, 1, 2, and $3$
states are well above $S^{\textrm{(min)}}$, indicating that
the particle angular momentum is substantially entangled with the
total angular momentum.

\subsection{Decoherence}\label{co}

The impurity causes decoherence, that is, it washes out the interference
pattern of pure wave function. In order to quantify the washing out we
introduced the decoherence measures (\ref{eq:delta_K}) and (\ref{eq:delta_SSS}).
They compare the probability densities $P$ with the purified densities $P_2$
(By squaring, the small eigenvalues $p_m$ are suppressed relative to
the large ones.)

Figure~\ref{f:P_P2_135Pr} compares $P(K)$ with $P_2(K)$.
The difference between the curves indicates the missing coherence. The
deviations of $P(K)$ from $P_2(K)$ are small, where $P_2(K)$ has more pronounced
minima and maxima than $P(K)$. The visual impression  of good partial coherence
is reflected by $\Delta_K$ in Fig.~\ref{f:DeltaC_135Pr}, which shows the total
area between the two curves in Fig.~\ref{f:P_P2_135Pr}. The decoherence measure
$\Delta_K$ is between 0.1 and 0.3 for all $I$. We found that $\Delta_K$ stays
below 0.33 for all PTR states while the entropy approaches 0.80
(=1.99 in natural units) and the purity decreases to 0.15 with
increasing excitation energy.

Figure~\ref{f:SSS_135Pr} compares $P(\phi_J)$ with $P_2(\phi_J)$. The
deviations are small as well, where $P_2(\phi_J)$ has more pronounced minima
and maxima than $P(\phi_J)$. Figure~\ref{f:DeltaC_135Pr} shows
$\Delta_{\textrm{SSS}}$ which measures corresponding absolute deviation,
i.e., the area between the curves. The good partial coherence
corresponds to $\Delta_{\textrm{SSS}}$ stays between 0.1 and 0.3.
Like $\Delta_{K}$ it stays below 0.27 for all PTR states.

\begin{figure*}[!th]
  \begin{center}
    \includegraphics[width=0.92\linewidth]{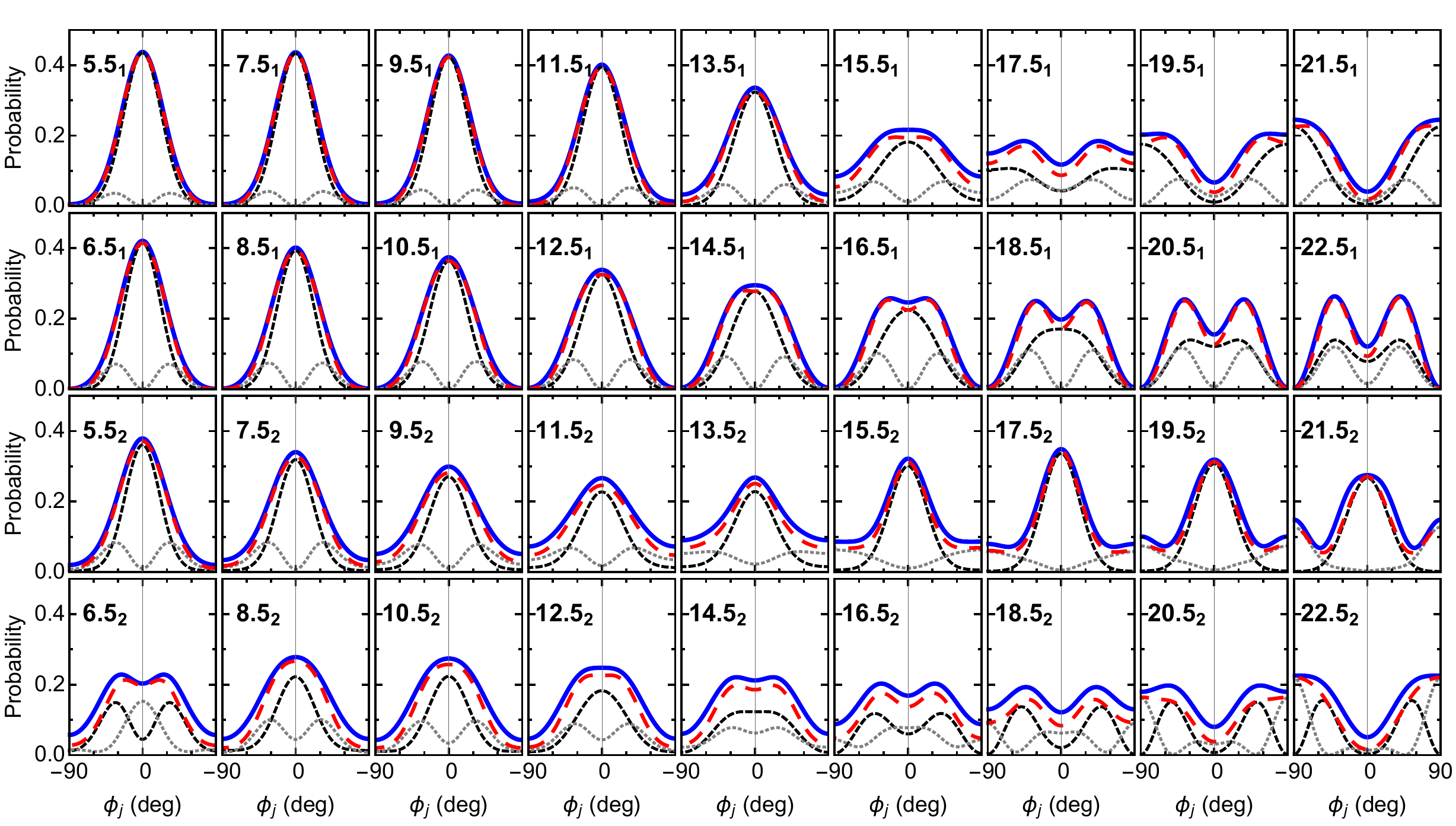}
    \caption{(Color online) Decomposition the SSS probability density
     distributions $P(\phi_j)$, $P_{12}(\phi_j)$, $P_{34}(\phi_j)$, and
     $P_{1234}(\phi_j)$ of the proton particle angular momentum
    for the PTR $n=0$-$3$ states in $^{135}$Pr. Same color coding as in
    Fig.~\ref{f:SSS_DM_total_135Pr}.}\label{f:SSS_DM_proton_135Pr}
  \end{center}
\end{figure*}

\subsection{Nature of decoherence}\label{deco}

According to Eqs.~(\ref{eq:SD-J}) and (\ref{eq:SD-j}), the matrix
elements of the reduced density matrix $\rho_{KK'}$ and $\rho_{kk'}$
can be written as the sum of the product of the eigenvalues and
the normalized eigenvectors. Each term represents a pair of pure states
in the $\bm{J}$- and $\bm{j}$-subspaces, which are orthonormal.
The PTR states can be interpreted as the incoherent sums of the
normalized probability distributions of these pairs of states in
the subspaces, which are weighted by their eigenvalues.

In most cases the oscillations of the
probability distributions of the respective sub-matrices
reflect their mutual orthogonality. However,
the SSS representations of the eigenvectors are complex. Only the different
oscillations of the squares of their real parts reflect that two
states are orthogonal. The same holds for their imaginary parts.
If the amplitudes of the oscillations of the real and imaginary parts
are similar and their zeros differently located, they may get washed
out in sum, which represents the probability distribution.

Figure~\ref{f:Eigenvalue_135Pr} shows that the sum of the first and
second pair of eigenvalues exhausts most of the trace of the reduced
density matrices, which is one. This indicates that $\rho_{KK'}$ and
$\rho_{kk'}$ are dominated by the pairs eigenvectors of the first four
eigenstates. Because of the degeneracy of the eigenvalues, we define
the following coherent sub-matrices
\begin{align}
\rho_{KK'}^{(m)}&=p_m C^{(m)}_{IK}C^{(m)}_{IK'},\\
 \rho_{KK'}^{(12)}&=\rho_{KK'}^{(1)}+\rho_{KK'}^{(2)},\\
 \rho_{KK'}^{(34)}&=\rho_{KK'}^{(3)}+\rho_{KK'}^{(4)},\\
 \rho_{KK'}^{(1234)}&=\rho_{KK'}^{(12)}+\rho_{KK'}^{(34)},
\end{align}
in which the upper indexes $m$ denote the order
of the eigenvalues of the density matrix. Using these
sub-matrices, we calculate the corresponding $\bm{J}$-SSS plots
$P_{12}(\phi_J)$, $P_{34}(\phi_J)$, and $P_{1234}(\phi_J)$ for
the total angular momentum and corresponding $\bm{j}$-SSS plots
$P_{12}(\phi_j)$, $P_{34}(\phi_j)$, and $P_{1234}(\phi_j)$
for the particle angular momentum. In addition, it is clear that
the non-zero eigenvalues of $\rho_{KK'}^{(12)}$ are $p_1$ and $p_2$,
while the non-zero eigenvalues of $\rho_{KK'}^{(1234)}$ are $p_1$,
$p_2$, $p_3$, and $p_4$. Correspondingly, we can calculate the
vN entropy for these sub-matrices according to the
definition (\ref{eq:eq1})
\begin{align}
 S^{(12)} &=-p_1\ln p_1-p_2\ln p_2,\\
 S^{(1234)} &=-p_1\ln p_1-p_2\ln p_2-p_3\ln p_3-p_4\ln p_4.
\end{align}
Obviously, $S^{(1234)}$ is larger than $S^{(12)}$.

Figure~\ref{f:SSS_DM_total_135Pr} compares the probability $P(\phi_J)$ of
the total angular momentum with $P_{12}(\phi_J)$, $P_{34}(K)$, and
$P_{1234}(K)$, Fig.~\ref{f:PK_DM_total_135Pr} compares the  probability
$P(K)$ of the total angular momentum with $P_{12}(K)$, $P_{34}(K)$,
and $P_{1234}(\phi_J)$, and Fig.~\ref{f:SSS_DM_proton_135Pr} compares
the SSS probability $P(\phi_j)$ of the proton particle angular momentum
with $P_{12}(\phi_j)$, $P_{34}(\phi_j)$, and $P_{1234}(\phi_j)$ for
the PTR $n=0$-$3$ states in $^{135}$Pr. As seen,
to good approximation the full PTR probabilities can be interpreted
as the incoherent combinations of the contributions from
the first two pairs of eigenvectors of the reduced density matrix.

\begin{figure*}[!th]
  \begin{center}
    \includegraphics[width=0.90\linewidth]{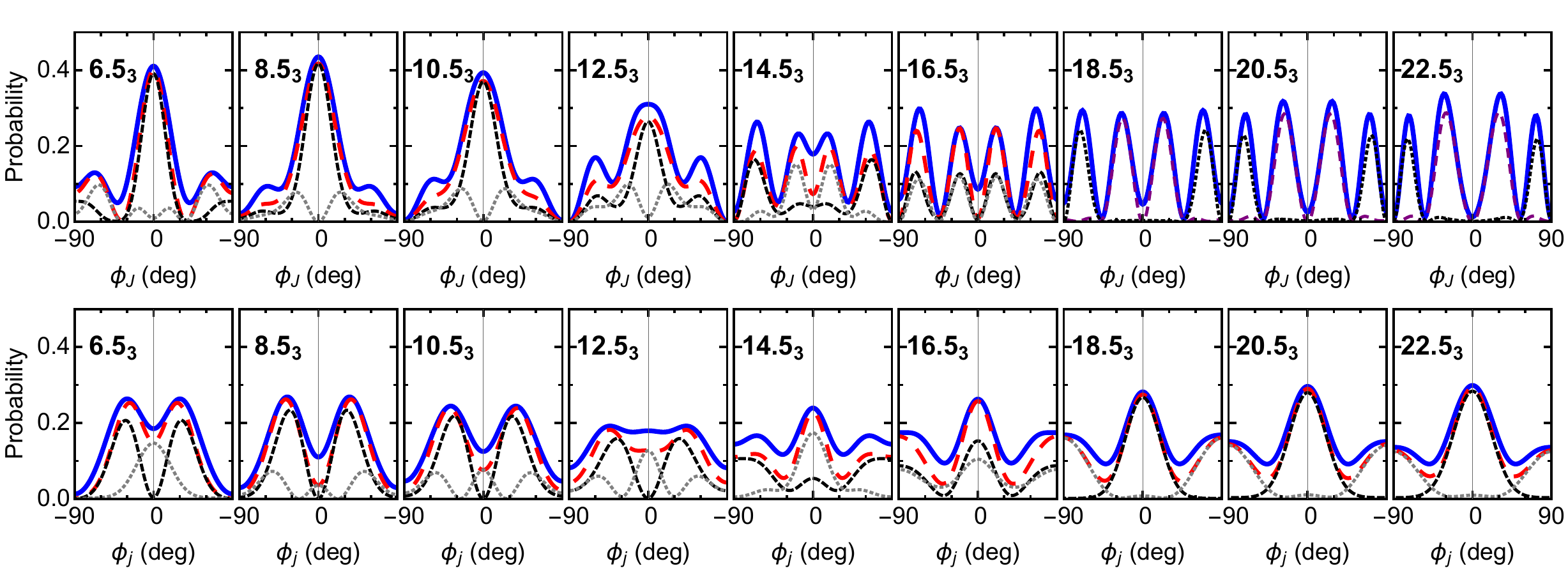}
    \caption{(Color online) Eigenstate decomposition the probability
     density of the total $P(\phi_J)$, $P_{12}(\phi_J)$, $P_{34}(\phi_J)$,
     and $P_{1234}(\phi_J)$  (upper panels) and proton
     particle $P(\phi_j)$, $P_{12}(\phi_j)$, $P_{34}(\phi_j)$,
     and $P_{1234}(\phi_j)$ (lower panels) angular momenta of the SP
     band in $^{135}$Pr. In the last three panels of the
     upper row only $P_{12}(\phi_J)$ and $P_{34}(\phi_J)$ are shown.
     Same color coding as in Fig.~\ref{f:SSS_DM_total_135Pr}.}\label{f:SSS_DM_SP_135Pr}
  \end{center}
\end{figure*}

The distributions $P_{12}(\phi_J)$ and $P_{12}(K)$ are related to
each other by a Fourier transform of their respective sub-density matrices.
This is reflected by similar oscillation pattern. In case of harmonic
oscillations they would agree when scaled by the oscillator lengths.
There is a difference in the display because $K$ is a discrete variable
while $\phi_J$ is a continuous variable. (See discussions below of
the density matrices in the complete discrete $\phi_J$ representation.)
The same Fourier-transform relation holds between $P_{34}(\phi_J)$
and $P_{34}(K)$.

The results of $S^{(12)}$ and $S^{(1234)}$ are further included in
Fig.~\ref{f:Ew_S_Pu_135Pr}(b). As seen in Fig.~\ref{f:Ew_S_Pu_135Pr}(b),
$S^{(12)}$ ($=0.74$ in natural units) is very close to $S^{\textrm{(min)}}$
($=\ln{2}=0.69$ in natural units) and $S^{(1234)}$ is restricted to a narrow
band around 0.5 ($=1.2$ in natural units). This explains why the decoherence
measures remain roughly constant with increasing $I$. The ``noise" from the
$m>4$ eigenstates does not generate much decoherence like thermal noise does
not disturb the interference pattern of electromagnetic waves in a medium.
In the following we discuss in detail the interpretation of the lowest
four PTR states in $^{135}$Pr.

\paragraph{{$n=0$} states:}

For the $I\leq 31/2$ yrast states the $\bm{j}$-SSS probability
$P_{12}(\phi_j)$ of the particle shows a bump at $\phi_j=0^\circ$, which
indicates that the proton is in its lowest state of a potential centered at
$\phi_j=0^\circ$. The width of the  bump increases as the potential becomes
softer with $I$. For $I=35/2$, the bump becomes unstable and is
shifted to $\phi_j=\pm 90^\circ$ with further increase of $I$, which indicates
that the proton is in the lowest state of a potential centered there.

The $\bm{J}$-SSS probability $P_{12}(\phi_J)$
of the total angular momentum can be interpreted  as representing
the lowest states of an effective collective Hamiltonian in the
$\bm{J}$-degree of freedom with a potential that changes from being
being centered at $\phi_J=0^\circ$ to being centered at $\phi_J=\pm 90^\circ$.

The SSS $\bm{j}$-probability $P_{34}(\phi_j)$ has the double
hump structure of the second state of the particle in
a potential that changes around  $I=31/2$ from being
centered at $\phi_j=0^\circ$ to being centered at $\phi_j=\pm 90^\circ $.
For $I<31/2$ the distribution $P_{34}(\phi_j)$ has a zero at $\phi_j=0^\circ$, while
for $I>31/2$ it has zeroes at $\phi_j=\pm 90^\circ$.
Around $I=31/2$ there are only minima at $\phi_j=0^\circ,~\pm 90^\circ$
corresponding to the transitional character of the potential.

The SSS $\bm{J}$-probability $P_{34}(\phi_J)$  has the double hump structure
of the second state of the total angular momentum in the effective collective
potential. It corrects the $P_{12}(\phi_J)$ term by widening the peaks,
flattening their apexes and enhancing the dips at $\phi_J=0^\circ$
in the flip region.

\paragraph{{$n=1$} states:}

For $I\leq 29/2$  the $\bm{j}$-SSS probability
$P_{12}(\phi_j)$ of the particle is similar to the one of the $n=0$
state. The bump at $\phi_j=0^\circ$ indicates that the proton is in its
lowest state of a potential centered there. The width of the bump increases as the potential
becomes softer with $I$. At larger  $I$ values a dip at $\phi_j=0^\circ$ develops,
while the zeroes at $\phi_j=\pm 90^\circ$ remain. The SSS $\bm{j}$-distribution
$P_{34}(\phi_j)$ has a double hump structure that is similar to the $n=0$ state.

The $\bm{J}$-SSS probability $P_{12}(\phi_J)$ of the total angular momentum
represents to good approximation the second states of the collective Hamiltonian
in the $\bm{J}$-degree of freedom with a potential that changes
from being centered at $\phi_J=0^\circ$ to being centered at $\phi_J=\pm 90^\circ$.

The SSS $\bm{J}$-distribution $P_{34}(\phi_J)$ changes with $I$ from
an $n=3$ like triple-peak structure in a potential centered at $\phi_J=0^\circ$ to
the two separated peaks at large $I$, which are similar to the ones
of the $n=0$ states in the panel above. The orthogonality of
the states 1, 2, 3, 4 becomes only apparent when one plots the densities
of their real and imaginary parts separately.

\paragraph{{$n=2$} states:}

The $\bm{j}$-SSS distributions $P_{12}(\phi_j)$ have a peak at $\phi_j=0^\circ$
similar to the $n=0,1$ states in the TW wobbling regime. At variance
with them, the peak stays about the same up to the largest $I$ values.
The reaction to the Coriolis force appears as the change of the $P_{34}(\phi_j)$
distribution from two peaks located near $\phi_j=\pm 45^\circ$ to two peaks
at $\phi_j=\pm 90^\circ$ while the value at $\phi_j=0^\circ$ remains always zero.

The $\bm{J}$-SSS distributions $P_{12}(\phi_J)$ determine the character of the total
distribution. In the TW wobbling region $I\leq 23/2$  it has three peaks
at $\phi_J=0^\circ, \approx \pm 45^\circ$, which reflect the typical nodal
structure of a collective state in a potential centered at $\phi_J=0^\circ$.
The $P_{34}(\phi_J)$ term leads to some modification, which is similar
to the corrections to the $n=0,1$ states. At larger $I$ values, the $P_{12}(\phi_J)$
distribution changes into a LW-like $n=2$ structure with peaks at
$\phi_J=\pm 90^\circ,~\approx \pm 45^\circ $. The total function $P$
is similar because $P_{34}$ is small.

\paragraph{{$n=3$} states:}

For $I=17/2,~21/2$ the $\bm{j}$-SSS distributions $P_{12}(\phi_j)$
have peak at $\phi_j=0^\circ$. Like the PTR states $n=0,~1,~2$,
the $P_{34}(\phi_j)$ distributions have two peaks, but with
lager weight. The sum $P_{1234}(\phi_j)$ gives a broader
$\phi_j=0^\circ$ peak. The $\bm{J}$-SSS distributions $P_{12}(\phi_J)$
have the typical $n=3$ nodal structure with zeros at
$\phi_J=0^\circ, \approx \pm 35^\circ$ and maxima at
$\phi_J\approx \pm 25^\circ,~\approx \pm 65^\circ$.
The relatively large incoherent contribution $P_{34}(\phi_J)$
considerably washes out the oscillation of the total distribution.

At larger $I$, the $\bm{j}$-SSS distributions $P_{12}(\phi_j)$ and $P_{34}(\phi_j)$
contribute with similar weight. The distributions $P(\phi_j)$ develop an
increasing minimum at $\phi_j=0^\circ$, which indicates the gradual alignment of
the particle with the $m$-axis. The $\bm{J}$-SSS distributions $P_{12}(\phi_J)$
and $P_{34}(\phi_J)$ contribute with similar weight as well. As the result
of their incoherent combination, $P(\phi_J)$ becomes localized at
$\phi_J=\pm 90^\circ$ for $I=45/2$.

\paragraph{SP states:}

Figure~\ref{f:SSS_DM_SP_135Pr} compares $P(\phi)$ with the $P_{12}(\phi)$,
$P_{34}(\phi)$, and $P_{1234}(\phi)$ for the total and proton
particle angular momenta of SP band in $^{135}$Pr,
which is built on the state $13/2_3$ (see Fig.~\ref{f:Ew_S_Pu_135Pr}).
As common, the name ``SP" is used to denote the first
excited state of the odd proton in the rotating potential. The excitation
corresponds to a certain reorientation of $\bm{j}$ away from the $s$ axis,
which is reflected by the appearance of two peaks symmetric $\phi_j=0^\circ$
in the $\bm{j}$-probability density $P(\phi_j)$ when $I \leq 23/2$. For these $I$
values the $\bm{J}$-probability $P(\phi_J)$ has a pronounced  peak at $\phi_J=0^\circ$,
which indicates that the nucleus is uniformly rotating about the $s$ axis.

As seen in Fig.~\ref{f:Ew_S_Pu_135Pr}, the states $13/2_2$ and $13/2_3$
are close to each other. We discussed in our previous study~\cite{Q.B.Chen2022EPJA}
that the two states represent a mixture of a pure $n=2$ TW state and
a pure SP state. The mixture is clearly seen as the dip in the $\bm{j}$-distribution
of the $13/2_2$ state.

For $I\geq 25/2$, the $\bm{j}$-distributions develop a peak at $\phi_j=0^\circ$
similar to the $n=2$ states $(I-1)_2$. Along with this, a minimum at
$\phi_j=0^\circ$ appears in the $n=3$ distributions, which indicates
the reorientation of $\bm{j}$ towards the $m$ axis.

For $I\geq 25/2$, the $\bm{J}$-distribution changes to an oscillation
with zeroes at $\phi_J=0^\circ$, $\pm 60^\circ$, $\pm 90^\circ$,
which corresponds to a $n=3$ wobbling mode without much modification
by the odd particle. As already discussed in \textit{(d)},
the $\bm{J}$-distributions of the $n=3$ states change to rotation about
the $m$ axis. It seems that the SP band and the $n=3$ band interchange
their character over the $I\geq 25/2$ region.

\subsection{Two-quasiparticle bands}

The TW mode based on the $(1h_{11/2})^2$ two-quasiproton in $^{130}$Ba
has been studied in Ref.~\cite{Q.B.Chen2019PRC_v1} in the framework
of the PTR model, where the details are given. As described in the
Sec.~\ref{sec:plots}, the reduced two-quasiparticle density matrix
is obtained by Eq.~(\ref{eq:RDM2qp_proton}), where the index
$k_{12}=k_1+k_2$ and the sum runs over all projections $K$ and all
couplings of the two quasipartilcles to $\bm{j}_{12}$. The reduced
density matrix for the total angular moment is obtained by
Eq.~(\ref{eq:RDM2qp_total}), where the sum runs over all projections
$k_1$ and $k_2$ of the two quasiparticles.

\begin{figure}[!ht]
  \begin{center}
    \includegraphics[width=0.80 \linewidth]{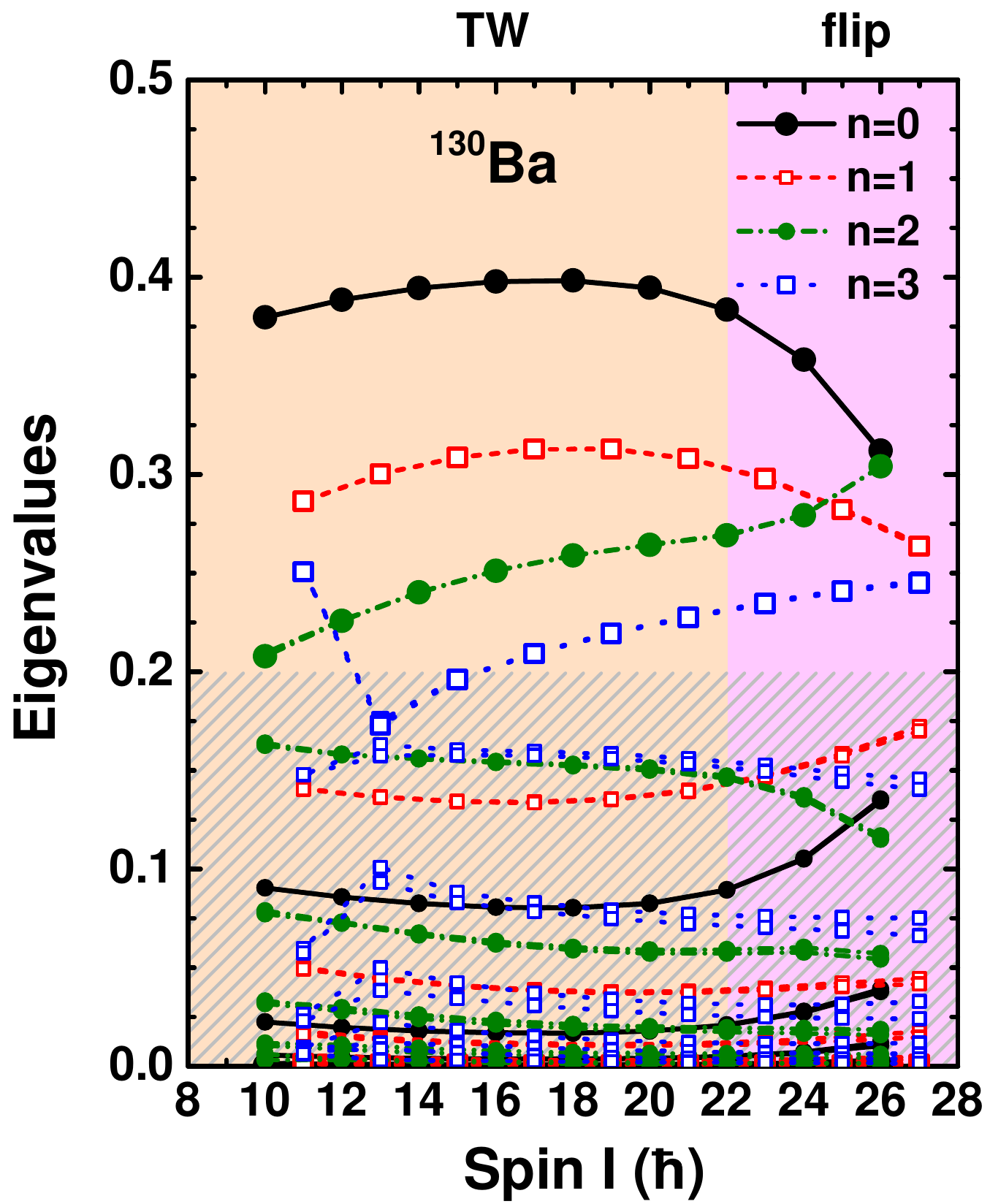}
    \caption{Eigenvalues of the reduced density matrices of $^{130}$Ba
    as functions of spin for the $n=0$, $1$, $2$, and $3$ states.
    The slant shadow denotes the eigenvalues below 0.2. In contrast
    to the case of $^{135}$Pr in Fig.~\ref{f:Eigenvalue_135Pr}, the paired
    eigenvalues $(p_1, p_2)$, $(p_3, p_4)$, .... are not the same.
    However, the differences are within the size of the
    symbols.}\label{f:Eigenvalue_130Ba}
  \end{center}
\end{figure}

\begin{figure}[!th]
  \begin{center}
   \includegraphics[width=0.80\linewidth]{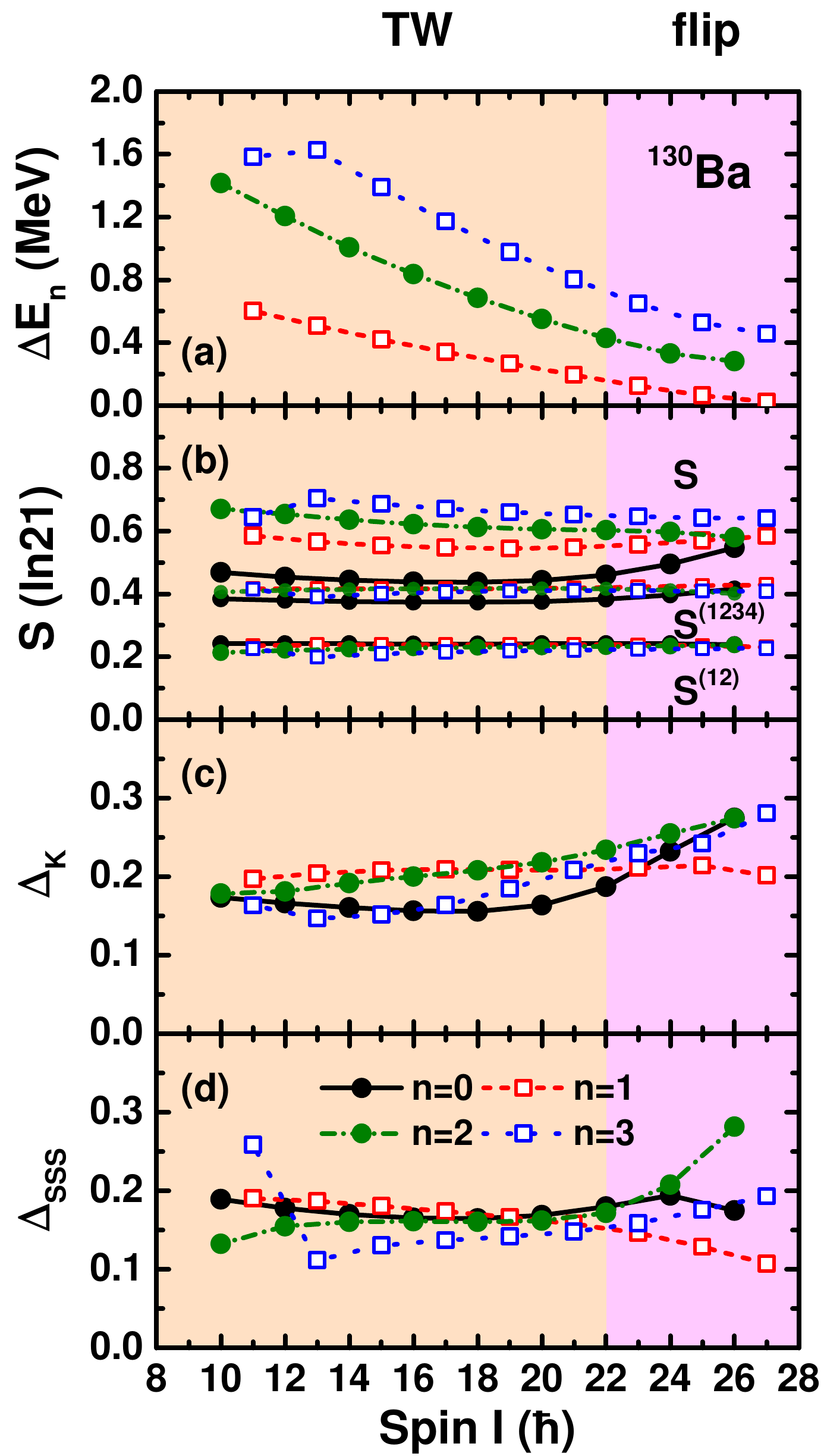}
    \caption{(Color online) (a) Excitation energies of the wobbling bands
    in $^{130}$Ba with wobbling numbers $n=1$, 2, and 3 with respect to the
    zero wobbling energy $\bar{E}_{\textrm{yrast}}(I)$ calculated by the PTR.
    (b) The entropy $S$, $S^{(1234)}$, and $S^{(12)}$ of the states
    in unit of $\ln 21$, where 21 is the dimension of the
    two-qp $\pi(h_{11/2})^2$ space.  (c) The absolute deviations
    $\Delta_K$ of the $K$ distributions $P(K)$.
    (d) The absolute deviations $\Delta_{\textrm{SSS}}$ of the SSS plots
    $P(\phi_J)$. The different backgrounds delineate the regions of
    the TW and flip modes.}\label{f:Ew_S_Delta_130Ba}
  \end{center}
\end{figure}

\begin{figure*}[!th]
  \begin{center}
    \includegraphics[width=0.90\linewidth]{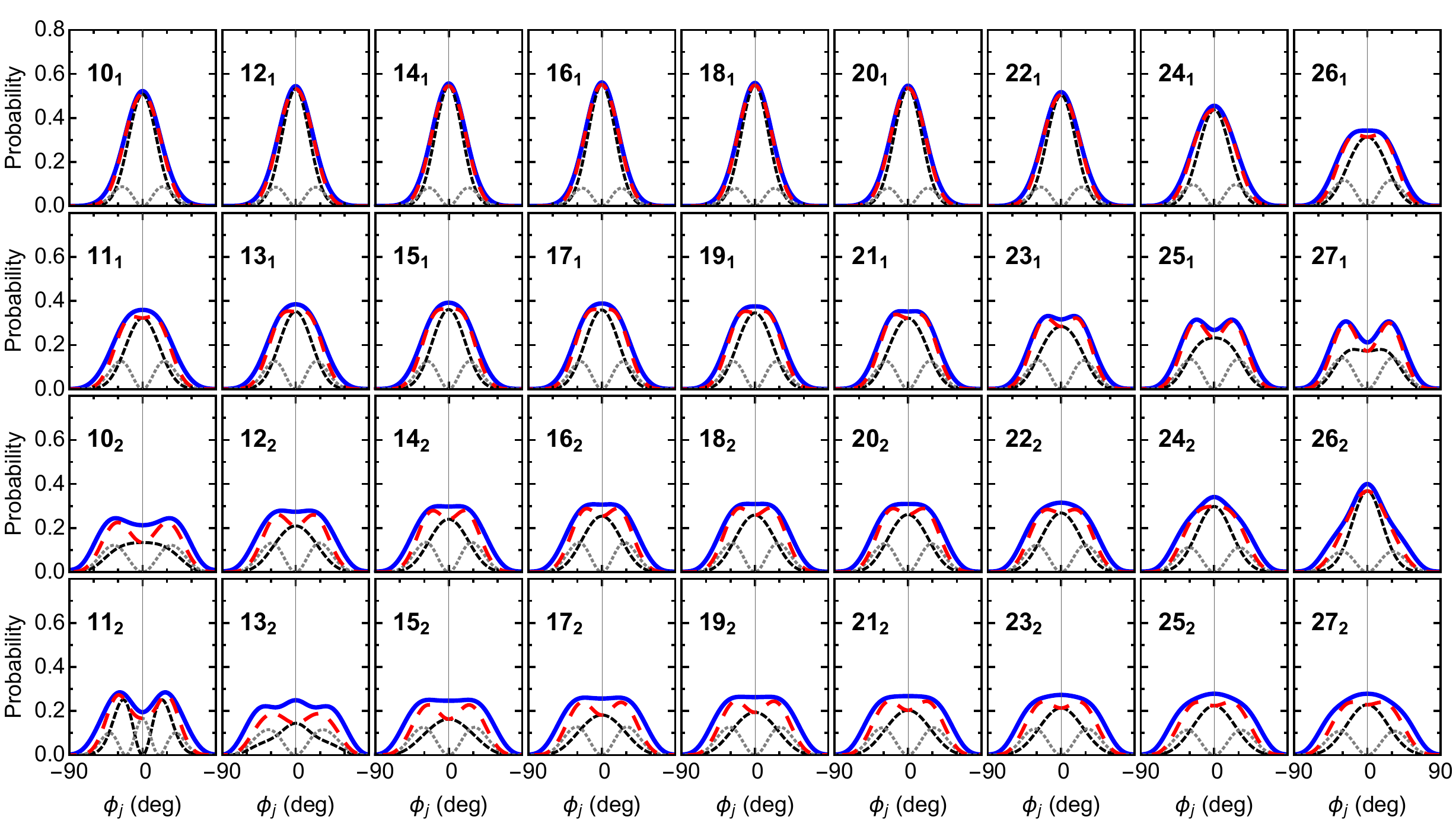}
    \caption{(Color online) Eigenstate decomposition of the SSS probability
    density $P(\phi_j)$, $P_{12}(\phi_j)$, $P_{34}(\phi_j)$, and
    $P_{1234}(\phi_j)$ of the two-proton angular
    momentum for the PTR states $n=0$-$3$ states in $^{130}$Ba.
    Same color coding as in Fig.~\ref{f:SSS_DM_total_135Pr}.}\label{f:SSS_DM_proton_130Ba}
  \end{center}
\end{figure*}

\begin{figure*}[!ht]
  \begin{center}
    \includegraphics[width=0.90\linewidth]{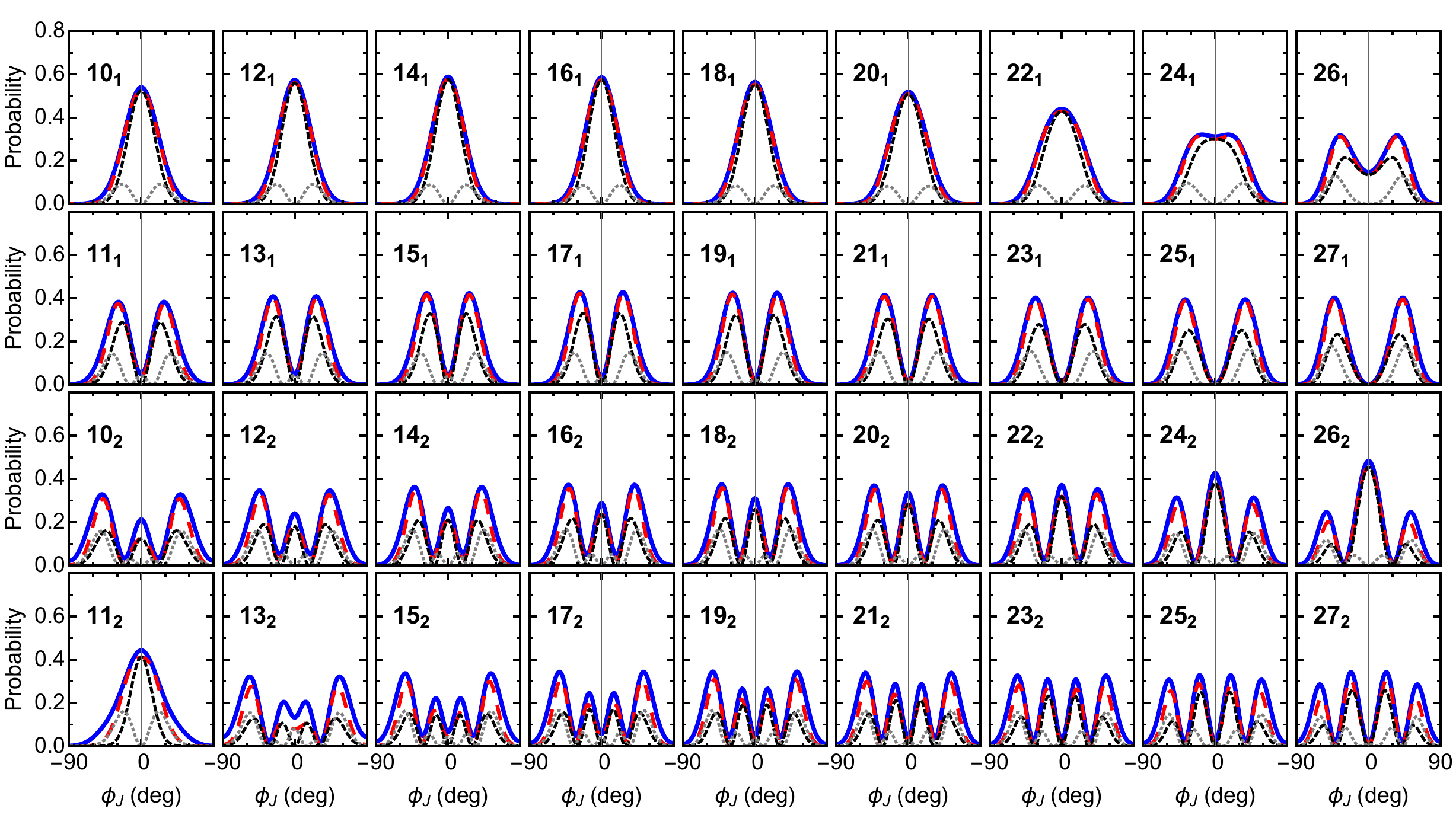}
    \caption{(Color online) Eigenstate decomposition of the SSS probability
    density $P(\phi_J)$, $P_{12}(\phi_J)$, $P_{34}(\phi_J)$, and
    $P_{1234}(\phi_J)$ of the total angular momentum for
    the PTR $n=0$-$3$ states in $^{130}$Ba. Same color coding
    as in Fig.~\ref{f:SSS_DM_total_135Pr}.}\label{f:SSS_DM_total_130Ba}
  \end{center}
\end{figure*}

Figure~\ref{f:Eigenvalue_130Ba} displays the eigenvalues of the reduced
density matrix as functions of spin for the $n=0$, $1$, $2$, and $3$ states
in $^{130}$Ba. Unlike the case of $^{135}$Pr, there are no
degenerate pairs $p_1=p_2$, $p_3=p_4$, $...$ of the eigenvalues of the
reduced density matrices in $^{130}$Ba because the dimensions of
even-$K$ and odd-$K$ bases are different. However, the spectra
consist of pairs of very close eigenvalues $p_1\approx p_2$,
$p_3\approx p_4$, $...$, which are not distinguishable in the figure.
Like in the case of $^{135}$Pr, the first two pairs of eigenvalues
dominate the spectrum for all $n=0$, 1, 2, and 3 states.

Figure~\ref{f:Ew_S_Delta_130Ba} shows the excitation energies
$\Delta E_{n}(I)$, the vN entropy $S$, $S^{(12)}$, and $S^{(1234)}$,
as well as the decoherence measures $\Delta_K$
and $\Delta_{\textrm{SSS}}$ of the wobbling bands in $^{130}$Ba
with wobbling numbers $n=0$, 1, 2, and 3. As discussed in our
previous work~\cite{Q.B.Chen2024PRC_v1}, the TW regime in
$^{130}$Ba extends to $I = 22$, beyond which the regime
transitions into the flip regime. Accordingly, the energy
spectrum evolves from the equidistant harmonic TW pattern at
low $I$ values to $\Delta E_{n=1}(I) \approx 0$ at $I = 26$,
signifying the onset of TW instability.

\begin{figure*}[!th]
  \begin{center}
    \includegraphics[width=0.95\linewidth]{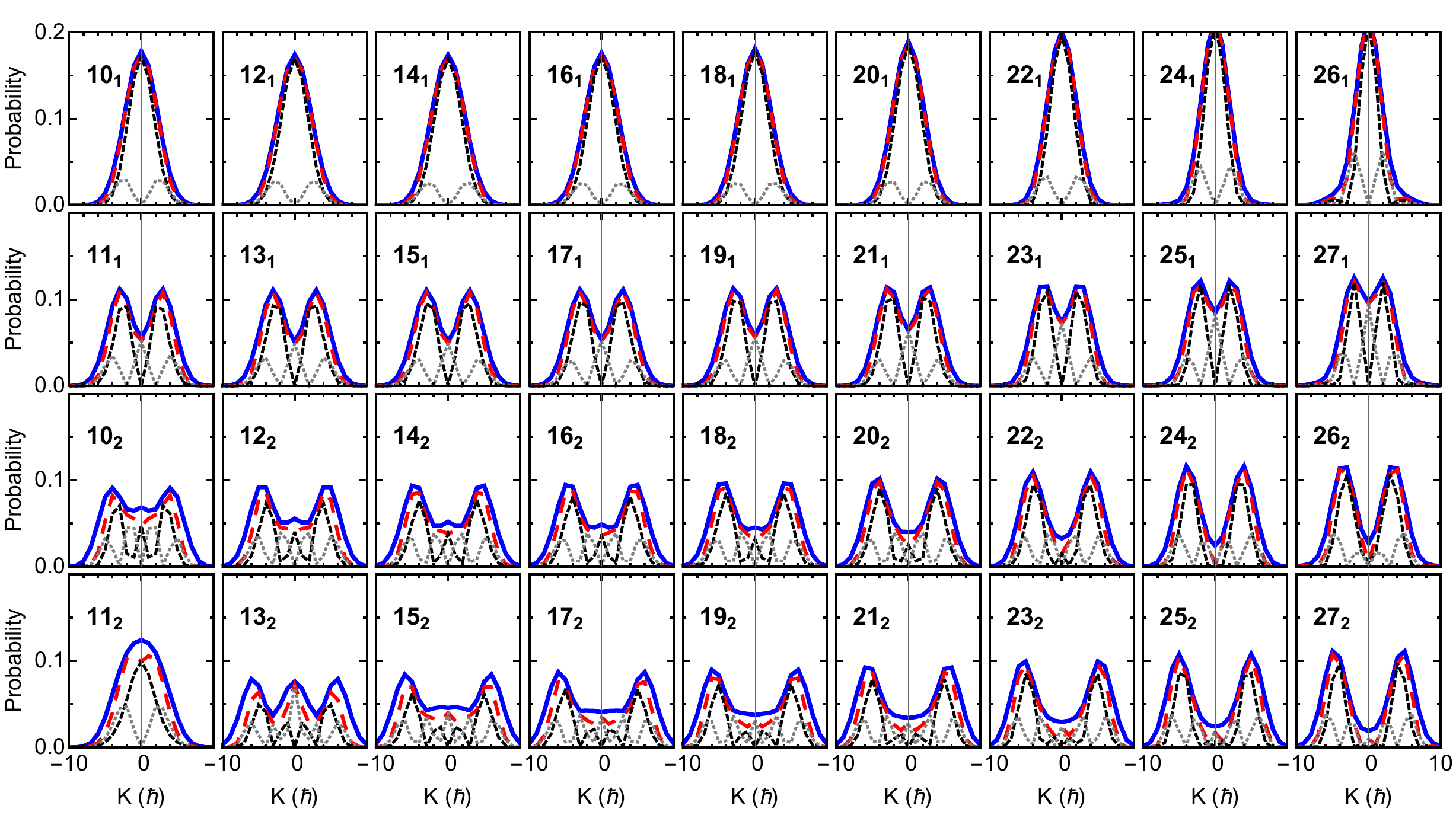}
    \caption{(Color online) Decomposition the probability distribution
    of the total angular momentum projection on the $l$ axis $P(K)$,
    $P_{12}(K)$, $P_{34}(K)$, and $P_{1234}(K)$
    for the $n=0$-$3$ states of $^{130}$Ba calculated by PTR.
    Only the region $-10\leq K\leq 10$ is shown. Same color
    coding as in Fig.~\ref{f:SSS_DM_total_135Pr}.}\label{f:PK_130Ba}
  \end{center}
\end{figure*}

The vN entropy $S^{(12)}$ of the four PTR states is about 0.25
($=0.74$ in natural units), which is very close to the value of
$^{135}$Pr. The vN entropy $S^{(1234)}$ is confined to a narrow
band around 0.4 ($=1.2$ in natural units) which agrees with
the band in $^{135}$Pr. We attribute the agreement to same number of
eigenstates (2 and 4, respectively) taken into account in the sub-matrices.
Compared with the TW region in $^{135}$Pr, the full vN entropy $S$
shows more clearly the expected increasing order with $n$, which reflects
the larger number of states in two-quasiparticle basis than in the
one-quasiparticle basis (12 vs. 21, respectively). Like in $^{135}$Pr,
the four states come together near 0.6 with increasing $I$, and their
trend indicates that the order becomes scrambled.

The function $S(I)$ reflects the $I$ dependence of the eigenvalues
$p_m$ in Fig.~\ref{f:Eigenvalue_130Ba}. The approach of $(p_1, p_2)$
and $(p_3, p_4)$ at large $I$ for the $n=0$ state is seen as
the increase of the entropy in Fig.~\ref{f:Ew_S_Delta_130Ba}(b).
For the $n=1$ state the distancing of $(p_1, p_2)$ from $(p_3, p_4)$
is seen as the decrease of the entropy. For the $n=3$ state,
there is an abrupt decrease of $(p_1, p_2)$ from $I=11$ to $I=13$.
This is attributed to the fact that the $11_2$ state has
a SP structure, while the $13_2$ state has
$n=3$ character. The structural change is reflected by the
entropy increase in Fig.~\ref{f:Ew_S_Delta_130Ba}(b).

In analogy to $^{135}$Pr, the incoherence measures $\Delta_K$
and $\Delta_{\textrm{SSS}}$ fall into a interval between about
0.15 and 0.35, and do not show a systematic $I$ and $n$ dependence.
These values suggest that the density matrix exhibits partial coherence.

Figures ~\ref{f:SSS_DM_proton_130Ba} and ~\ref{f:SSS_DM_total_130Ba}
show, respectively, the eigenstate decomposition of the $\bm{j}$-SSS
$P(\phi_j)$ and $\bm{J}$-SSS $P(\phi_J)$ distributions for the states
in $^{130}$Ba. Similar to $^{135}$Pr, the function $P_{1234}(\phi_J)$
well approximates the total distribution $P(\phi_J)$.

Figure~\ref{f:SSS_DM_proton_130Ba} shows that, as for $^{135}$Pr, the
probability densities $P_{12}(\phi_j)$ have a peak at $\phi_j=0^\circ$,
indicating the alignment of the proton pair with the $s$ axis, and
$P_{34}(\phi_j)$ are the double-peak distributions, which account for the
realignment of the proton pair. The contribution of $P_{34}(\phi_j)$
becomes more important with the wobbling number $n$. The ``noise" from
the $m>4$ eigenstates washes out the structure of $P_{1234}(\phi_j)$
to some extent.

The SP structure of the PTR state $11_2$ is recognized
as the double hump of the leading term $P_{12}(\phi_j)$ and the triple hump of
$P_{34}(\phi_j)$ (compare with Fig.~\ref{f:SSS_DM_SP_135Pr}, states $8.5_3$,
$10.5_3$, $12.5_3$). The function $P(\phi_J)$ in Fig.~\ref{f:SSS_DM_total_130Ba}
has one peak at $\phi_J=0^\circ$ that is characteristic for the signature
partner state.

As illustrated in Fig.~\ref{f:SSS_DM_total_130Ba},
the $\bm{J}$-SSS probability $P(\phi_J)$ for yrast states of the total
angular momentum exhibits a pronounced peak at $\phi_J = 0^\circ$ for
$I \leq 22$. This peak suggests that the total angular momentum is
predominantly in the lowest potential state aligned with $\phi_J = 0^\circ$
(c.f. Fig.~10 of Ref.~\cite{Q.B.Chen2024PRC_v1}). As the
spin $I$ increases, the potential becomes increasingly softer,
resulting in a broadening of this peak. For $I = 24$, this peak
becomes unstable and shifts towards $\phi_J = \pm 90^\circ$,
signifying the onset of a TW instability.

The functions $P(\phi_J)$ for the remaining PTR states show oscillation
that are typical for $n=1$, 2, and 3 states in a potential with a center
that shift from $\phi_J=0$ to $\pm 90^\circ$. The corresponding shift
of maximal amplitude from $\phi_J=0$ to $\pm 90^\circ$ is the reason
why in Fig.~\ref{f:SSS_DM_proton_130Ba} for the $n=2$ and 3 states
$P_{12}(\phi_j)$ is comparable with $P_{34}(\phi_j)$ at low $I$
while $P_{12}(\phi_j)$ is dominates $P_{34}(\phi_j)$ at large $I$.
This, at first thought, unexpected trend reflects that $\bm{j}$
adiabatically follows $\bm{J}$ (see next section).

Furthermore, Fig.~\ref{f:PK_130Ba} show the decomposition of the
$P(K)$ distributions for the PTR $n=0$-$3$ states in $^{130}$Ba.
The function $P_{1234}(K)$ well reproduces the total distribution
$P(K)$. With increasing $n$ or $I$, $P_{34}(K)$ plays more and more
important role.

\section{Entanglement and adiabatic approximation}

In Ref.~\cite{Q.B.Chen2024PRC_v1}, we derived a collective
Hamiltonian (CH) for the $\bm{J}$-degree of freedom by applying
the classical adiabatic approximation to the $\bm{j}$-degree of freedom.
This detour via a classical Hamiltonian and its re-quantization results
in a collective wave function with a pure density matrix. In this section
we discuss the relation with the concept of coherence based on
the eigenstate decomposition. The adiabatic approximation relies on
the a fast time scale for $\bm{j}$ compared to a slow time scale
for $\bm{J}$. That is, the lower the wobbling excitations the
better the description by the CH.

The reduced density matrix $\rho_{\phi_J,\phi_J'}$  is approximated
 by a pure density matrix $\rho_{\phi_J,\phi_J'}^{(\textrm{CH})}$
generated from the collective eigenfunction of the effective CH in
the $\bm{J}$-degree of freedom. The coupling to the  $\bm{j}$-degree of
freedom is taken into account assuming that $\bm{j}$ follows $\bm{J}$
in an adiabatic way, which is an alternative to describing the entanglement
of the two degrees of freedom by means of the Schmidt decomposition.
In detail, this means that the total angular momentum operator
$\bm{\hat J}$ in the PTR Hamiltonian (\ref{eq:HPTR}) is
replaced by the c-number $\bm{J}$. The adiabatic energy $E(\bm{J})$
is calculated by either diagonalizing the parametric PTR Hamitonian
$H_{\textrm{PTR}}(\bm{J},\bm{\hat j})$ and select the lowest eigenvalue
(tilted axis cranking) or simply take the minimum with respect to
$\bm{j}$ for fixed $\bm{J}$ of the classical energy
$E(\bm{J},\bm{j})$. The adiabatic energy $E(\phi_J,J_3)$ has the form of a narrow valley
along  $J_3=0$, which is approximated by
\begin{align}\label{eq:Ead}
 E(\phi_J,J_3)=  \frac{J_3^2}{2B(\phi_J)}+V(\phi_J)
\end{align}
in deriving the CH by re-quantization $E(\phi_J,J_3)\rightarrow
H_{\textrm{CH}}(\phi_J,\hat J_3)$.

Figures 5 and 6 of Ref.~\cite{Q.B.Chen2024PRC_v1} compare the approximate
CH energies and $E2$ transition probabilities for $^{135}$Pr with the
the PTR values. Figures 11 and 12 of Ref.~\cite{Q.B.Chen2024PRC_v1}
and Fig.~\ref{f:BE2_130Ba} provide the same comparison
for $^{130}$Ba. Generally, the CH approximation works quite well for $^{130}$Ba
for all $n=0$, 1, 2, and 3 states into the flip region. For $^{135}$Pr it gives
a fair description for the $n=0$, 1, and 2 states below the flip region.

\subsection{Comparison of the probability densities}\label{sec:P-CH}

The CH approximates the PTR states by a wave function in the
$\bm{J}$-degree of freedom, which corresponds to a pure density matrix. In order
to assess the quality, we compare the pertaining probability distribution
$P_{\textrm{CH}}$ with the full distribution $P$ and the distribution
calculated from the pure sub-density matrix with the largest
probability $P_{12}/(p_1+p_2)$. According to the Schmidt decomposition,
the latter is the ``best" representation as a product wave function
in the $\bm{J}$ and $\bm{j}$ degrees of freedom.

As the CH couples only matrix elements $\Delta K=0$ and $\pm 2$, the eigenstates
appear in pairs. In the case of odd-$A$ nucleus, the two eigenstates have
$K=I-\textrm{even}$ or $K=I-1-\textrm{even}$ and the two eigenvalues
are the same. In the case of even-$A$ nucleus, the two eigenstates have
even-$K$ or odd-$K$ and the two eigenvalues are nearly the same.
In the following figures we display the mean value of the two
probability distributions $P_{\textrm{CH}}=\left(P_{\textrm{CH}}^{(1)}
+P_{\textrm{CH}}^{(2)}\right)/2$.

\begin{figure*}[!ht]
  \begin{center}
    \includegraphics[width=0.90\linewidth]{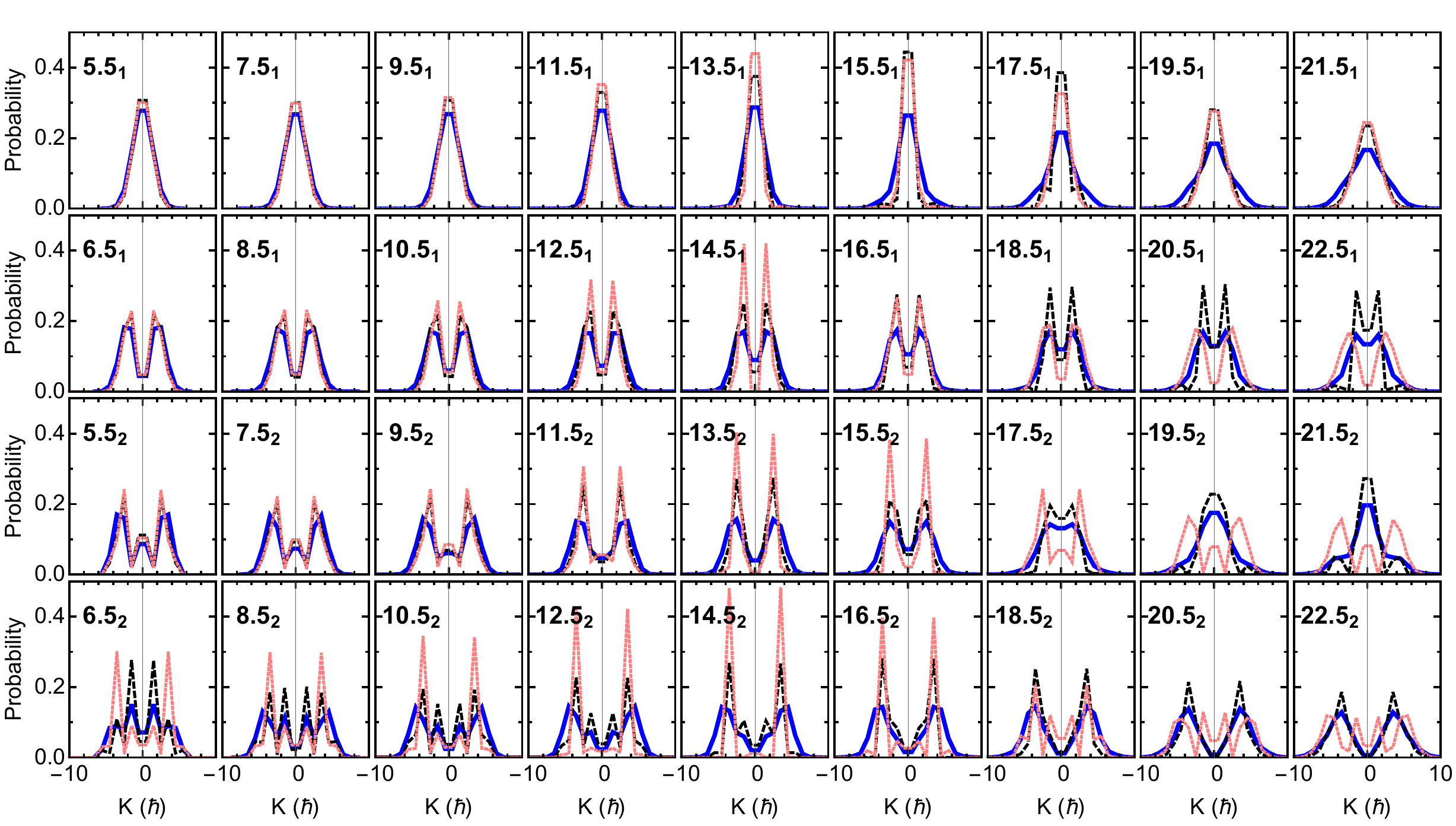}
    \caption{(Color online) Probability distribution of the total angular
     momentum projection on the $l$ axis $P(K)$ (thick blue
     full curves) and normalized $P_{12}(K)/(p_{1}+p_{2})$ (black
     short dashed) obtained by PTR in comparison with the $P_{\textrm{CH}}(K)$
     obtained from the collective Hamiltonian (thin pink dotted)~\cite{Q.B.Chen2024PRC_v1}
     for the PTR states of $^{135}$Pr. Only the region $-9.5\leq K\leq 9.5$
    is shown. }\label{f:PK_P12_CH_135Pr}
  \end{center}
\end{figure*}

Figure~\ref{f:PK_P12_CH_135Pr} compares the full distribution
$P(K)$ and the normalized distribution $P_{12}(K)/(p_1+p_2)$
with the  probability $P_{\textrm{CH}}(K)$ obtained from
the adiabatic CH for the $n=0$, 1, 2, and 3 states of $^{135}$Pr.
As seen, $P_{\textrm{CH}}(K)$ agrees  well with $P_{12}(K)$ for
the $n=0$ states in the first row. For the $n=1$ states
$13/2_1$-$33/2_1$ it also agrees reasonably well with
$P_{12}(K)/(p_1+p_2)$ being somewhat too spiky in the flip region.
For the states $37/2_1$-$45/2_1$ the distribution $P_{\textrm{CH}}(K)$
differs from $P(K)$ and $P_{12}(K)/(p_1+p_2)$. Also for the $n=2$
states $11/2_2$-$31/2_2$ the distribution $P_{\textrm{CH}}(K)$
reasonably well reproduces $P_{12}(K)/(p_1+p_2)$ being somewhat
too spiky in the flip region. It differs from $P(K)$ and $P_{12}(K)/(p_1+p_2)$
for the states $35/2_2$-$43/2_2$. The state $13/2_2$ has SP character
and cannot be described by the CH. For the states $17/2_2$-$25/2_2$
the distribution $P_{\textrm{CH}}(K)$ roughly traces $P_{12}(K)/(p_1+p_2)$,
which has $n=3$ character. The adiabatic approximation fails for larger $I_2$.

\begin{figure*}[!ht]
  \begin{center}
    \includegraphics[width=0.90\linewidth]{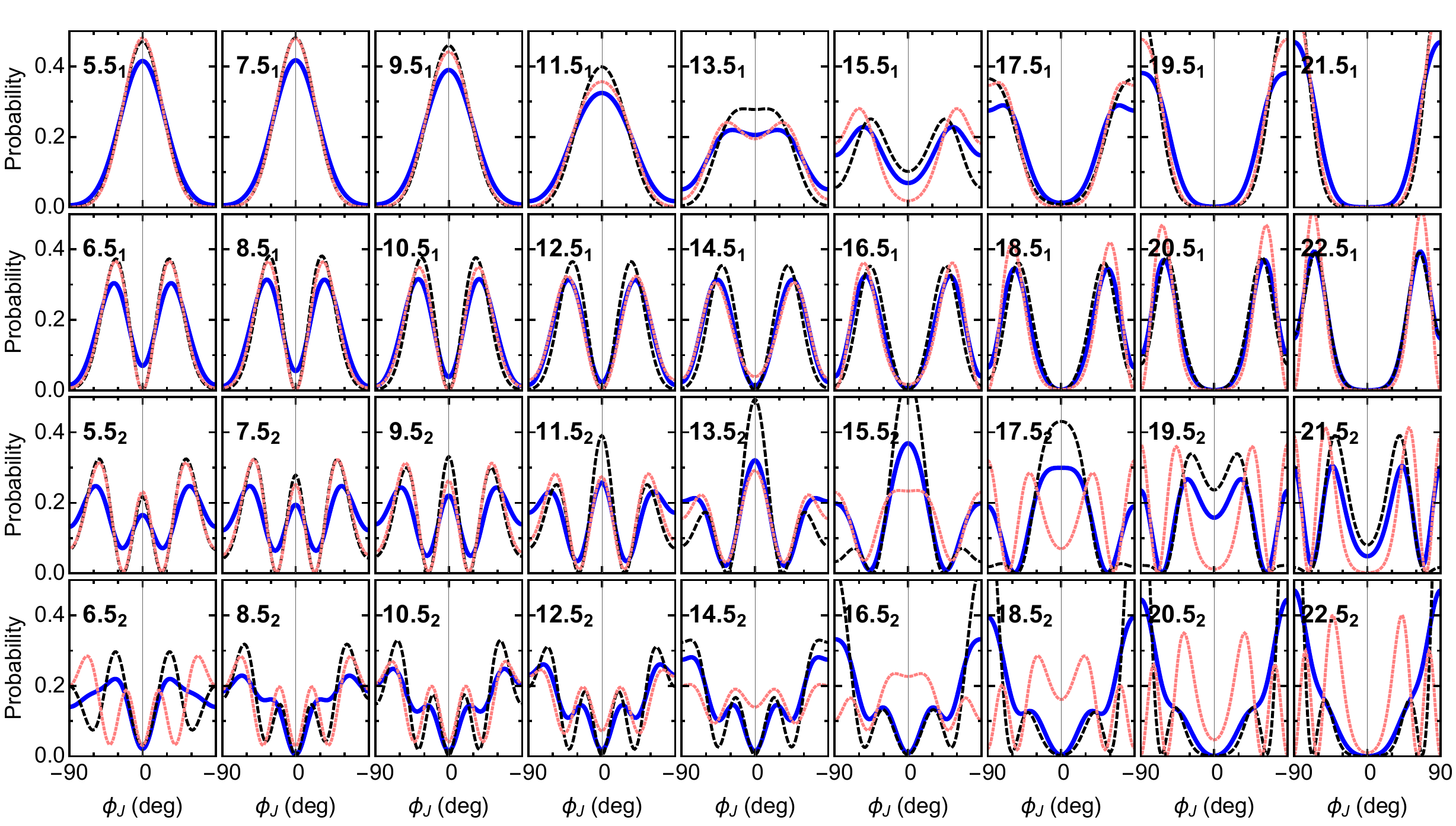}
    \caption{(Color online) The SSS probability density $P(\phi_J)$ (thick blue
     full curves) and normalized SSS $P_{12}(\phi_J)/(p_{1}+p_{2})$ (black
     short dashed) of the total angular momenta in comparison with the
     SSS probability density $P_{\textrm{CH}}(\phi_J)$ obtained from
     the collective Hamiltonian (thin pink dotted)~\cite{Q.B.Chen2024PRC_v1}
     for the PTR states of $^{135}$Pr.}\label{f:SSS_rho12Nor_CH_135Pr}
  \end{center}
\end{figure*}

Figure~\ref{f:SSS_rho12Nor_CH_135Pr} compares the full distribution
$P(\phi_J)$ and the normalized distribution $P_{12}(\phi_J)/(p_1+p_2)$
with the SSS probability $P_{\textrm{CH}}(\phi_J)$ obtained from
the adiabatic CH for the $n=0$, 1, 2, and 3 states of $^{135}$Pr.

As seen in Fig.~\ref{f:SSS_rho12Nor_CH_135Pr}, $P_{\textrm{CH}}(\phi_J)$ of the
$n=0$ states $11/2_1$ and $15/2_1$ agrees very well with $P_{12}(\phi_J)$,
from which it deviates increasing  $I$ such that it approaches $P(\phi_J)$
for the flip state $27/2_1$.  For the state $31/2_1$ above the flip region
$P_{\textrm{CH}}(\phi_J)$ deviates from $P(\phi_J)$ such that
it approaches $P_{12}(\phi_J)$ with increasing $I$.

The wobbling energy $\Delta E_1$ is minimal for $I=27/2$, that is, the adiabatic
approximation works best. The function $P_{\textrm{CH}}(\phi_J)$ exhibits
a more pronounced dip at $\phi_J=0^\circ$ than $P(\phi_J)$.
The reason for the difference is the following. The adiabatic classical potential
accounts for the response the particle to the Coriolis force for each orientation
of the total angular momentum $\bm{J}$. The term $\rho_{KK'}^{(12)}$
is selected such that the $\bm{j}$- and $\bm{J}$-degrees of
freedom are decoupled as best as possible, which corresponds to a state of the
particle with an average orientation $\bm{j}=(\sqrt{\langle \hat{j}^2_1\rangle},
\sqrt{\langle \hat{j}^2_2\rangle},\sqrt{\langle \hat{j}^2_3\rangle})$.

For the $n=1$ states in the second row of Fig.~\ref{f:SSS_rho12Nor_CH_135Pr}
one observes a similar $I$ dependence. For the state $13/2_1$ the function
$P_{\textrm{CH}}(\phi_J)$ agrees with $P_{12}(\phi_J)$. With increasing
$I$ it changes to the full $P(\phi_J)$ for the flip state $29/2_1$.
Above the flip region $P_{\textrm{CH}}(\phi_J)$ progressively deviates from both
$P_{12}(\phi_J)$ and $P(\phi_J)$, which indicates that the adiabatic approximation
becomes problematic.

The $n=2$ states in the third row of Fig.~\ref{f:SSS_rho12Nor_CH_135Pr}
show the same $I$ dependence of $P_{\textrm{CH}}(\phi_J)$ from $P_{12}(\phi_J)$
for $11/2_2$ to $P(\phi_J)$ for $27/2_2$. Above the flip region
$P_{\textrm{CH}}(\phi_J)$ strongly  deviates from both $P_{12}(\phi_J)$
and $P(\phi_J)$, which indicates that the adiabatic approximation fails.
As discussed above, the incoherent term $P_{34}(\phi_J)$
increases with the wobbling number $n$ and so the difference
between $P_{\textrm{CH}}(\phi_J)$ and $P(\phi_J)$ at low spin.

For the states $17/2_2$, $21/2_2$, $25/2_2$, and $29/2_2$ in the lowest
row of Fig.~\ref{f:SSS_rho12Nor_CH_135Pr}, $P(\phi_J)$ has a washed-out
shape of  $P_{\textrm{CH}}(\phi_J)$, which is expected because the
incoherent terms $P_{34}(\phi_J)$ are the largest. As discussed, the
state $13/2_2$ has SP structure. The CH generates a collective $n=3$-type
state with the characteristic four maxima. The adiabatic approximation
fails for $I>29/2$.

In general, the distributions $P_{\textrm{CH}}(K)$ are more spiky than
$P_{12}(K)/(p_{1}+p_{2})$ while the distributions $P_{\textrm{CH}}(\phi_J)$
are smoother than $P_{12}(\phi_J)/(p_{1}+p_{2})$, which reflects their
relation by a Fourier transform.

\begin{figure*}[!ht]
  \begin{center}
    \includegraphics[width=0.90\linewidth]{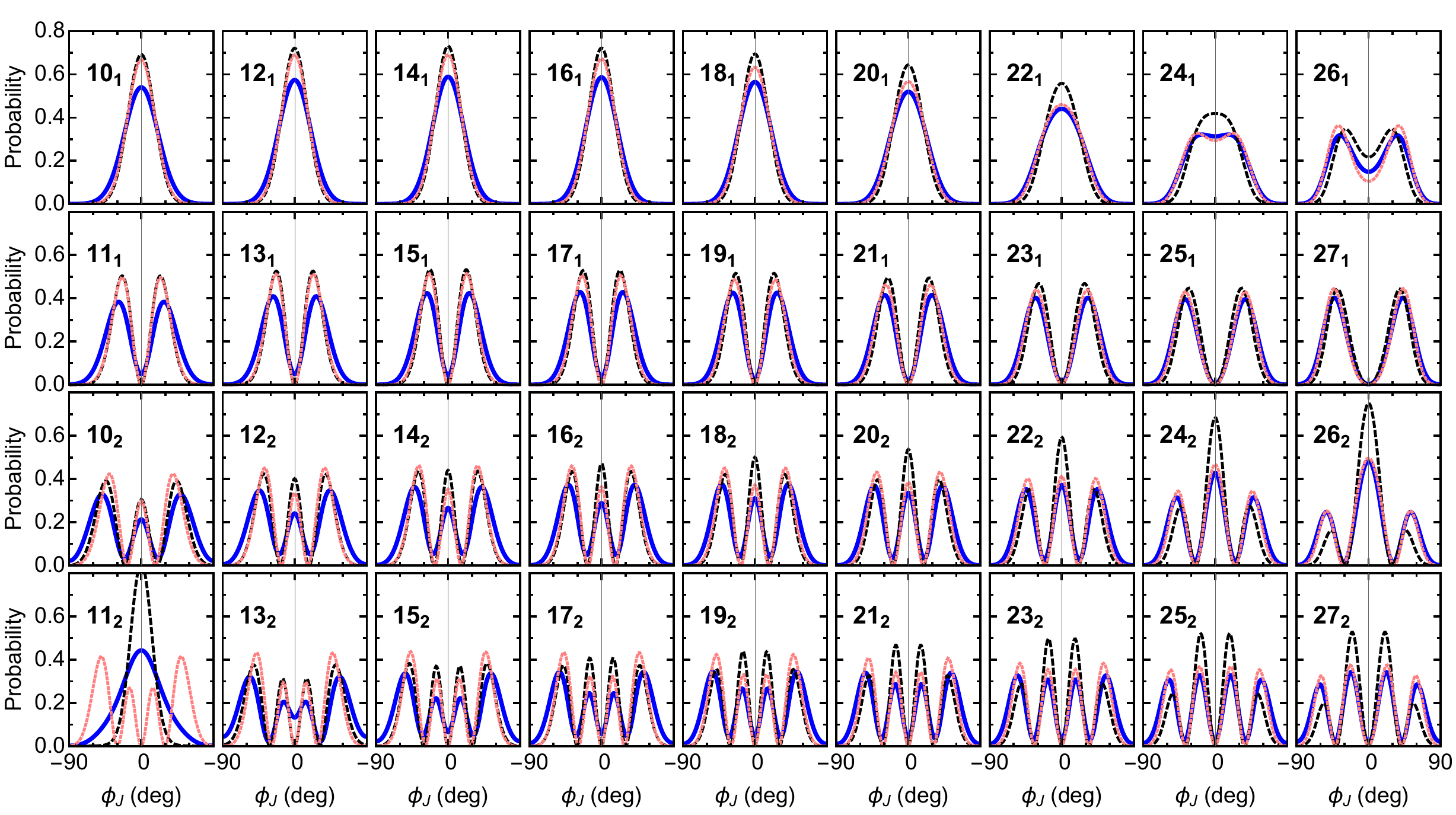}
    \caption{(Color online) The SSS probability density $P(\phi_J)$ (thick blue
     full curves) and normalized SSS $P_{12}(\phi_J)/(p_{1}+p_{2})$ (black
     short dashed) of the total angular momenta in comparison with the
     SSS probability density $P_{\textrm{CH}}(\phi_J)$ obtained from
     the collective Hamiltonian (thin pink dotted)~\cite{Q.B.Chen2024PRC_v1}
     for the PTR states in $^{130}$Ba.}\label{f:SSS_rho12Nor_CH_130Ba}
  \end{center}
\end{figure*}

\begin{figure*}[!ht]
  \begin{center}
    \includegraphics[width=0.90\linewidth]{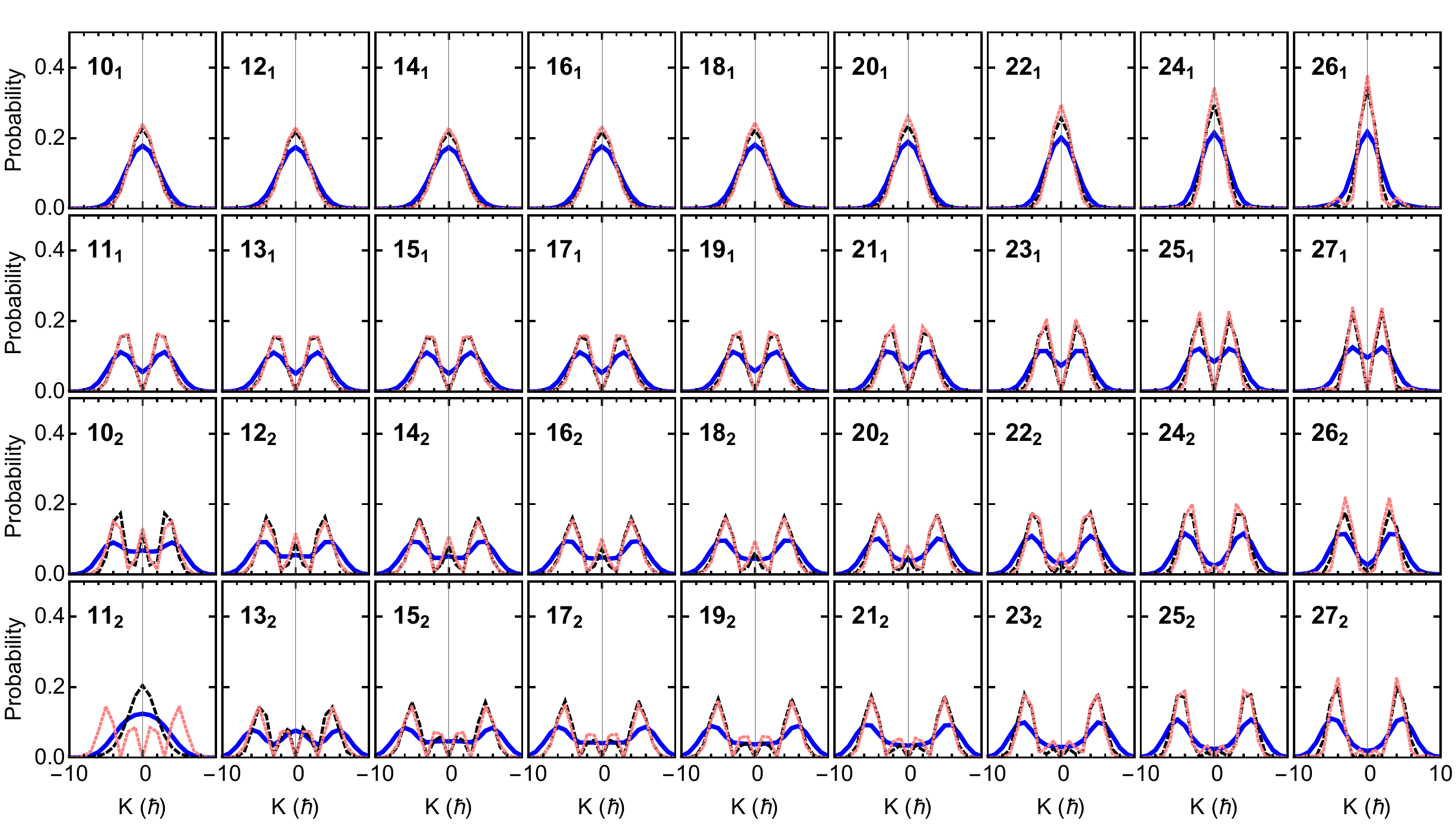}
    \caption{(Color online) Probability distribution of the total angular
     momentum projection on the $l$ axis $P(K)$ (thick blue
     full curves) and normalized $P_{12}(K)/(p_{1}+p_{2})$ (black
     short dashed) obtained by PTR in comparison with the $P_{\textrm{CH}}(K)$
     obtained from the collective Hamiltonian (thin pink
     dotted)~\cite{Q.B.Chen2024PRC_v1} for the PTR states of
     $^{130}$Ba. Only the region $-10\leq K\leq 10$
    is shown. }\label{f:PK_P12_CH_130Ba}
  \end{center}
\end{figure*}

For $^{130}$Ba the functions $P_{\textrm{CH}}(\phi_J)$
in Fig.~\ref{f:SSS_rho12Nor_CH_130Ba} characterize the mode, as $n=0$, 1, 2,
and 3 TW excitations, where $n$ counts the respective number of zeroes.
As illustrated in Fig.~10 of Ref.~\cite{Q.B.Chen2024PRC_v1}, the adiabatic
potential of the CH is approximately quadratic around $\phi_J=0^\circ$ at
low $I$ and develops a flat bottom with increasing $I$. The peak heights of
$P_{\textrm{CH}}(\phi_J)$ reflect the change of the adiabatic potential.
In the harmonic TW region the maxima's height increase with $\vert \phi\vert$
like for the corresponding Hermite polynomials. For $I=24,~25$ the flip
region is reached. The pattern changes to equal heights with the tendency
of higher peaks near $\phi_J=0^\circ$, which suggest the transition to
the LW regime.

Figure~\ref{f:SSS_rho12Nor_CH_130Ba} compares the full distribution
$P(\phi_J)$ and the normalized distribution $P_{12}(\phi_J)/(p_1+p_2)$
with the SSS probability $P_{\textrm{CH}}(\phi_J)$ obtained from
the adiabatic CH for $^{130}$Ba. Except the SP state $11_2$, which has
a SP structure, the $P_{\textrm{CH}}(\phi_J)$ agree rather well
with the scaled $\bm{J}$-SSS distributions \mbox{$P_{12}(\phi_J)/(p_1+p_2)$}
for low $I$. Analog to one-quasiparticle cases in $^{135}$Pr,
for $I>17$ the functions $P_{\textrm{CH}}(\phi_J)$ notably
deviate from the scaled probability densities $P_{12}(\phi_J)/(p_1+p_2)$
and approach the full density $P(\phi_J)$ in the flip region,
where the adiabatic approximation works best.

As discussed in the Appendix~\ref{app2}, the $E2$ transition probabilities
calculation can be approximated by replacing the trace of the
product of an intrinsic quadrupole operator with the density matrix by the
integral over the product of the classical quadrupole moments (\ref{eq:Q0Q2class})
with the probability densities $P(\phi_J)$,
$P_{12}(\phi_J)/(p_1+p_2)$, and $P_{\textrm{CH}}(\phi_J)$.
This explains why the CH transition probabilities in Fig.~\ref{f:BE2_130Ba}
change in the same way with $I$ as $P_{\textrm{CH}}(\phi_J)$.
At low $I$, where the adiabatic approximation is
less accurate, they agree with the values by $P_{12}(\phi_J)/(p_1+p_2)$
and for large $I$, where the adiabatic approximation is
more accurate, they agree with the PTR values by $P(\phi_J)$.

Figure~\ref{f:PK_P12_CH_130Ba} compares the full distribution
$P(K)$ and the normalized distribution $P_{12}(K)/(p_1+p_2)$
with the probability $P_{\textrm{CH}}(K)$ obtained from
the adiabatic CH for $^{130}$Ba. Similar to $^{135}$Pr,
$P_{\textrm{CH}}(K)$ agrees well with $P_{12}(K)/(p_1+p_2)$
for all $I$. The features of wobbling excitations with $n=0$, 1,
2, and 3 are reflected by the number of the nodes
in the $P_{\textrm{CH}}(K)$ plots. Only for $11_2$ state,
which has a SP structure, $P_{\textrm{CH}}(K)$ fails to
reproduce $P_{12}(K)/(p_1+p_2)$. The $E2$ matrix elements
do not depend on the basis they are calculated from.
This is not at variance with the fact that $P_{\textrm{CH}}(K)$ agrees
with $P_{12}(K)/(p_1+p_2)$ for all $I$ while $P_{\textrm{CH}}(\phi_J)$
deviates from it. The non-diagonal matrix elements of the
operator $\hat Q_{\pm 2}$ generate the differences in
a non-obvious way (see details in the Appendix~\ref{app1}).

The quadratic approximation (\ref{eq:Ead}) of the adiabatic energy
explains the different $I$ dependence of $P_{\textrm{CH}}$ in comparison with
$P$ and $P_{12}$ in Figs.~\ref{f:SSS_rho12Nor_CH_130Ba} and \ref{f:PK_P12_CH_130Ba}.
It maps out the potential $V(\phi_J)$ in detail while the kinetic term
is approximated to be quadratic in $\hat J_3$. The CH provides low-lying
wave functions in the $\phi_J$-degree of freedom that are the better the smaller
their kinetic energy is. As a consequence, $P_{\textrm{CH}}(\phi_J)$
approaches $P(\phi_J)$. This is not the case for the kinetic term which
does not fully map out $E(\phi_J,J_3)$ in $J_3$ direction. Moreover,
the SSS basis states are maximal localized in $\phi_J$ and maximal
uncertain in $K$ (constant weight).

\subsection{Comparison of the density matrices}\label{sec:DM-CH}

In calculating the reduced density matrix, the tracing out of
the $\bm{j}$-degree of freedom destroys to some extend
the coherence of the complete PTR wave function.
The properties of the $\bm{J}$-degrees of freedom are described
by the reduced density matrix, which cannot be further simplified.
As we discussed in the previous section, it represents
an ensemble of quantal states in the $\bm{J}$-subspace with the
probability $p_m$. These states are entangled with pertaining quantal
states $\bm{j}$-subspace, which appear with the same probability $p_m$.
That is, the combined system can be interpreted as an ensemble,
where $\bm{J}$ and $\bm{j}$ appear in pairs of quantal
states in the respective subspaces.

The description of the PTR system in terms of the effective CH
accounts for the entanglement by means of the  approximation
that $\bm{j}$ adiabatically follows $\bm{J}$, which implies
coherence. Accordingly the  pure density matrix $\rho^{\textrm{(CH)}}$
represents the coupled system. In the Appendix~\ref{app1},
we compare $\rho^{(\textrm{CH})}$ with the mixed matrix
$\rho$ and the pure sub-matrix $\rho^{(12)}$ for the PTR
states $10_1$ and $26_2$ in $^{130}$Ba in order to
illustrate their relations.

The density matrices $\rho_{KK'}$ and $\rho_{\phi_J,\phi_J'}$ are related to
each other by the Fourier transform between the $K$ and $\phi_J$ representations.
However, their diagonal matrix elements are not related by a simple Fourier
transform, because the basis change involves the non-diagonal matrix elements as well.
As discussed, the diagonal matrix elements $\rho^{(\textrm{CH})}_{KK}$ approximate
$\rho^{(12)}_{KK}/(p_1+p_2)$ of the pure sub-density with the largest weight in
$\rho_{KK'}$, which loosely speaking realizes as good as possible one factor of the product
wave function (with  $\rho^{(12)}_{kk'}/(p_1+p_2)$ being the other factor).
When the diagonal matrix elements
of $\rho^{(\textrm{CH})}_{\phi_J,\phi_J'}$ approach the
diagonal matrix elements of $\rho_{\phi_J,\phi_J'}$,
the non-diagonal matrix elements of the two matrices remain different.
However, these matrix elements are insignificant in calculating the
$E2$ transition matrix elements as discussed in the Appendix \ref{app2}.
An analog  consideration applies to the $M1$ operator, which involves
the adiabatic response of $\bm{j}$ explicitly. Hence, the concept
of a coherent wave function is appropriate as long as
this kind of quasi-local in $\phi_J$ operators are of interest.
Operators that involve highly non-diagonal matrix elements
are not amenable to the adiabatic approximation.

\begin{figure}[!th]
  \begin{center}
    \includegraphics[width=0.95\linewidth]{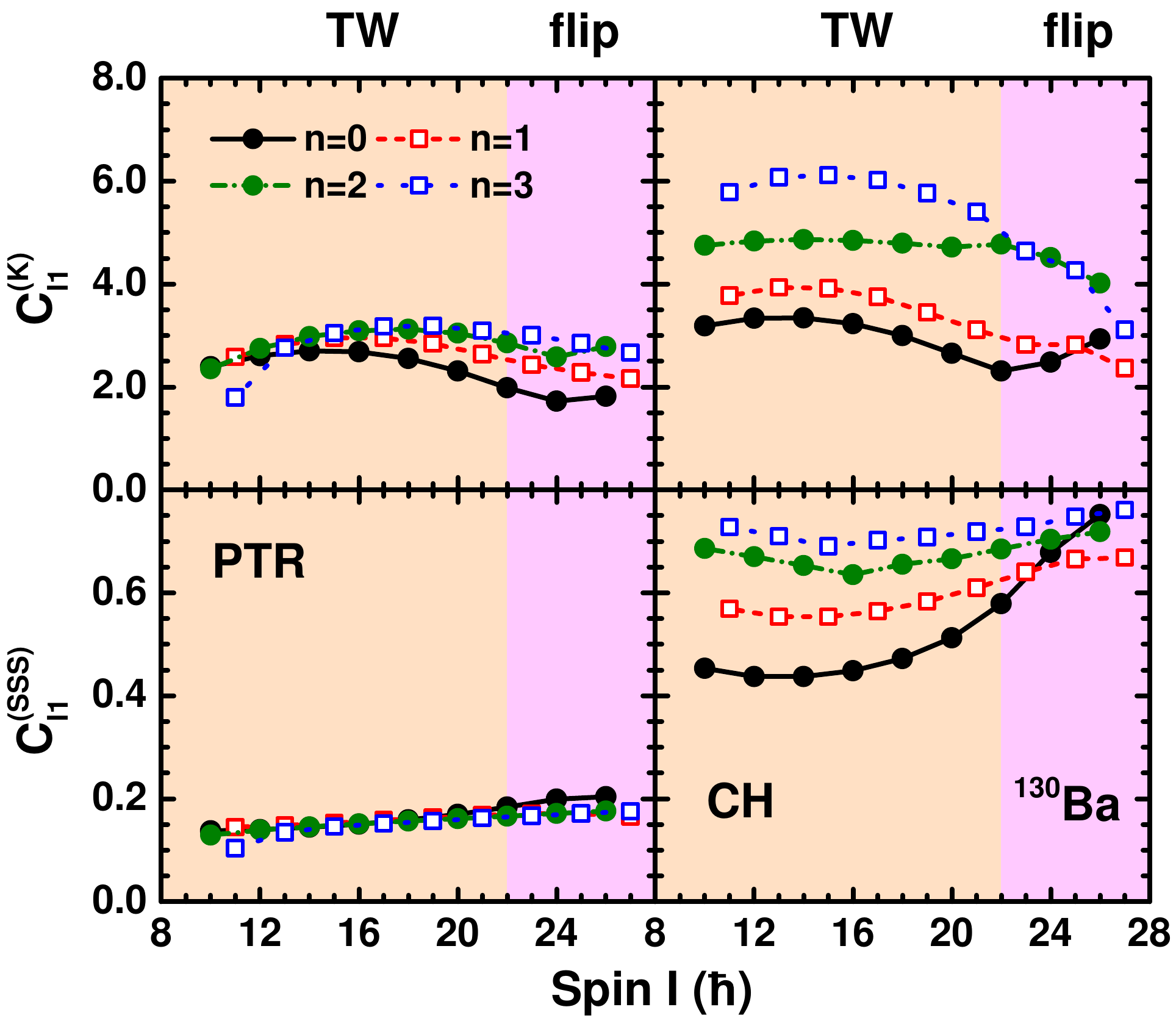}
    \caption{(Color online) Calculated $C_{l_1}^{(K)}$ and $C_{l_1}^{(\textrm{SSS})}$
    of the density matrices $\rho_{KK^\prime}$ and $\rho_{\phi_m,\phi_n}$
    as functions of spin by PTR and CH for the $n=0$-$3$ states in $^{130}$Ba.}
    \label{f:Cl1_130Ba}
  \end{center}
\end{figure}

To characterize the coherence properties of the density matrices of
PTR and CH, Fig.~\ref{f:Cl1_130Ba} further displays the $l_1$ norm
$C_{l_1}$ of density matrices $\rho_{KK^\prime}$ and $\rho_{\phi_m,\phi_n}$
as functions of spin $I$  for $n=0$-$3$ calculated from the PTR and
CH states in $^{130}$Ba. In general, the $C_{l_1}$ of CH is larger
than that of PTR, indicating that a larger number of non-diagonal
matrix elements is needed to generate the pure density matrix of a
coherent eigenstae of the CH. The $C_{l_1}$ norm increases with adding
wobbling quanta $n$ to the CH eigenstates, while
 $C_{l_1}$ of the PTR states does not change much.

\subsection{Comparison of electromagnetic properties}\label{sec:EM}

The $E2$ transition probabilities are obtained by tracing the
product of an intrinsic quadrupole operator (\ref{eq:Q0Q2}) with the
density matrix. The examples of density matrix and quadrupole matrix
elements are given in the Appendixes~\ref{app1} and \ref{app2},
respectively, for $10_1$ and $26_2$ states of $^{130}\textrm{Ba}$.

\begin{figure}[!ht]
  \begin{center}
    \includegraphics[width=0.85 \linewidth]{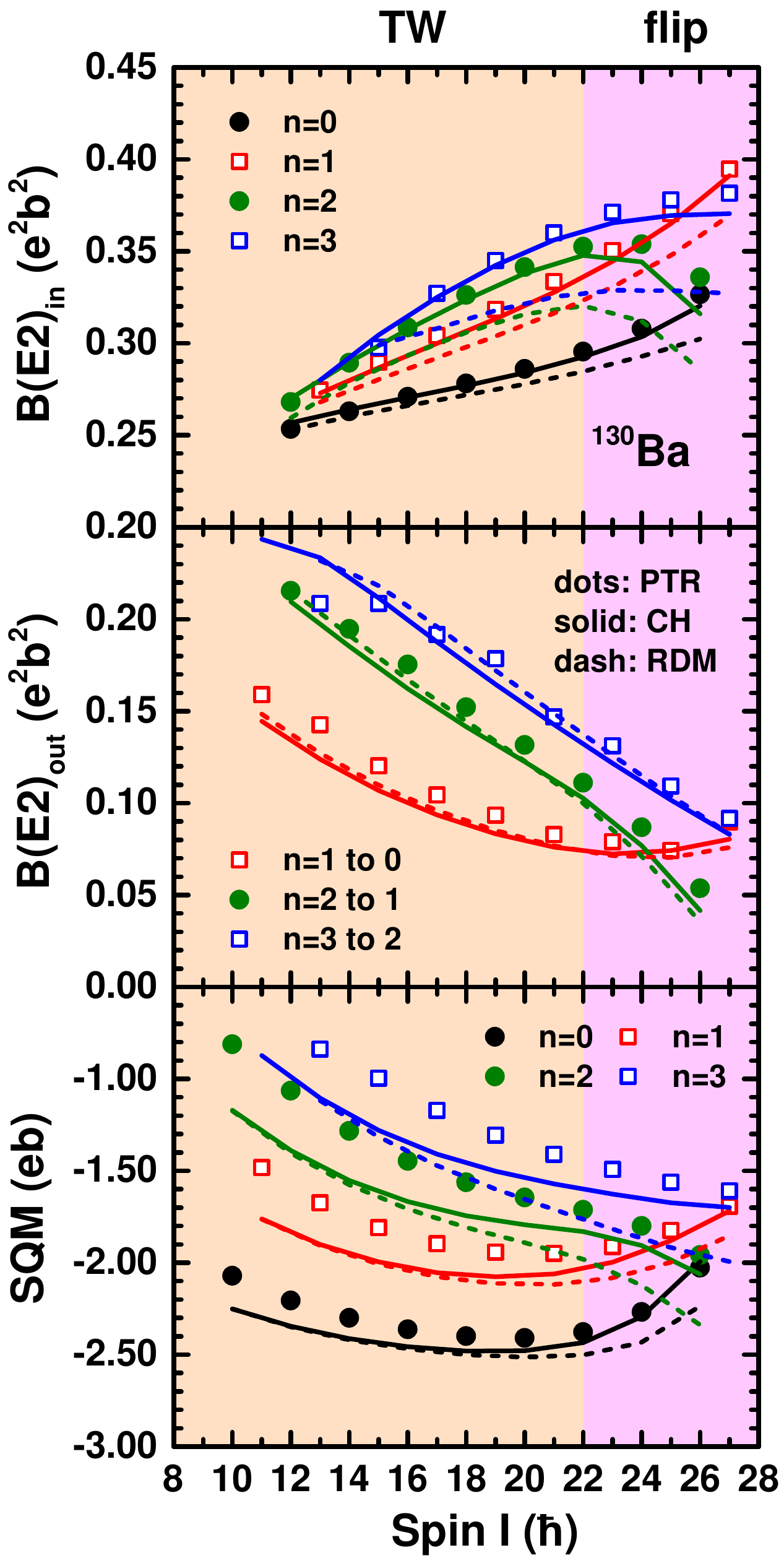}
    \caption{(Color online) Upper panel: In-band
    $B(E2)_{\textrm{in}}(I \to I-2)$ calculated by the PTR, CH, and
    sub-density matrix ($\rho^{(1)}$ or $\rho^{(2)}$, labeled as RDM)
    for the $n=0$, 1, 2, and 3 bands in $^{130}$Ba. Middle panel:
    Inter-band $B(E2)_{\textrm{out}}(I \to I-1)$.
    Lower panel: Spectroscopic quadrupole moments $Q(I)$.}
    \label{f:BE2_130Ba}
  \end{center}
\end{figure}

Figure~\ref{f:BE2_130Ba} shows the electromagnetic transition probabilities
and spectroscopic quadrupole moments (SQMs). The upper panel illustrates the in-band
$B(E2)_{\textrm{in}}(I \to I-2)$ transition probabilities for the $n=0$, 1, 2,
and 3 bands. The middle panel presents the inter-band
$B(E2)_{\textrm{out}}(I \to I-1)$ transition probabilities among these bands.
The strong collective $\Delta I = 1$ $E2$ transitions associated
with the wobbling motion are noted.
For $B(E2)_{\textrm{in}}(I \to I-2)$ the CH values  well agree with the PTR ones
 while for $B(E2)_{\textrm{out}}(I \to I-1)$ the
CH values are slightly lower than the PTR ones.
The results derived from the
normalized sub-density matrix ($\rho^{(1)}$ or $\rho^{(2)}$,
referred to as RDM),
for  $B(E2)_{\textrm{out}}(I \to I-1)$ are close to the CH values for all $I$.
For the $n=0$ and 1 bands, the
RDM $B(E2)_{\textrm{in}}(I \to I-2)$ values are somewhat lower
than those from PTR and CH, with progressively larger deviations
observed for the $n=2$ and 3 bands. This trend suggests that
contributions from higher-order density matrices, particularly
$\rho^{(3)}$ and $\rho^{(4)}$, become increasingly significant
in the higher $n$ states.

The SQMs in the bottom panel increase with $n$. Both CH
and RDM tend to underestimate the PTR results in the
low-spin region, where they are close together. The deviations from the
PTR values grow with $n$, which indicates that $\rho^{(3)}$
and $\rho^{(4)}$, become increasingly significant in the higher $n$ states.
As the spin increases differences between these models emerge,
with CH predictions converging to the PTR results. As already discussed
in the preceding sections, the relations between the PTR, CH and RDM results
can be elucidated by examining the SSS plots $P(\phi_J)$
presented in Fig.~\ref{f:SSS_rho12Nor_CH_130Ba}.


\section{Summary and Conclusions}

We investigated the entanglement between the total and the
quasi particle angular momenta, using the PTR model
studies~\cite{Q.B.Chen2022EPJA, Q.B.Chen2024PRC_v1} of $^{135}$Pr and
$^{130}$Ba as examples for one and two quasiparticles coupled
to a triaxial rotor. The study was carried out from two perspectives.
The first starts from the bi-partition of the coupled system into
the two subsystems, which are described by their respective
reduced density matrices. The entanglement is quantified by
the von Neumann (vN) entropy. It is found that the vN entropy
$S$ increases with excitation energy, that is, the particle and total
angular momenta become more entangled with the wobbling number $n$.
The $S$ growths with the value of $I$ in the region of transverse
wobbling (TW), because the Coriolis interaction becomes stronger.
Above the critical spin, where the TW mode becomes unstable,
$S$ does not change much with $I$, and the states group around
$S/S^{(\textrm{max})}=0.7$ with no specific order, which is
attributed to the smallness of the system.

Finite entropy implies a loss of coherence, that is the subsystems cannot be
completely described by a wave function of their own. In order to characterize
the decoherence we decomposed the reduced density matrices into their eigenstates,
each of which representing a pure density matrix (or equivalently the
wave function) of a  rotor in the presence of one or two quasiparticles in
a fixed  quantum state, which can be viewed as the frozen alignment
(FA) states of Ref.~\cite{Frauendorf2014PRC} with an individual
effective particle angular momentum that is not completely alignde
with the short axis.

We compared the probability distributions with respect to the angular
momentum projection $K$ onto the long axis and of the angle $\phi_J$
of the angular momentum in the short-medium plane (spin squeezed states-SSS)
of the full reduced density matrix with the pertaining distributions
calculated from the sub-density matrices for each eigenstate. It turned
out that to good approximation the full PTR probability distributions
can be interpreted as the incoherent combination of the contributions
from the first two pairs of eigenvectors of the reduced density matrix.
The two states in each pair are even and odd linear combinations to
the angular momentum vector with respect of the short-medium plane
of the triaxial shape. Their probability distributions are almost identical.

We introduced  the decoherence measures $\Delta_K$ and $\Delta_{\textrm{SSS}}$
which compare the probability distributions of the reduced density matrix with
the ``purified" distributions calculated from the square of the reduced
density matrix. Their relative values scatter between 0.1 to 0.2
at low spin and between 0.1 and 0.3 at high spin. The values indicate
a partially coherent density matrix, which is consistent with the plots
of the distributions for each state.

The second perspective starts with the eigenstates of the collective
Hamiltonian (CH) introduced in Ref.~\cite{Q.B.Chen2024PRC_v1},
which accounts for the entanglement by assuming particle angular
momentum $\bm{j}$ adiabatically follows the total angular momentum
$\bm{J}$ such that the total energy is as small as possible.
The $K$ and SSS probability distributions from the PTR reduced
density matrix and sub-density matrixes were compared with
the pertaining distributions from the CH eigenstates.

It is found that for $^{135}$Pr the adiabatic approximation works
well for the $n=$ 0 and 1 wobbling bands, marginally
for $n=2$ band up to instability of the TW mode. Above it fails
like for for $n=3$ band, which indicates that adiabaticity
progressively deteriorates with $n$. When the CH provides a reasonable
description, its distributions agree approximately with the scaled
distributions calculated from the eigenstate pair of the reduced
density matrix with the largest probability. This means that the collective
wave function cannot provide a better description than the first
pair of eigenstates. The remaining incoherent contributions become
increasingly  important with $n$.

For $^{130}$Ba it is found that  the CH provides a fair description of
all wobbling bands $n=$0, 1, 2, 3, where the deviations of the CH dstributions from
the PTR ones increase  with $n$ as well.
Within the considered spin range, which extends only
to the $I$ values where the TW mode become unstable,
the adiabatic approximation is more accurate for two quasiparticles coupled to
the rotor than for one.  The distributions
$P_{\textrm{CH}}(K)$ agree with the corresponding scaled distributions
calculated from the pair of sub-matrices with the largest weight in the
PTR density matrix, which is found for $^{135}$Pr alike when
the adiabatic approximation is applicable.
The SSS distributions $P_{\textrm{CH}}(\phi_J)$ show a remarkable spin dependence.
At low $I$ they are also close to the scaled distributions with
the maximal weight.  With increasing spin $P_{\textrm{CH}}(\phi_J)$
approaches the distribution $P(\phi_J)$ calculated from the complete
reduced density.

In the TW regime the accuracy of the adiabatic approximation improves with
spin because the wobbling energies go down. The smaller the excitation energies ---
the slower the collective motion --- the better the adiabaticity.
We attribute the spin dependence of $P_{\textrm{CH}}(\phi_J)$ to this improvement
of the approximation. The different $I$ dependence of $P_{\textrm{CH}}(K)$
is explained by the supposition that the motion is slow in $\phi_J$ and
corresponding SSS basis states are as narrow as possible, which implies that
they are distributed over the full $K$ range.

The electric $E2$ transition probabilities
and spectroscopic quadrupole moments reflect the spin dependence of
$P_{\textrm{CH}}(\phi_J)$. At low $I$ they agree with the values
from the pair of sub-density  matrices with the largest weight,
and they approach the PTR values from the complete reduced density
matrix with increasing $I$. This is expected because the quadrupole operator
is nearly local on the $\phi_J$ representation.

The conclusions obtained in the present study are to some extend
specific to the PTR model. Nevertheless, many of the observed features
result from the relative small dimensions of the entangled Hilbert spaces
should apply to other coupled systems with comparable dimensions.
It will be interesting to extend the present study to other exotic
rotational modes, e.g., nuclear low-energy quadrupole modes~\cite{Bohr1975}
and chiral rotation~\cite{Frauendorf1997NPA} in the near future.

\section*{Acknowledgements}

One of the authors (QBC) thanks Professor Lei Ma at
East China Normal University for helpful discussions
on the concept of entanglement entropy. This work was
supported by the National Natural Science Foundation of
China under Grant No.~12205103.


\appendix

\section{Structure of the density matrices}\label{app1}

\begin{figure}[ht]
  \begin{center}
    \includegraphics[width=0.82 \linewidth]{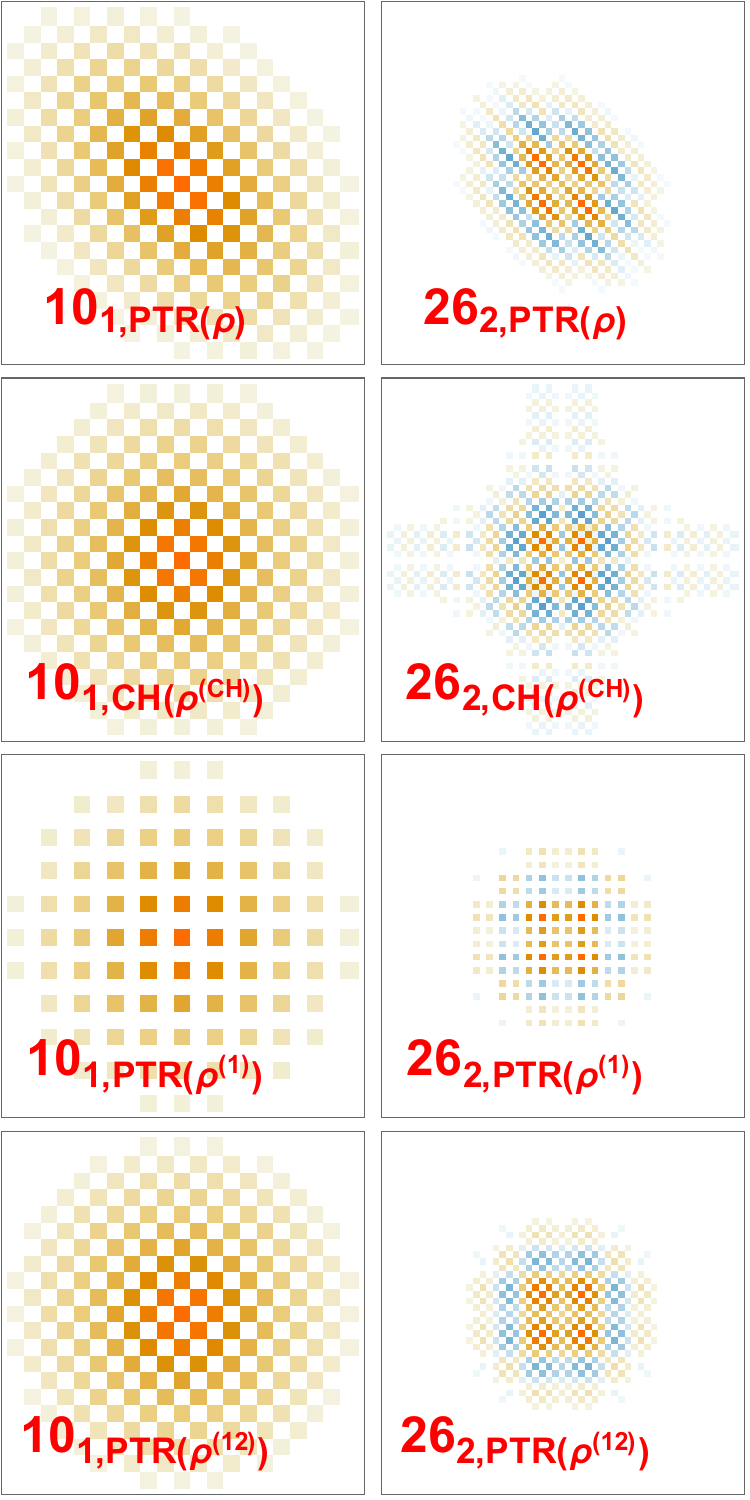}
    \caption{(Color online) Matrix plots of $\rho_{KK^\prime}$
    for the $10_1$ and $26_2$ states in $^{130}$Ba. First row:
    full density matrices $\rho_{KK^\prime}$ of PTR
    for the $10_1$ (left, $C_{l_1}=2.39$, $r_{C_{l_1}}
    =C_{l_1}/C_{l_1}^{(\textrm{max})}=C_{l_1}/2I
    =11.95\%$) and $26_2$ (right, $2.79$, $5.37\%$) states.
    Second row: density matrices $\rho_{KK^\prime}^{\textrm{(CH)}}$
    (combination of even- and odd-$K$ solutions) of CH for the
    $10_1$ (left, $3.19$, $15.95\%$) and $26_2$ (right, $3.94$, $7.58\%$)
    states. Third row: sub-density matrices $\rho_{KK^\prime}^{(1)}$
    of PTR for the $10_1$ (left, $1.27$, $6.35\%$) and
    $26_2$ (right, $1.39$, $2.68\%$) states.
    Fourth row: sub-density matrices $\rho_{KK^\prime}^{(12)}$
    of PTR for the $10_1$ (left, $2.54$,
    $12.70\%$) and $26_2$ (right, $2.81$,
    $5.40\%$) states. For each plot,
    the $K$ values are ordered as $K=-I$, $-I+1$, ..., $+I$
    for both rows and columns. The color scale shows the zero values
    as white, the positive values as red, and the negative
    values as blue.}
    \label{f:rho_K_PTR_CH_130Ba}
  \end{center}
\end{figure}

\begin{figure}[ht]
  \begin{center}
    \includegraphics[width=0.82 \linewidth]{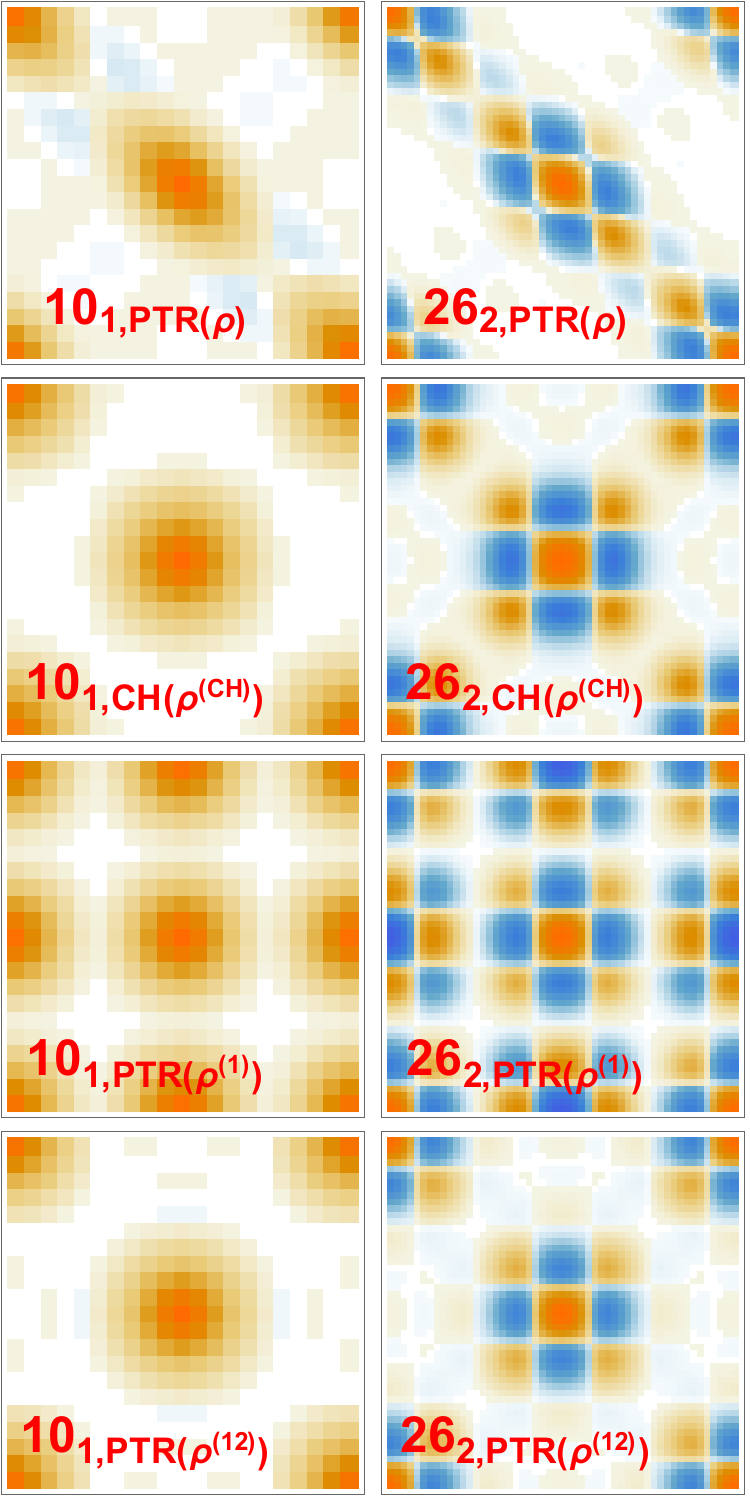}
    \caption{(Color online) Matrix plots of $\rho_{\phi_m,\phi_n}$
    for the $10_1$ and $26_2$ states in $^{130}$Ba. First row: generated
    from the full density matrix $\rho_{KK^\prime}$ of PTR for the
    $10_1$ (left, $C_{l_1}=2.75$, $r_{C_{l_1}}=C_{l_1}/C_{l_1}^{(\textrm{max})}
    =C_{l_1}/2I=13.75\%$) and $26_2$ (right, $9.16$, $17.62\%$) states.
    Second row: generated from the density matrix $\rho_{KK^\prime}^{\textrm{(CH)}}$
    (combination of even- and odd-$K$ solutions) of CH for the $10_1$ (left,
    $4.03$, $20.15\%$) and $26_2$ (right, $18.22$, $35.04\%$)
    states. Third row: generated from the sub-density matrix
    $\rho_{KK^\prime}^{(1)}$ of PTR for the $10_1$ (left, $3.25$, $16.25\%$)
    and $26_2$ (right, $9.16$, $17.62\%$) states.
    Fourth row: generated from the sub-density matrix
    $\rho_{KK^\prime}^{(12)}$ of PTR for the $10_1$ (left, $2.86$,
    $14.30\%$) and $26_2$ (right, $8.86$, $17.05\%$) states.
    For each plot, the angles $\phi_n=2n\pi/(2I+1)$ are ordered as $n=-I$,
    $-I+1$, ..., $+I$ for both rows and columns. The color coding
    is the same as Fig.~\ref{f:rho_K_PTR_CH_130Ba}.}
    \label{f:rho_SSS_PTR_CH_130Ba}
  \end{center}
\end{figure}

In this Appendix, we compare $\rho^{(\textrm{CH})}$ with
the mixed matrix $\rho$ and the pure sub-matrix $\rho^{(12)}$.
The density matrices $\rho_{KK'}$ and $\rho_{\phi_J,\phi_J'}$ are
related to each other by the Fourier transform between the $K$ and $\phi_J$
representations. The matrix plots in Figs.~\ref{f:rho_K_PTR_CH_130Ba}
and \ref{f:rho_SSS_PTR_CH_130Ba} illustrate the relation
for the PTR states $10_1$ and $26_2$ of $^{130}$Ba. For
each panel in Fig.~\ref{f:rho_K_PTR_CH_130Ba}, the $K$ values
are ordered as $K=-I$, $-I+1$, ..., $+I$ for both rows and columns.
There are no matrix elements that connect even-$K$ basis states with
odd-$K$ ones, which reflects the $\textrm{D}_2$ symmetry of the PTR
Hamiltonian and of the CH. Accordingly, the eigenstates of the reduced
density matrix appear in pairs of even-$K$ and odd-$K$ with nearly the same
probability $p_m\approx p_{m+1}$. The third row's panels show
$\rho_{K K^\prime}^{(1)}$ of the first eigenstate of the PTR reduced
density matrix, which includes only even-$K$ components. The second
solution with odd-$K$ is added in the fourth row's panels which
show $\rho_{K K^\prime}^{(12)}$.
As expected from the density of the plots,
its $C_{l_1}$ norm is twice of that of $\rho_{K K^\prime}^{(1)}$.
The second row's panels show  $\rho_{K K^\prime}^{(\textrm{CH})}$, which is
the combination of the  even-$K$ and odd-$K$ eigenstates of the CH.
The matrix is nearly the same as $\rho_{K K^\prime}^{(12)}$, which holds for
the diagonal shown in Fig. \ref{f:PK_P12_CH_130Ba} as well. The two
matrices differ from $\rho_{K K^\prime}$ shown in the in the
first row's panels, which includes $\rho_{K K^\prime}^{(34)}$ and the higher terms.

In our previous study \cite{Q.B.Chen2024PRC_v1}, we
considered only the even-$K$ solutions. The odd-$K$ solutions have
very similar energies and $E2$ matrix elements. Thus, to compare
the results of PTR, the combination of even-$K$ and odd-$K$ solutions with
equal weight of 1/2 for $10_1$ and $26_2$ states are shown in the second row's panels of
Fig.~\ref{f:rho_K_PTR_CH_130Ba}.

As discussed in Ref.~\cite{Q.B.Chen2024PRC_v1}, the discrete SSS
states form an orthonormal set, which is comprised of $2I+1$ angles
of the angular momentum with respect to the 3-axis.
The Eq.~(\ref{eq:K-SSS}) transforms the basis from $|K\rangle$ states
to the discrete SSS states $|\phi_n\rangle$. Figure~\ref{f:rho_SSS_PTR_CH_130Ba}
shows matrix plots of the density matrices $\rho_{\phi_m, \phi_n}$
generated from the corresponding $\rho_{KK^\prime}$ in Fig.~\ref{f:rho_K_PTR_CH_130Ba}.
These matrices are calculated using Eq.~(\ref{eq:rhophiphi}), where
the discrete angles in the plots are defined as $\phi_n=2n\pi/(2I+1)$,
with the order of $n=-I$, $-I+1$, ..., $+I$ for rows and columns. The
diagonal matrix elements of the respective matrices are a discrete
selection of the pertaining functions $P(\phi_J)$ and $P_{\textrm{CH}}(\phi_J)$
in Fig.~\ref{f:SSS_rho12Nor_CH_130Ba}. The matrix elements in
the corners of the panels reflect the symmetry of the PTR solutions.

The plot of the first eigenstate of the density matrix $\rho_{\phi_m, \phi_n}^{(1)}$
(third row) well agrees with the plot of first eigenstate of the CH, which
both involves only the even-$K$ basic states. We do not show the plot of
the second eigenstate of the density matrix. It well agrees with
the plot of second eigenstate of the CH (not shown), which involve only
odd-$K$ basis states. The plots agree with the ones generated from
the even-$K$ matrices, except that the matrix elements in regions around
$(\phi_m, \phi_n)=(0, \pm \pi)$ and $(\pm \pi,0)$ have the opposite sign.
These regions cancel each other when adding $\rho_{\phi_m, \phi_n}^{(2)}$
to $\rho_{\phi_m, \phi_n}^{(1)}$, which makes
$\rho_{\phi_m, \phi_n}^{(12)}$ (fourth row) more sparse than
$\rho_{\phi_m, \phi_n}^{(1)}$ (third row). Correspondingly,
the $l_1$ norm $C_{l_1}$ of $\rho_{\phi_m, \phi_n}^{(12)}$ is smaller than
that of $\rho_{\phi_m, \phi_n}^{(1)}$, and the entropy increases from 0 to
$2\times\ln 2/\ln 21=0.46$. The matrix $\rho_{\phi_m, \phi_n}^{(12)}$
is nearly identical with $\rho_{\phi_m, \phi_n}^{(\textrm{CH})}$ (second row),
which holds for the diagonal in Fig.~\ref{f:SSS_rho12Nor_CH_130Ba} as well.
The reduced density matrix of the PTR state $10_1$ (left of the first row)
is also more sparse than the one generated from CH eigenstate (left of
the second row), which corresponds to its larger entropy of $0.5\times\ln21$.
The reduction is generated by the terms $\rho_{\phi_m, \phi_n}^{(34)}$
and higher.

As seen in Fig.~\ref{f:SSS_rho12Nor_CH_130Ba}, for $I=10$ the diagonal
matrix elements  \mbox{$\rho_{\phi_n,\phi_n}^{(12)}/(p_1+p_2)$}
and \mbox{$\rho^{(\textrm{CH})}_{\phi_n,\phi_n}$} are approximately equal.
Both deviate from the full distribution $\rho_{\phi_n,\phi_n}$,
mainly by \mbox{$\rho_{\phi_n,\phi_n}^{(34)}/(p_3+p_4)$}. With increasing $I$,
the diagonal matrix elements \mbox{$\rho^{\textrm{(CH)}}_{\phi_n,\phi_n}$}
deviate from \mbox{$\rho_{\phi_n,\phi_n}^{(12)}/(p_1+p_2)$}
and approach $\rho_{\phi_n,\phi_n}$.

The sub-matrices of the  PTR state $26_2$ in the right  column of
Fig.~\ref{f:rho_SSS_PTR_CH_130Ba} merge in an analog way as discussed
for the matrices in the left column. Although the diagonal matrix element
matrix elements of  \mbox{$\rho^{\textrm{(CH)}}_{\phi_m,\phi_n}$} are very
close to the ones of $\rho_{\phi_m,\phi_n}$ (see Fig.~\ref{f:SSS_rho12Nor_CH_130Ba}),
the full matrices differ from each other. The elements of reduced
density matrix (right of the first row) are concentrated around
the diagonal, which reflects the relative large entropy of
$0.6\times\ln21$. The density matrix of the corresponding
paired eigenstates of the CH (right of the second row) has
elements that are more spread over its range, which is needed
to satisfy the purity condition $\rho^2=\rho$ for each of
the two eigenstates. The entropy is $2\ln2=0.46\ln21$, because
a pair of matrices is combined.

\section{Quadrupole moment matrix elements}
\label{app2}

In this Appendix, we provide the results of quadrupole moment
matrix elements.

\begin{figure}[!ht]
  \begin{center}
    \includegraphics[width=0.90 \linewidth]{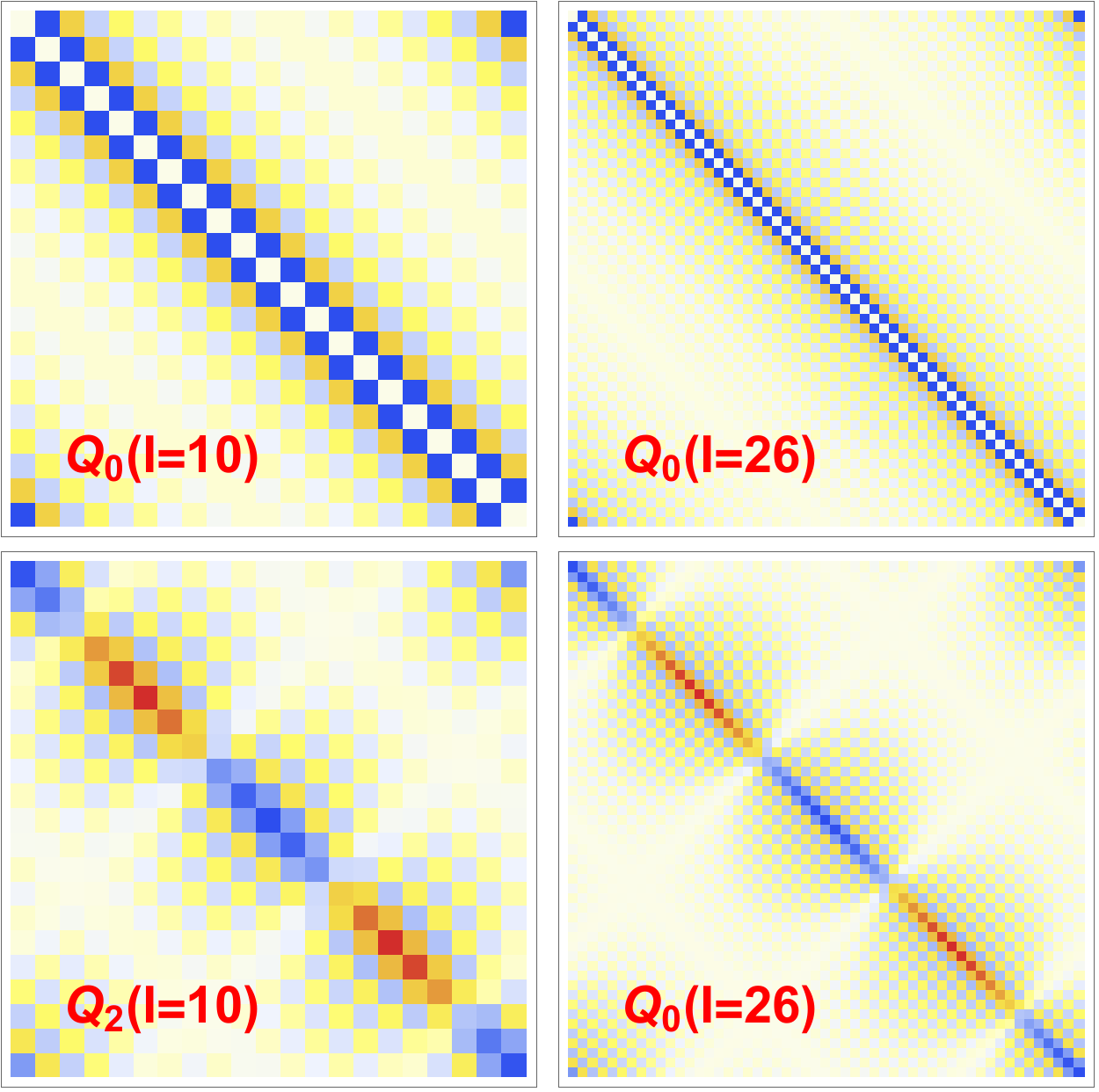}
    \caption{(Color online) Matrix elements of $\hat{Q}_{0}$ (upper panels)
    and $\hat{Q}_{2}+\hat{Q}_{-2}$ (lower panels) operators on the basis
    $|\phi_n\rangle$ with $\phi_n=2n\pi/(2I+1)$ and $n=-I$, $-I+1$, ..., $+I$
    for $I=10$ (left panels) and $26$ (right panels) in $^{130}$Ba. The color
    scale shows the zero values as white, the positive values as
    red, orange, and yellow, and the negative values as blue and light blue.}
    \label{f:Q0_Q2}
  \end{center}
\end{figure}

The matrix elements of quadrupole operators $\hat{Q}_{0}$ and
$\hat{Q}_{2}+\hat{Q}_{-2}$ on the basis of $|\phi_n\rangle$ are shown
in Fig.~\ref{f:Q0_Q2}. In the $K$-basis, the transition matrix elements
of $\hat{Q}_{0}$ and $\hat{Q}_{2}+\hat{Q}_{-2}$ are calculated
as~\cite{Bohr1975}
\begin{align}
 &\langle I^\prime, K^\prime|\hat{Q}_{0}|I,K\rangle
  =Q_{20}^\prime
  \langle IK20|I^\prime K^\prime \rangle, \quad K^\prime=K, \\
 &\langle I^\prime, K^\prime|\hat{Q}_{2}+\hat{Q}_{-2}|I,K\rangle
  =Q_{22}^\prime
  \Big[\langle IK22|I^\prime K^\prime\rangle
   +\langle IK2-2|I^\prime K^\prime\rangle\Big], \notag\\
   &\quad\quad\quad\quad\quad\quad\quad\quad\quad\quad\quad\quad\quad\quad\quad
   K^\prime=K\pm 2,
\end{align}
with the intrinsic quadrupole moment $Q_{20}^\prime=Q_0^\prime\cos\gamma$
and $Q_{22}^\prime=Q_0^\prime\sin\gamma/\sqrt{2}$
($Q_0^\prime$ is an empirical quadrupole moment that
is related to the axial deformation $\beta$). In addition,
$I$ is the initial spin value and the final spin $I^\prime$
can be taken as $I$, $I\pm 1$, and $I\pm 2$.
In the discrete SSS representation $|\phi_n\rangle$, which is
complete orthonormal, we need transform the above matrix elements
to the SSS basis using the following relationship~\cite{Q.B.Chen2024PRC_v1}
\begin{align}
 |I, \phi_n \rangle &=\frac{1}{\sqrt{2I+1}}\sum_{K=-I}^I e^{iK\phi_n}|I, K\rangle.
\end{align}
In detail, the matrix elements of quadrupole operators
on the basis of $|\phi_n\rangle$ are calculated as
\begin{align}\label{eq:Q0Q2}
 &\quad \langle I^\prime, \phi_m|\hat{Q}_{0}|I, \phi_n\rangle \notag\\
 &=\frac{1}{\sqrt{2I^\prime+1}}\frac{1}{\sqrt{2I+1}}Q_{20}^\prime
  \sum_{K} e^{i(K\phi_n-K \phi_m)}\langle IK20|I^\prime K\rangle,\\
 &\quad \langle I^\prime, \phi_m|\hat{Q}_{2}+\hat{Q}_{-2}|I, \phi_n\rangle \notag\\
 &=\frac{1}{\sqrt{2I^\prime+1}}\frac{1}{\sqrt{2I+1}}Q_{22}^\prime
  \sum_{K}\Big[
  e^{i[K\phi_n-(K+2) \phi_m]}\langle IK22|I^\prime K+2\rangle\notag\\
  &\quad +e^{i[K\phi_n-(K-2) \phi_m]}\langle IK2-2|I^\prime K-2\rangle\Big].
\end{align}
Since the matrix elements for the cases of
$I^\prime=I$, $I\pm 1$, and $I\pm 2$ look similar,
Fig.~\ref{f:Q0_Q2} only shows the matrix elements for the case of
$I^\prime=I$. For each plot, the angles $\phi_n=2n\pi/(2I+1)$ are
ordered as $n=-I$, $-I+1$, ..., $+I$ for both rows and columns.
Figure~\ref{f:Q0_Q2} illustrates that the matrix elements of
$\hat{Q}_{0}$ and $\hat{Q}_{2}+\hat{Q}_{-2}$ on the basis of $|\phi_n\rangle$
are much more denser than those on the basis of $|K\rangle$, where
the $\hat{Q}_{0}$ and $\hat{Q}_{2}+\hat{Q}_{-2}$ have matrix elements
only when $K^\prime=K$ and $K^\prime=K\pm 2$, respectively.

However, it should be noted that the only matrix elements
close to the diagonal are significant when evaluating
the trace of the product of a quadrupole matrix with a density
matrix, because  those further away have much smaller
values and display an oscillatory behavior. To a good approximation
one can replace the traces by integrals over the various
probability densities times the classical values
\begin{align}\label{eq:Q0Q2class}
 \hat{Q_0}=-Q^\prime_{20}/2, \quad
 \hat{Q}_{2}+\hat{Q}_{-2} =Q^\prime_{22}\sqrt{3/2}\cos2\phi_J.
\end{align}

The fact that the near-diagonal matrix elements of
$\rho^{\textrm{(CH)}}_{\phi_m,\phi_n}$ are close to the ones of
$\rho_{\phi_m,\phi_n}$ provides an intuitive interpretation of
structure of the state. However, the CH wave function describes
the $E2$ matrix elements between the PTR states only with
the accuracy of $P_{12}(K)/(p_1+p_2)$ because the intrinsic
quadrupole operator $\hat Q_0$ has only diagonal matrix elements,
and the matrix elements of $\hat Q_2+\hat Q_{-2}$
are restricted to the region $\Delta K=\pm 2$ around the diagonal
of $\rho_{K K^\prime}^{(12)}$. The $E2$ matrix elements between the PTR
states do not depend on the chosen basis. The closeness of near-diagonal
matrix matrix elements of $\rho^{\textrm{(CH)}}_{\phi_m,\phi_n}$ to
the ones of $\rho_{\phi_m,\phi_n}$ is offset by wide distribution of the
matrices $\hat Q_0$ and $\hat Q_2+\hat Q_{-2}$ in
the $\phi_n$ basis when calculating
$\mathrm{Tr}\left(\hat Q_0 \rho^{(\textrm{CH})}\right)$
and $\mathrm{Tr}\left[(\hat Q_{2}+ \hat Q_{-2})\rho^{(\textrm{CH})}\right]$.
The expressions and an illustration are given in the Appendix~\ref{app1}.
Nevertheless as seen in Fig.~\ref{f:Q0_Q2}, the eigenstates of
the CH quite well reproduce the $B(E2)$ values of the PTR states.


%

\end{CJK}

\end{document}